\documentclass{optica-article}

\journal{opticajournal} 

\articletype{Research Article}

\usepackage{amsmath}
\usepackage{cases}
\usepackage{empheq}
\usepackage{subfig}
\usepackage{bm}
\usepackage{color}
\usepackage{xspace}
\usepackage{overpic}
\usepackage{tikz}

\newcommand{\openfoam}{Open\nolinebreak\hspace{-.2em}{\color{blue}\Large$\nabla$}\nolinebreak\hspace{-.2em}FOAM\textsuperscript{\textregistered}\xspace}

\begin{document}
\usetikzlibrary{math}

\title{Stabilized POD Reduced Order Models for convection-dominated incompressible flows
}

\author{Pierfrancesco Siena,\authormark{1,*} Michele Girfoglio,\authormark{1,*} Annalisa Quaini,\authormark{2,*} and Gianluigi Rozza \authormark{1,*}}

\address{\authormark{1}SISSA, International School for Advanced Studies, Mathematics Area, mathLab, via Bonomea 265, I-34136 Trieste, Italy \\
\authormark{2}University of Houston, Department of Mathematics, 3551 Cullen Blvd, 77204, Houston TX, USA
}

\email{\authormark{*}psiena@sissa.it, mgirfogl@sissa.it, aquaini@central.uh.edu,grozza@sissa.it} 


\begin{abstract*} 
We present a comparative computational study of two stabilized Reduced Order Models (ROMs) for the simulation of convection-dominated incompressible flow (Reynolds number of the order of a few thousands). Representative solutions in the parameter space, which includes
either time only or time and Reynolds number,
are computed with a Finite Volume method and used to generate a reduced basis via
Proper Orthogonal Decomposition (POD). Galerkin projection of the Navier-Stokes equations onto the reduced space is used to compute the ROM solution. To ensure computational efficiency,
the number of POD modes is truncated 
and ROM solution accuracy is recovered through two stabilization
methods: i) adding a global constant artificial viscosity to the reduced dimensional model, and ii) adding a different value of artificial viscosity for the different POD modes.
We test the stabilized ROMs for fluid flow in an idealized
medical device consisting of a conical convergent, a narrow throat, and a sudden expansion. Both stabilization methods significantly
improve the ROM solution accuracy
over a standard (non-stabilized) POD-Galerkin model. 
\end{abstract*}

\section{Introduction}

For a few decades, Reduced Order Models (ROMs)  have become the 
methodology 
of choice to reduce the computational cost when traditional 
computational techniques, e.g., Finite Element methods and Finite Volume methods, 
have to be carried out for several parameter values, 
as is the case in uncertainty quantification, optimal control, and inverse problems. See, e.g., \cite{peter2021modelvol1,benner2020modelvol2,benner2020modelvol3,hesthaven2016certified,rozza2008reduced}
for reviews on reduced order modeling. 
ROMs rely on traditional computational methods, also called Full Order Models (FOMs),
to build a lower-dimensional approximation of the solution space 
that captures the essential behavior of the solutions in a given parameter space.
In this paper, the parameters will be time and the Reynolds number, which is 
a parameter related to the physics of the problem.
The construction of the lower-dimensional approximation
is done in a preliminary phase, called offline, during which 
one generates a database of several FOM solutions associated to given times and physical parameter values. 
The database of FOM solutions 
is used to generate a reduced basis, which is (hopefully much) 
smaller than the high-dimensional FOM basis but still preserves the essential features of the system. 
One of the most used techniques to generate the reduced bases space 
is Proper Orthogonal Decomposition (POD), which extracts the dominant modes from the FOM database. 
In a second phase, called {online}, one uses this reduced basis 
to quickly compute the 
solution for newly specified times and parameter values. Note that, 
while the offline phase is 
performed only once, 
the online phase is performed as many times as needed.

ROMs are very efficient surrogate models when 
the number of reduced basis functions is small (i.e., $\mathcal{O}$(10)), as is typically the case for diffusion-dominated flows. Unfortunately, though, a large number of basis functions is needed to capture the essential features of convection-dominated flows. If one 
retains such a large number of modes for the ROM to be accurate, then
the computational efficiency suffers. If the number of modes is otherwise kept low, a severe loss of information
hinders the accurate reconstruction of the solution. 
In fact, projection-based ROMs of turbulent flows are affected by energy stability problems
related to the fact that 
{POD} retains the modes biased toward large,
high-energy scales, while the turbulent kinetic energy is dissipated at the level of
the small scales.  A possible way to tackle this challenging problem 
is to introduce dissipation via a closure model \cite{wang2012proper, AHLS88, ahmed2021closures} or stabilization~\cite{carlberg2011efficient,grimberg2020stability,baiges2012explicit,reyes2020projection, parish2020adjoint}.


We consider two stabilization techniques that introduce
artificial viscosity in the online phase of a 
POD-Galerkin ROM of the incompressible Navier-Stokes equations. 
By POD-Galerkin ROM, it is meant that that projection of the system equations
onto the reduced space
is used to compute the ROM solution during the online phase. 
The first stabilization method we consider is Heisenberg stabilization \cite{bergmann2009enablers,san2015stabilized}, 
where a global constant artificial viscosity is added in the reduced dimensional model. The second method is a variation of the first, where a different value
of artificial viscosity is used for the different modes \cite{cazemier1997proper,san2015stabilized}. 
We call this method
coefficient-dependent viscosity where coefficient refers to the POD
mode index. Specifically, a small amount of dissipation is introduced to 
the lowest index mode and it increases proportionally to the 
mode index for every other POD mode. So far, these approaches have been tested on academic problems: 2D square
cylinder wake flow in laminar regime, 2D lid-driven cavity flow in turbulent regime and mid-latitude simplified oceanic flow (a 2D problem too).

We test the proposed stabilized POD-Galerkin ROMs 
on a realistic 3D problem: flow at different Reynolds numbers 
through a nozzle that
contains all the features commonly encountered in medical devices (flow contraction and expansion,
recirculation zones, etc.). See Fig.~\ref{fig:fda}.
This benchmark test was originally proposed by the U.S Food and Drug 
Administration (FDA), which asked three independent laboratories 
to perform flow visualization experiments
on fabricated nozzles for different flow rates up to Reynolds number 6500
\cite{hariharan}. 
This led to the creation of measurement datasets that can 
be used to validate Computational Fluid Dynamics (CFD) simulations. 
As FOM, we use a Finite Volume method discussed in 
\cite{girfoglio2019finite} and implemented in  
\openfoam \cite{weller1998tensorial,jasak1996error}.
The stabilized POD-Galerkin ROMs are implemented
in ITHACA-FV \cite{Stabile2017CAIM,stabile2018finite}, an open-source finite volume C++ library\footnote{\url{https://ithaca-fv.github.io/ITHACA-FV/)}}, and compared for Reynolds numbers 2000 (transitional regime), 3500, 5000, and 6500 (progressively more developed turbulent regimes). 

The rest of the paper is organized as follows. Sec.~\ref{sec:fom} states 
the Navier-Stokes equations and briefly describes the Finite Volume method used as FOM. 
Sec. \ref{sec:ROM} presents the proposed stabilized POD-Galerkin ROMs. 
Sec.~\ref{sec:num_res} presents numerical results obtained 
for the FDA benchmark.
Finally, Sec.~\ref{sec:concl} summarises the conclusions.

\section{Full order model}
\label{sec:fom}
\subsection{Problem formulation}
The  flow of an incompressible, viscous, and Newtonian fluid in domain $\Omega$ and over time interval $(t_0, T)$ is modeled with the {parametrized} time-dependent incompressible Navier-Stokes equations:
{\small
\begin{empheq}[left=\empheqlbrace]{alignat=4}
       \frac{\partial \bm{u}(\bm{x}, t, \bm \mu)}{\partial t} + \nabla \cdot (\bm{u}(\bm{x}, t, \bm \mu) \otimes \bm{u}(\bm{x}, t, \bm \mu)) - \nu \Delta\bm{u}(\bm{x}, t, \bm \mu) + \nabla p(\bm{x}, t, \bm \mu) & = & 0 & \quad \mbox{ in } \Omega \times (t_0, T), \label{eq:NS1} \\
      \nabla \cdot \bm{u}(\bm{x}, t, \bm \mu) & = & 0 & \quad \mbox{ in } \Omega \times (t_0, T), \label{eq:NS2}
\end{empheq}}
where $\bm{u}(\bm{x}, t, \bm \mu)$ is the velocity vector, $p(\bm{x}, t, \bm \mu)$ is the pressure, $\nu$ is the kinematic viscosity, and {$\bm \mu$ is the parameter vector}, i.e., a vector containing all the physical parameters, other than time and viscosity, the problem depends upon. {In the rest of the paper, for a lighter notation, we will omit the spatial-temporal dependence of the variables and the dependence on the parameters $\bm\mu$ as well.} 
To characterize the flow regime, we introduce
the Reynolds number, which is defined as:
\begin{equation}\label{eq:Re}
    Re = \frac{U L}{\nu},
\end{equation}
where $U$ and $L$ are characteristic macroscopic speed and length for the flow. For an internal flow in a cylindrical pipe, $U$ is the mean sectional
velocity and $L$ is the pipe diameter. For large Reynolds numbers, inertial forces are dominant over viscous
forces and vice versa for small Reynolds numbers.

Let $\Gamma_i$ be the inlet boundary, $\Gamma_w$ the wall, and $\Gamma_o$ the outlet, such that $\Gamma_i \cup \Gamma_w \cup \Gamma_o = \partial\Omega$ and $\Gamma_i \cap \Gamma_w \cap \Gamma_o = \emptyset$. 
We impose a non-homogeneous
{Dirichlet boundary conditions at the inlet, no-slip conditions on the wall and a homogeneous Neumann condition at the outlet:}
\begin{equation}
\begin{cases}
    \bm u = \bm u_D &\quad \text{in $\Gamma_i\times (t_0, T)$}, \\
    \bm u  = \bm 0 &\quad \text{in $\Gamma_w\times (t_0, T)$}, \\
    (2\mu \nabla \bm u - p I)\bm n = \bm 0 &\quad \text{in $\Gamma_o\times (t_0, T)$}.
\end{cases}
\label{BC0}
\end{equation}
In \eqref{BC0}, $\bm u_D$ is a given inlet velocity profile and $\bm n$ is the unit normal outward vector to the boundary. For the results in Sec.~\ref{sec:num_res}, $\bm u_D = \bm u_D (\bm x)$, i.e., the velocity profile is space dependent and time independent. 

\subsubsection{Finite volume approximation}
To discretize problem \eqref{eq:NS1}-\eqref{eq:NS2} in space, we adopt a Finite Volume (FV) method. For this purpose, we divide domain $\Omega$ into control volumes $\Omega_i$, for $i = 1,\dots,N_h$, such that $\bigcup_{i=1}^{N_h} \Omega_i=\Omega$ and $ \Omega_i\cap\Omega_j=\emptyset$ for $i,j = 1,\dots,N_h$ and $i\ne j$.

By integrating equation \eqref{eq:NS1} over control volume $ \Omega_i$ and applying the Gauss-divergence theorem, we obtain:
\begin{equation}
    \int_{\Omega_i} \frac{\partial \bm{u}}{\partial t} d\Omega +  \int_{\partial\Omega_i} (\bm{u} \otimes \bm{u}) \cdot d\bm{A} - \nu \int_{\partial\Omega_i} \nabla \bm{u} \cdot d\bm{A} + \int_{\partial\Omega_i} p d\bm{A} = 0,
\end{equation}
{where $\partial\Omega_i$ is the boundary of the control volume $\Omega_i$, $d\Omega$ is an infinitesimal volume element, and $d\bm{A}$ is an infinitesimal surface element}. 
Boundary terms are approximated with the sum  
on each face $j$ of cell $\Omega_i$ as follows:
\begin{align}
    & \int_{\partial\Omega_i} p d\bm{A} \approx \sum_{j} p_j \bm{A_j}, \quad & \mbox{(Gradient term)} \\
    & \int_{\partial\Omega_i} (\bm{u} \otimes \bm{u}) \cdot d\bm{A} \approx \sum_{j} (\bm{u_j} \otimes \bm{u_j}) \cdot \bm{A_j} = \sum_{j} (\bm{u_j} \cdot \bm{A_j}) \bm{u_j}, \quad & \mbox{(Convective term)} \\
    & \int_{\partial\Omega_i} \bm{\nabla u} \cdot d\bm{A} \approx \sum_{j} (\bm{\nabla u})_j \cdot \bm{A}_j, \quad & \mbox{(Diffusion term)}
\end{align}
where $p_j$, $\bm{u_j}$, and $(\bm{\nabla u})_j$ denote the pressure, the velocity and the gradient of the velocity at the centroid of the face $j$ and $\bm{A_j}$ is the outward surface vector of the face $j$. The values at the face centroids are obtained by linear interpolation of values from cell centres to face centres. {A \emph{Gauss linear} second order accurate scheme is used both for convective and diffusion terms.}  

For the time discretization, we adopt an equi-spaced grid with time step $\Delta t = {(T-t_0)}/{N_T}$, where $N_T$ is the number of time intervals in the grid. At each time $t^n=t_0+n\Delta t$ for $n=1,\dots,N_T$, the time derivative of $\bm{u}$ is approximated with a backward second order scheme:
\begin{equation}
    \frac{\partial \bm{u}}{\partial t} \approx \frac{1}{\Delta t}\left( \frac{3}{2}\bm{u}^{n+1} -2 \bm{u}^{n} + \frac{1}{2}\bm{u}^{n-1}\right),
\end{equation}
where $\bm{u}^n$ is the approximation of the velocity at the time step $t^n$.

Let $\mathbf{b}^{n+1}={(4\mathbf{u}^n-\mathbf{u}^{n-1})}/{(2\Delta t)}$. 
Then, the discretized form of eq.~\eqref{eq:NS1} is:
\begin{equation}
\frac{3}{2\Delta t} \bm{u}_i^{n+1}+\sum_j \phi_j \bm{u}_{i,j}^{n+1}-\nu\sum_j(\nabla \bm{u}_i^{n+1})_j\cdot \bm{A}_j+\sum_j p_{i,j}^{n+1}\bm{A}_j=\bm{b}_i^{n+1}, \quad \phi_j = \bm{u}_j^{n} \cdot \bm{A}_j,
\label{FV_form}
\end{equation}
where $\bm{u}_i^{n+1}$ and $\bm{b}_i^{n+1}$ are the average velocity and the source term in $\Omega_i$, and  $\bm{u}_{i,j}^{n+1}$ and $p_{i,j}^{n+1}$ the velocity and pressure at the centroid of face $j$ normalized by the volume of $\Omega_i$.  
{Note that we adopt a first order extrapolation for the convective velocity although we use a backward differentiation formula of order 2 for the time discretization of problem \eqref{eq:NS1}-\eqref{eq:NS2} because this is what OpenFOAM solvers do.} 

The discrete version of eq.~\eqref{eq:NS2} is derived 
from the semi-discretized form of \eqref{FV_form}, where the pressure term is in continuous form while all other terms are in discrete form. The application of the divergence free constraint and the Gauss-divergence theorem leads to the following equation:
\begin{equation}
\sum_j (\nabla p^{n+1})_j\cdot \bm{A}_j = \sum_j  \Big( -\sum_j \phi_j \bm{u}_{j}^{n+1} + \nu \sum_j(\nabla \bm{u}^{n+1})_j\cdot \bm{A}_j + \bm{b}^{n+1} \Big)_j \cdot \bm{A_j},
\label{FV_form1}
\end{equation}
where $\phi_j$ is the convective flux associated to the velocity $\bm{u}^{n}$ through face $j$ of the control volume $\Omega_i$ defined in \eqref{FV_form}.
More details about the derivation of the equations can be found in \cite{girfoglio2019finite,siena2023fast}. 

The above discrete scheme is implemented in the C++ library \openfoam.
The Pressure Implicit with Splitting of Operators (PISO) algorithm \cite{issa1986solution} is used to decouple the computation of the velocity from the computation of the pressure in coupled problem  \eqref{FV_form}-\eqref{FV_form1}. 
For this reason, we need to modify boundary conditions \eqref{BC0} as follows:
\begin{equation}
\begin{cases}
    \bm u = \bm u_D &\quad \text{in $\Gamma_i\times (t_0, T)$}, \\
    \bm u = \bm 0 &\quad \text{in $\Gamma_w\times (t_0, T)$}, \\
    (\nabla\bm u)\cdot\bm n = \bm 0 &\quad \text{in $\Gamma_o\times (t_0, T)$}, \\
    (\nabla p) \cdot \bm n= \bm 0 &\quad \text{in $(\Gamma_i\cup\Gamma_w)\times (t_0, T)$}, \\
    p = 0 &\quad \text{in $\Gamma_o\times (t_0, T)$}.
\end{cases}
\label{BC}
\end{equation}

The method described in this subsection assumes that the mesh for space discretization is fine enough to capture all the flow structures, i.e., eddies and vortices. When such a refined mesh cannot be afforded with the given computing facilities and one is forced to use a coarser mesh, then an alternative approach, like Large Eddy Simulation (LES), should be used. For the specific benchmark problem considered in Sec.~\ref{sec:num_res}, one easy to implement LES method is presented in \cite{bertagna2016deconvolution,girfoglio2019finite}.

\section{Reduced order model}\label{sec:ROM}

The ROM adopted for this work is the POD-Galerkin approach {\cite{quarteroni2011certified,quarteroni2015reduced,hesthaven2016certified,rowley2004model,iollo2000stability}} with stabilization techniques to recover the effect of the neglected modes \cite{cazemier1997proper,bergmann2009enablers,san2015stabilized}. 
{Here, we briefly recall the main ideas of this methodology. 
}. 

We assume that the solution $(\bm u, p)$ to problem \eqref{eq:NS1}-\eqref{eq:NS2}
can be approximated 
as a linear combination of basis functions $\left(\bm \phi_i(\bm x), \psi_i(\bm x)\right)$ that depend on space only and coefficients $\left(a_i(t, \bm\mu), b_i(t, \bm\mu) \right)$ that depend on time {and $\bm\mu$} only:
\begin{align}
    \bm u (\bm x, t, \bm\mu) \approx \bm u_{\text{r}}(\bm x, t, \bm\mu) = \sum_{i=1}^{N_{\bm u}} a_i(t, \bm\mu) \bm\phi_i(\bm x), 
    \label{u_rb} \\
    p (\bm x, t, \bm\mu) \approx p_{\text{r}}(\bm x, t, \bm\mu) = \sum_{i=1}^{N_p} b_i(t, \bm\mu) \psi_i(\bm x).
    \label{p_rb}
\end{align}
This assumption allows us to decouple the computations into an 
expensive phase (offline) to be performed only once and a cheap phase (online) to be performed for every new time and parameter of interest. 

The specific operations performed during each phase are:
\begin{description}
\item[Offline:] given a set of time instances and {parameters}, the full order solutions (also called snapshots) are computed and collected into a matrix. The lifting function method is adopted to treat non-homogeneous boundary conditions (see Sec.~\ref{sec:lift_func}). The POD algorithm  is used to extract the reduced basis space (see Sec.~\ref{sec:pod}). 
The Pressure Poisson Equation (PPE) approach is adopted to satisfy the inf-sup condition.

%
\item[Online:] Given a set of time {and parameter} values, the corresponding modal coefficients are computed by solving the reduced system. {For stabily, an artificial viscosity is added in that system with different approaches (see Sec.~\ref{sec:galerkin_projection}).} The approximated solution \eqref{u_rb}-\eqref{p_rb} is a linear combination between these coefficients and the POD reduced basis functions computed offline. 
\end{description}


\subsection{Proper orthogonal decomposition}
\label{sec:pod}

To generate a set of orthonormal basis functions in a least-square setting, we use the POD algorithm \cite{eckart1936approximation,hawkins1973generalized,bang2004greedy}. 
Below, we summarize how the algorithm works.

Given a discrete set of time instances $\{ t^1, \dots, t^{{N_T}} \} \subset (t_0, T)$ {and parameters $\{ \bm\mu^1, \dots, \bm\mu^{{N_{\mu}}} \}$}, the corresponding FOM solutions $\Phi=\{p,\bm u\}$ of problem \eqref{eq:NS1}-\eqref{eq:NS2} are stored 
as the columns of a snapshot matrix $S_{\Phi}$: 
{
\begin{equation*}
    \mathcal{S}_{\Phi} = \begin{Bmatrix}
    \Phi(t^1, \bm\mu^1), & \cdots &, \Phi(t^{N_T}, \bm\mu^1), \Phi(t^1, \bm\mu^2), & \cdots &, \Phi(t^{N_T}, \bm\mu^2), & \cdots &, \Phi(t^{N_T}, \bm\mu^{N_{\mu}})
    \end{Bmatrix}.
\end{equation*}}
Let $N_{\Phi} \ll \text{min}(N_{h}, {M})$ be the size of the reduced basis, with ${M = N_T\cdot N_{\mu}}$. We will explain in eq.~\eqref{eq:energy} how $N_{\Phi}$ is chosen.
The reduced basis  
is the solution {$\mathcal{V} = \{\bm\phi_1,\dots,\bm\phi_{N_{\bm \Phi}}\}$} 
of the optimization problem \cite{kunisch2002galerkin,quarteroni2015reduced}:
\begin{equation}
    \min_{\mathcal{V}} \Vert \mathcal{S}_{\Phi}- \mathcal{V}\mathcal{V}^T\mathcal{S}_{\Phi} \Vert \quad s.t. \quad \mathcal{V}^T\mathcal{V}=\mathcal{I}.
    \label{min}
\end{equation}
Problem \eqref{min} seeks to minimize the distance between the snapshots and their projection onto the reduced (POD) space. Note that it can be equivalently formulated as an eigenvalue problem
\cite{kunisch2002galerkin}, which is easier to handle:
\begin{equation}
    C_{\Phi} \bm c^{\Phi}_s = \lambda_s^{\Phi} \bm c_s^{\Phi}, \quad s = 1, \dots, {M}. \label{eq:eig}
\end{equation}
In \eqref{eq:eig}, $C_{\Phi}=\frac{1}{N_{T}}S_{\Phi}^TS_{\Phi}\in \mathbb{R}^{{M}\times {M}}$ is the correlation matrix related to the snapshots, and {$\bm c^{\Phi}_s$ and $\lambda_s^{\Phi}$ are the corresponding eigenvectors and eigenvalues}.
Then, the POD basis function are computed as follows: 

\begin{equation*}
    \bm\phi_i=\frac{1}{\sqrt{\lambda^{\Phi}_i}} S_{\Phi}\bm c^{\Phi}_i, \quad i = 1,\dots, N_{\Phi}.
\end{equation*} 

The dimension $N_{\Phi}$ is chosen by comparing the cumulative energy of the eigenvalues with a threshold $\delta$ selected by the user:
\begin{equation}
\frac{\sum_{i=1}^{N_{\Phi}}\lambda_i^{\Phi}}{\sum_{i=1}^{{M}}\lambda_i^{\Phi}} \ge \delta.
\label{eq:energy}
\end{equation}
{The error made by approximating the snapshots matrix $S_{\Phi}$ 
with the POD basis 
is given by the sum of the neglected singular values}
\cite{quarteroni2015reduced}: 
\begin{equation}
 \Vert \mathcal{S}_{\Phi}- \mathcal{V}\mathcal{V}^T\mathcal{S}_{\Phi} \Vert^2 = \sum_{i=N_{\Phi}+1}^{\text{min}(N_{h},{M})} \lambda_i^{\Phi}.
     \label{err}
\end{equation}
Therefore, by varying $N_{\Phi}$, the ROM solution can approximate the FOM solution with arbitrary accuracy. However, we note that there is no guarantee that the $N_{\Phi}$ needed to achieve the desired accuracy is small when compared to $N_{h}$ or ${M}$.

\subsection{Galerkin projection}
\label{sec:galerkin_projection}

The $L^2$ orthogonal projection of the momentum equation \eqref{eq:NS1} onto $\text{span}\{\bm\phi_1,\dots,\bm\phi_{N_{\bm u}}\}$ leads to   \cite{akhtar2009stability,bergmann2009enablers,lorenzi2016pod}:
\begin{equation}
    \big(\bm \phi_i, \partial_t \bm{u} + \nabla \cdot (\bm{u} \otimes \bm{u}) - \nabla \cdot (\nu \nabla \bm{u})+\nabla p \big)_{L^2(\Omega)} = 0, \quad \text{for }i = 1, \dots, N_{\bm u}.
    \label{proj_l2}
\end{equation}
If we plug \eqref{u_rb} and \eqref{p_rb} into \eqref{proj_l2} and take into account the orthonormality of the reduced basis, we get
the following ordinary differential equation in matrix form:
\begin{equation}
    \dot{\bm a} = \nu \bm B \bm a - \bm a^T \bm C \bm a - \bm K \bm b,
    \label{galerkin_u}
\end{equation}
where $\bm a = \{ a_i(t)\}_{i=1}^{N_u}$ and $\bm b = \{ b_i(t)\}_{i=1}^{N_p}$ are the ROM coefficients for velocity and pressure and: 
\begin{align}
    B_{ij} &= \big( \bm\phi_i, \Delta \bm\phi_j \big)_{L^2(\Omega)}, \quad \text{for } i, j = 1, \dots N_{\bm u},\\
    C_{ijk} &= \big( \bm\phi_i, \nabla \cdot (\bm\phi_j \otimes \bm\phi_k) \big)_{L^2(\Omega)}, \quad \text{for } i, j, k = 1, \dots N_{\bm u}\\
    K_{ij} &= \big( \bm\phi_i, \nabla \psi_j \big)_{L^2(\Omega)}, \quad \text{for } i= 1, \dots N_{\bm u} \text{ and } j =1, \dots, N_{p}.
\end{align}

To handle the divergence free constraint \eqref{eq:NS2}, we adopt the PPE approach {\cite{hesthaven2018non,guermond1998stability,stabile2018finite}}. Therefore, we replace eq.~\eqref{eq:NS2} with
\begin{align}
    &\Delta p = - \nabla \cdot ( \nabla \cdot (\bm{u} \otimes \bm{u})) \quad \text{in $\Omega$}, \label{poission1}  \\ 
    & \frac{\partial p}{\partial \bm n} = -\nu \bm n \cdot (\nabla \times \nabla \times \bm u)  \quad \text{on $\partial \Omega$}. \label{poission2} 
\end{align}
Formulation \eqref{poission1}-\eqref{poission2} is obtained by applying the divergence operator to the momentum equation and by exploiting the incompressibility constraint, under suitable hypotheses of regularity. See, e.g., \cite{saddam2017pod} for more details.

The projection of eq.~\eqref{poission1} onto $\text{span}\{\psi_1, \dots, \psi_{N_p}\}$ leads to: 
\begin{equation}
    \big(\nabla \psi_i,\nabla p \big)_{L^2(\Omega)} - \big(\psi_i,\nabla p \bm n \big)_{L^2(\partial \Omega)}= \big( \psi_i, \nabla \cdot (\nabla \cdot (\bm{u} \otimes \bm{u})) \big)_{L^2(\Omega)}, \quad \text{for }i = 1, \dots, N_{p},
    \label{proj_p}
\end{equation}
where the left-hand side is obtained integrating by parts and accounting for boundary condition \eqref{poission2}. If we plug \eqref{u_rb} and \eqref{p_rb} into \eqref{proj_p}, we obtain the matrix form: 
\begin{equation}
    \bm D \bm b - \bm N \bm b = \bm a^T \bm G \bm a,
    \label{galerkin_p}
\end{equation}
where
\begin{align}
    D_{ij}&=\big( \nabla \psi_i, \nabla \psi_j \big)_{L^2(\Omega)}, \quad \text{for } i, j = 1, \dots N_p,\\
    N_{ij}&=\big( \psi_i, \nabla \psi_j \bm n \big)_{L^2(\partial\Omega)}, \quad \text{for } i, j = 1, \dots N_p,\\
    G_{ijk}&=\big( \psi_i, \nabla \cdot(\nabla \cdot (\bm \phi_j\otimes\bm\phi_k)) \big)_{L^2(\Omega)}, \quad \text{for } i = 1, \dots N_p \text{ and } j, k = 1, \dots N_{\bm u}.
\end{align}

We note that all the matrices in \eqref{galerkin_u} and \eqref{galerkin_p} are computed once and for all during the offline stage. {Computational challenges associated with nonlinear terms ($\bm C$ and $\bm G$) can be addressed by limiting the number of POD basis functions. In fact, notice that the 
dimension of tensor $\bm C$ grows as $N_{\bm u}^3$, while dimension of tensor $\bm G$ grows as $N_{\bm u}^2 N_p$.} 
{Eq. \eqref{galerkin_u} and \eqref{galerkin_p} are solved with Powell's dog leg method \cite{powell1970hybrid},
an iterative optimization algorithm for the solution of non-linear problems.}

The ROM framework introduced so far is able to capture the dynamics of the system if the first {$N_{\bm u}$ and $N_p$} 
modes{, respectively for the velocity and the pressure,} is large enough to capture the FOM solutions with accuracy. For a convection dominated flow, $N_{\bm u}$ is typically very large, which leads to high computational cost. However, if ones keeps $N_{\bm u}$ e $N_p$ low to reduce the computational cost, the accuracy in the reconstruction of the flow at the reduced order level is severely compromised. One possible way to recover information lost to mode truncation is to introduce a  stabilization method. 

The easiest way to achieve stabilization is by adding
a global constant viscosity \cite{bergmann2009enablers,kalb2007intrinsic,wang2012proper}, i.e., the viscosity in the projected momentum equation \eqref{galerkin_u} is modified as follows:
\begin{equation}
    \nu \quad \Rightarrow \quad \nu (1 + \nu_a),
    \label{constant_kernel}
\end{equation}
where $\nu_a$ is {a scaling factor}, 
Although trial and error is not the best way to proceed to set $\nu_a$, we would like to remark that \eqref{constant_kernel} is performed only at the reduced order level, i.e., on a system whose size is contained. Hence, though cumbersome, the procedure is not computationally expensive. 

A possible easy improvement over \eqref{constant_kernel} is to modify the amount of added viscosity for each mode. The idea 
consists in replacing  the viscosity in the projected momentum equation \eqref{galerkin_u} as follows:
\begin{equation}
    \nu \quad \Rightarrow \quad \nu \left(1 + \nu_a\frac{k}{N_{\bm u}}\right),
    \label{linear_kernel}
\end{equation}
where $k$ is the modal index. This means that 
more diffusivity is added for the modes with less energy content in the system. 


\subsection{Lifting function method}
\label{sec:lift_func}

Non-homogeneous boundary conditions are introduced at reduced level with the lifting function (also called control function) method \cite{saddam2017pod,fick2018stabilized,graham1999optimal,gunzburger2007reduced}. This 
entails building a homogeneous reduced basis space by removing non-homogeneous boundary value from the collected snapshots. Then, the non-homogeneous boundary value is added again during the online phase, when the ROM solution is computed.

Let $\bm\chi(\bm x)$ be a
lifting function, i.e., a divergence free field (like the modes) that takes the value $\bm u_D (\bm x)$ where the non-homogeneous boundary condition is enforced. The snapshots are scaled as follows:
\begin{equation}
\bm u' (t^n, {\bm\mu^m}) = \bm u (t^n, {\bm\mu^m}) - \bm\chi(\bm x),
\end{equation}
where $\bm u (t^n, {\bm\mu^m})$ is the collected snapshot and $\bm u' (t^n, {\bm\mu^m})$ is the ``homogenized'' snapshot.
A common approach to compute the lifting function $\bm\chi$ is to solve a potential flow problem. This means that if $\Gamma_i$ is the boundary where the non-homogeneous Dirichlet condition is imposed, 
then $\bm\chi$ is found by solving the following system:
\begin{equation*}
    \begin{cases}
        \nabla \cdot \bm u = 0 & \quad \text{in $\Omega$}, \\
        \bm u = \nabla \bm\chi & \quad \text{in $\Omega$}, \\
        \bm\chi = \bm u_D (\bm x) & \quad \text{in $\Gamma_i$},\\
        \bm\chi = \bm 0 & \quad \text{in $\Gamma_w$},\\
        \nabla \bm\chi \cdot \bm n = 0 & \quad \text{in $\Gamma_o$}.
    \end{cases}
\end{equation*}

Snapshot $\bm u' (t^n, {\bm\mu^m})$, which satisfies homogeneous boundary conditions, is then used to generate the reduced basis as explain in Sec.~\ref{sec:pod}. Then, \eqref{u_rb} is updated with:
\begin{equation}
    \bm u (\bm x, t^n, {\bm\mu^m})\approx \bm u_{\text{r}}(\bm x, t^n, {\bm\mu^m}) =\bm\chi(\bm x)+ \sum_{i=1}^{N_{\bm u}} a_i(t^n, {\bm\mu^m}) \bm\phi_i(\bm x).
\end{equation}

\section{Numerical Results}\label{sec:num_res}

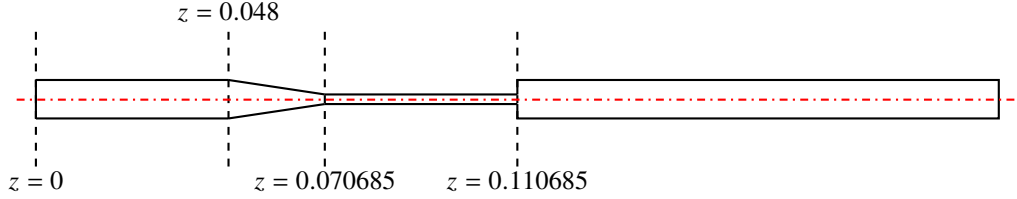
\begin{figure}[htb!]
\centering
\begin{tikzpicture}[thick,scale=64]

\draw (0,0.0015+0.0015) -- (0,0) -- (0.1,0) -- (0.1,0.008) -- (0,0.008) -- (0,0.0035+0.0015);
 \draw (-0.04,0.0015+0.0015) -- (0,0.0015+0.0015); 
 \draw (-0.04,0.0035+0.0015) -- (0,0.0035+0.0015);
 \draw (-0.04,0.0015+0.0015) -- (-0.06,0); 
 \draw (-0.04,0.0035+0.0015) -- (-0.06,0.008); 
 \draw (-0.10,0) -- (-0.06,0); 
 \draw (-0.10,0.008) -- (-0.06,0.008); 
 \draw (-0.10,0) -- (-0.10,0.008);

 \draw[dashed] (0,-0.01) -- (0,0.019);
 \node at (0,-0.013) (z0) {$z=0.110685$};

 \draw[dashed] (-0.04,-0.01) -- (-0.04,0.019);
 \node at (-0.04,-0.013) (z1) {$z=0.070685$};

\draw[dashed] (-0.06,-0.01) -- (-0.06,0.019);
\node at (-0.06,0.022) (z2) {$z=0.048$};

\draw[dashed] (-0.10,-0.01) -- (-0.10,0.019);
\node at (-0.10,-0.013) (z4) {$z=0$};

\draw[dash dot,red] (-0.104,0.0039) -- (0.104,0.0039);


\end{tikzpicture}
\caption{FDA benchmark scheme.}
\label{fig:fda}
\end{figure}

To test the ROM framework described in Sec.~\ref{sec:ROM}, we choose the FDA Benchmark 1, i.e., 3D flow in a convergent-divergent 
nozzle for low-to-moderate Reynolds numbers. 
The interest in the nozzle geometry, a cross-section of which is shown in Fig. \ref{fig:fda}, lies in the fact that its features  are commonly encountered in medical devices. The advantage of this benchmark is the availability of experimental measurements\footnote{\url{https://ncihub.cancer.gov/wiki/FDA_CFD}} against which one can compare the numerical results. The FDA collected the experimental data from three independent laboratories, some of which performed more than one trial, for a total of five data sets. 

\pgfmathsetmacro{\Lo}{1.44}
\pgfmathsetmacro{\Li}{0.048}
\pgfmathsetmacro{\Dt}{0.004}
\pgfmathsetmacro{\Di}{0.012}

We will consider Reynolds numbers in the throat of the nozzle equal to 2000, 3500, 5000 and 6500, i.e., transitional to turbulent regimes.
We will show that the FOM solutions compare well with the experimental data and then compare the ROM solutions with the 
FOM solutions.


The density and viscosity of the fluid are set to $\rho = 1056$ kg/m$^3$ and $\mu = 0.0035$ Pa/s, with $\nu=\mu/\rho=3.31\cdot 10^{-6}$ (Pa $\cdot$ m) / (s$\cdot$Kg), to match the fluid used in the experiments.
A Poiseuille velocity profile
is imposed at the inflow boundary: 
\begin{equation}
    \bm u_D (r, \theta, z) = \left (0, 0, 2 V_{\text{mean}}\left(1-\frac{r^2}{R_i^2}\right) \right),
\end{equation}
where $r$, $\theta$, and $z$ are the radial, the polar, and  axial coordinates and $V_{\text{mean}}$ is the mean inlet velocity magnitude to obtained the desired Reynolds number in the throat. Finally, 
$R_i$ is the inlet radius.

A DNS for the Reynolds numbers under consideration requires meshes with millions of cells. See, e.g., \cite{bertagna2016deconvolution,zmijanovic2017numerical}. In this paper, we consider a much coarser mesh whose features are reported in Table \ref{table_mesh}. 
In general, a simulation with this mesh 
would lead to non-physical oscillations because the mesh size
is too large to capture the small scale eddies. However, our FV method gives numerical results that compare well 
with the experimental measurements
even with a mesh much coarser than a DNS requires. 
A possible reason for the better performance of our FV method with respect to, e.g., the FE method in \cite{bertagna2016deconvolution}, could be the following: it yields exact conservation and thus
provides acceptable results despite the use of a coarse mesh.

\begin{table}[htb!]
\caption{Features of the mesh used for all the simulations.}
\centering
\begin{tabular}{ccccc}
\hline
\rowcolor{gray!20} 
 N cells & $h_{\text{min}}$ (m) & $h_{\text{max}}$ (m) & max non-orthogonality ($^\circ$) & max skewness\\
\hline
145845 & 4.3e-4 & 3.4e-3 & 20.6 & 1.036 \\
\hline
\end{tabular}
\label{table_mesh}
\end{table}

\begin{figure}
\centering
    \begin{overpic}[percent,height=.29\textwidth,grid=false]{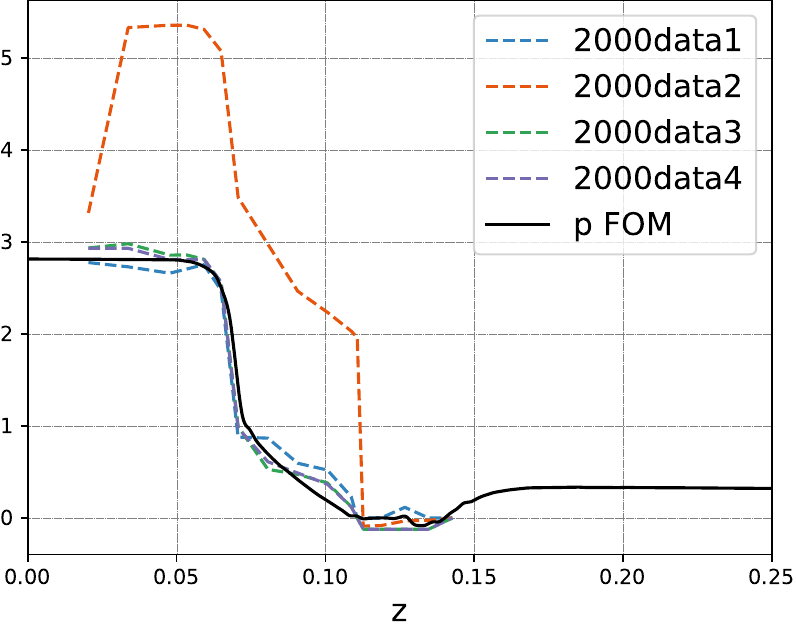}
    \put(3,82){Case $Re = 2000$, Pressure (m$^2$/s$^2$)}
    \end{overpic}
    \begin{overpic}[percent,height=.29\textwidth,grid=false]{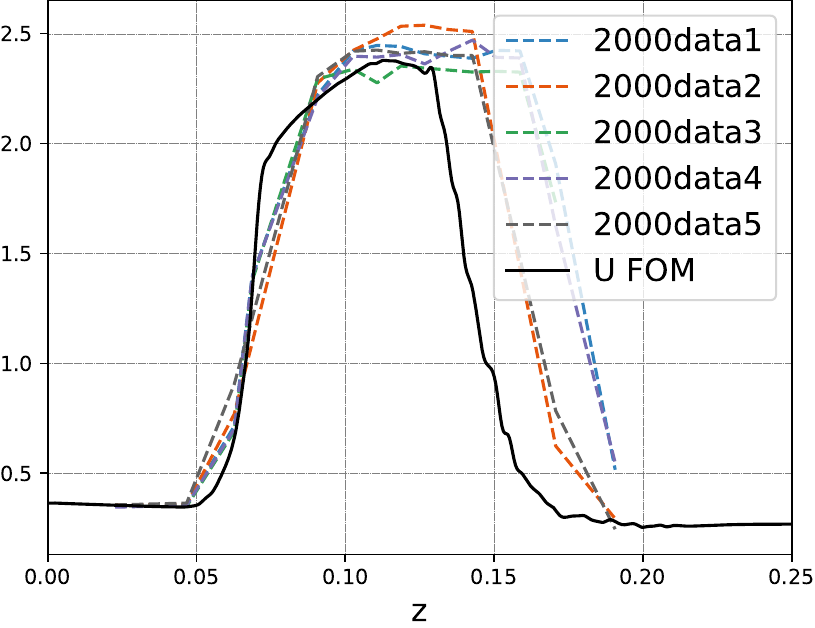}
    \put(9,80){Case $Re = 2000$, Velocity (m/s)}
    \end{overpic}\\
    \vskip .5cm
    \begin{overpic}[percent,height=.29\textwidth,grid=false]{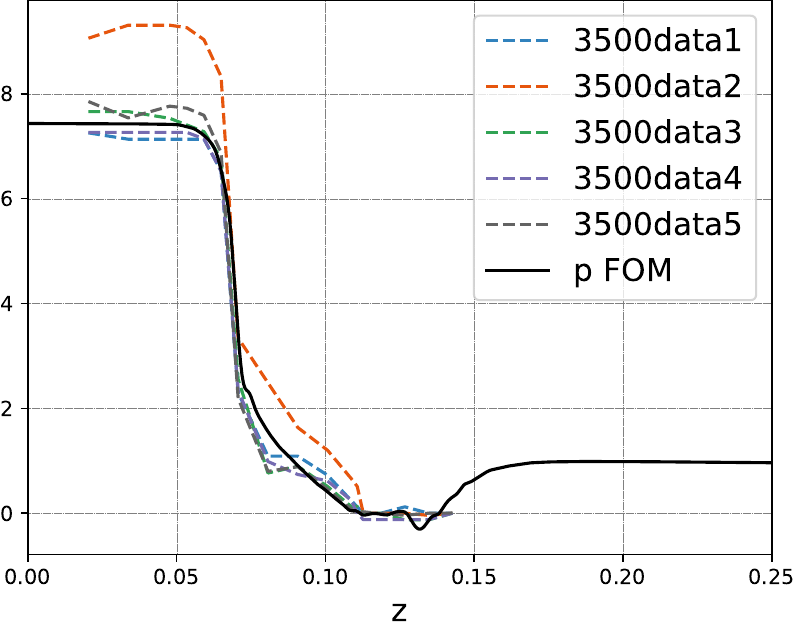}
    \put(3,82){Case $Re = 3500$, Pressure (m$^2$/s$^2$)}
    \end{overpic}
    \begin{overpic}[percent,height=.29\textwidth,grid=false]{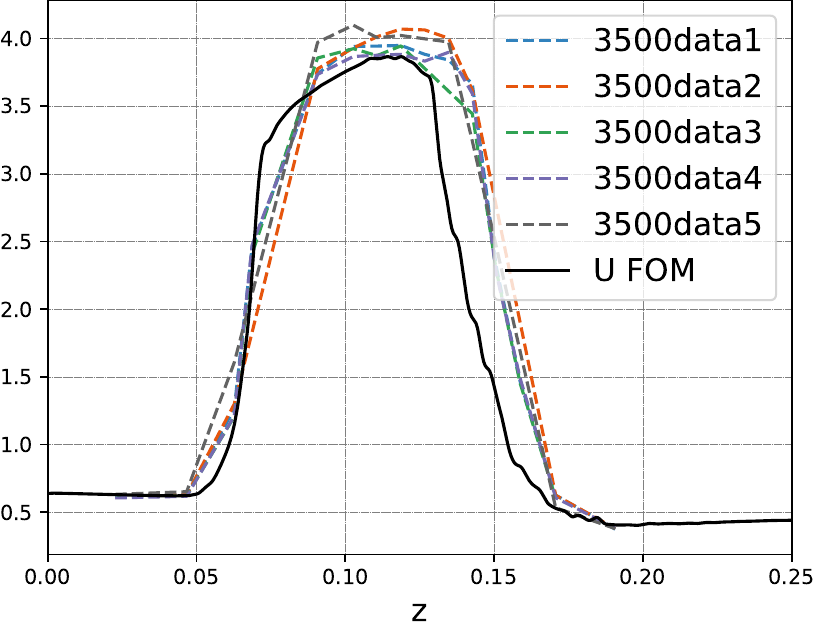}
    \put(9,80){Case $Re = 3500$, Velocity (m/s)}
    \end{overpic}\\
    \vskip .5cm
    \begin{overpic}[percent,height=.29\textwidth,grid=false]{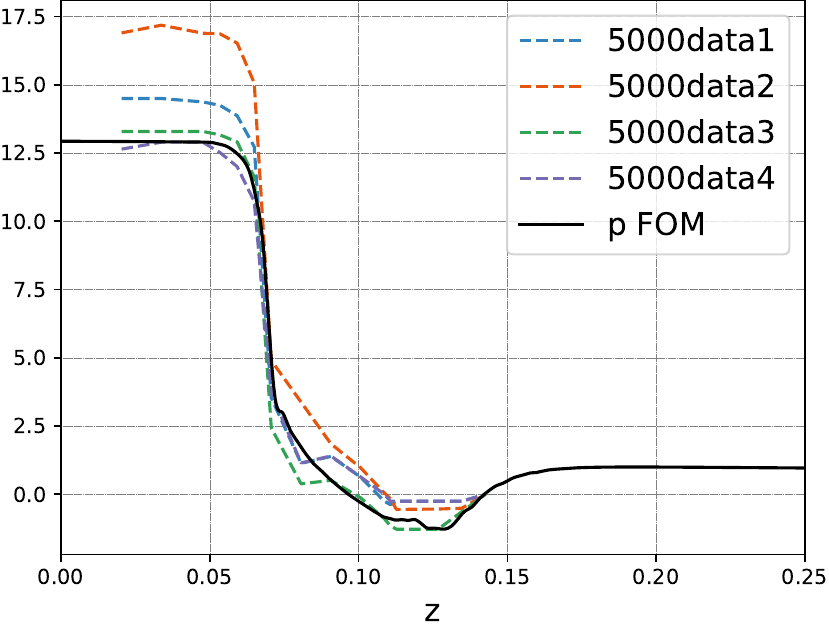}
    \put(3,79){Case $Re = 5000$, Pressure (m$^2$/s$^2$)}
    \end{overpic}
    \begin{overpic}[percent,height=.29\textwidth,grid=false]{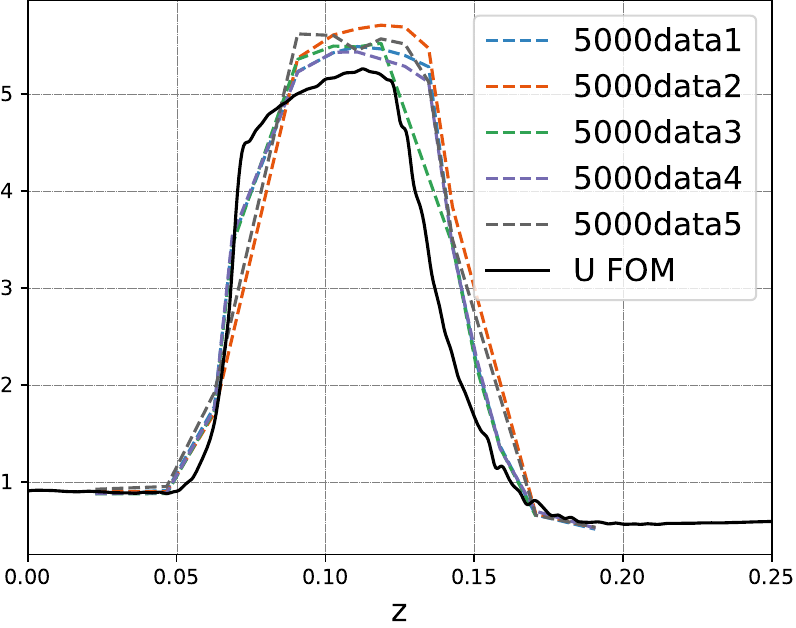}
    \put(9,82){Case $Re = 5000$, Velocity (m/s)}
    \end{overpic} \\
    \vskip .5cm
    \begin{overpic}[percent,height=.29\textwidth,grid=false]{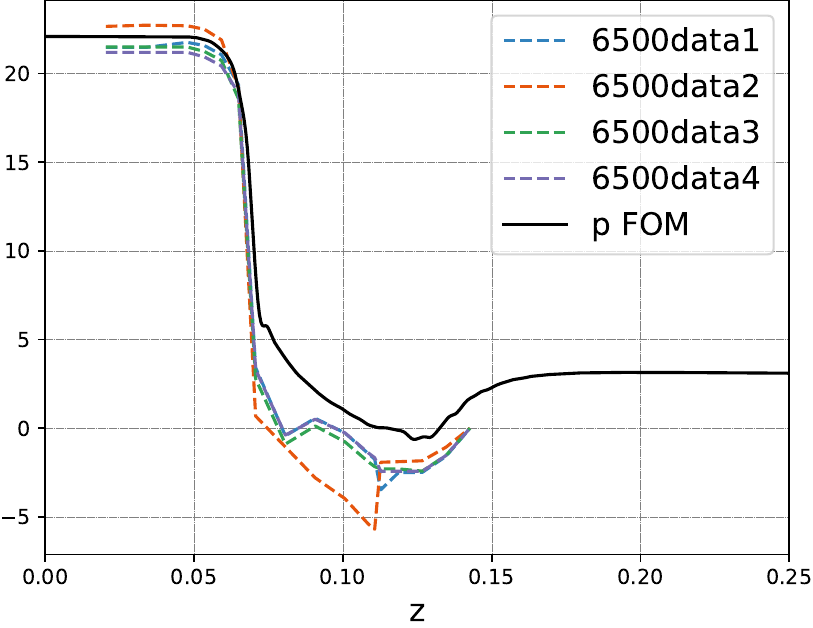}
    \put(3,80){Case $Re = 6500$, Pressure (m$^2$/s$^2$)}
    \end{overpic}
    \begin{overpic}[percent,height=.29\textwidth,grid=false]{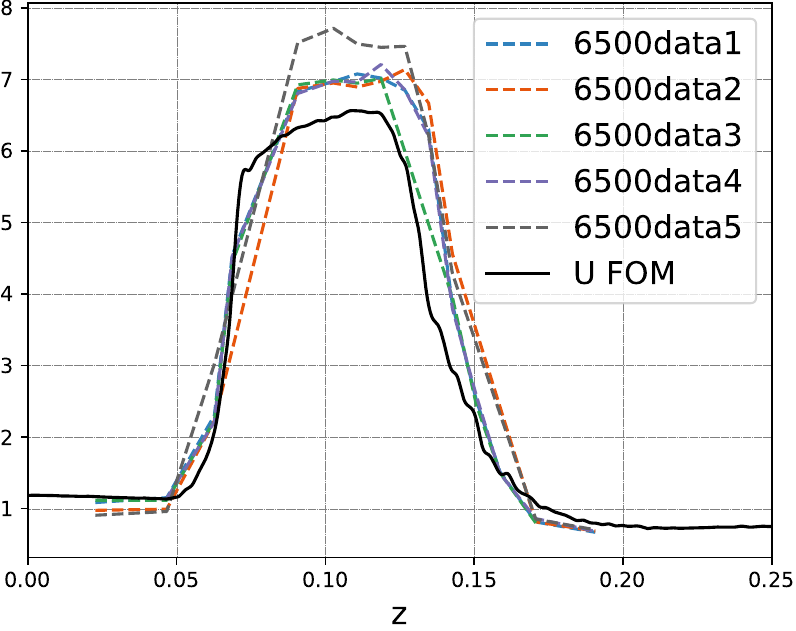}
    \put(5,82){Case $Re = 6500$, Velocity (m/s)}
    \end{overpic}
\caption{Comparison between experimental data (dashed lines) 
 and the FOM solution (solid line) for the time-averaged axial pressure (left) 
 and velocity (right) for different Reynolds numbers.}
\label{fig:dati_3500}
\end{figure}

Fig. \ref{fig:dati_3500} compares the results obtained with our coarse mesh and the experimental data for all the Reynolds number under consideration. Let us comment first on the comparison for the pressure 
on the left column in Fig.~\ref{fig:dati_3500}. 
We observe that for $Re=2000, 3500, 5000$ the full order pressure 
is in excellent agreement with most experimental data 
in the entrance channel, conical convergent, and throat. 
Note that no pressure data in the expansion channel are provided by the FDA
for any Reynolds number. 
{For $Re = 6500$, the computed pressure overestimates the measured pressure
in the {throat}
(bottom left panel in  Fig.~\ref{fig:dati_3500}).}
Now, let us look at the comparison for the axial velocity on the right column in Fig. \ref{fig:dati_3500}. 
For all three Reynolds numbers, we see a good agreement with the experimental data in the
entrance channel and conical convergent. However, the computed axial velocity tends to be lower than most data in the throat for all $Re$. This becomes more evident as the Reynolds number increases.
Instead, the length of the jet in the expansion channel is better predicted as 
the Reynolds number increases. One could improve these results with LES \cite{girfoglio2019finite,zmijanovic2017numerical,bertagna2016deconvolution,janiga2014large,delorme2013large,fehn2019modern} or RANS \cite{stewart2012assessment} models. 
Since the purpose of this paper is to show how to reconstruct a convection-dominated 
flow at the reduced level, the results given by our FV approach on a coarse mesh with 
no LES or RANS model are an equally good starting point.

In Sec.~\ref{sec:3500}, we present a thorough analysis of the results obtained with the methods presented in Sec.~\ref{sec:galerkin_projection} for $Re = 3500$, with time as the only parameter. Then, in Sec.~\ref{sec:2000_6500} we show that the conclusions drawn for the $Re = 3500$ case apply also for the transitional case 
($Re = 2000$) and the highest Reynolds
number case ($Re = 6500$). {In Sec.~\ref{sec:physical_param}, the Reynolds number is treated as a parameter, in addition to time.}

{The time interval of interest for all the simulations is $(t_0, T) = (0, 1.2)$ s, with the fluid at rest at $t_0$. 
For the FOM,  we set $\Delta t = 0.0001$ s and the CFL is $0.6$, as in \cite{girfoglio2019finite,zmijanovic2017numerical}. 
}

\subsection{Case $Re=3500$}\label{sec:3500}
In order to obtain the desired Reynolds number, we set $V_{\text{mean}} = 0.32225$ m/s. 
For the POD algorithm, we use with 120 equispaced snapshots in time interval $(0, 1.2)$ s.

\begin{figure}[htb!]
	\centering
 	\subfloat[][Cumulative eigenvalues.\label{fig:cum_eig}]{\includegraphics[width=.48\textwidth]{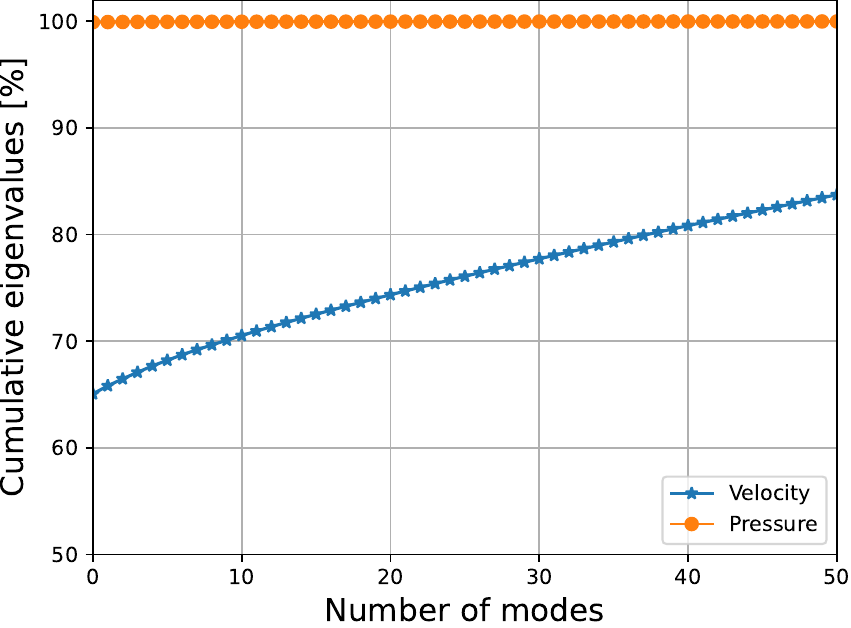}}
	\subfloat[][Mean relative error.\label{fig:err_Nmodes}]{\includegraphics[width=.48\textwidth]{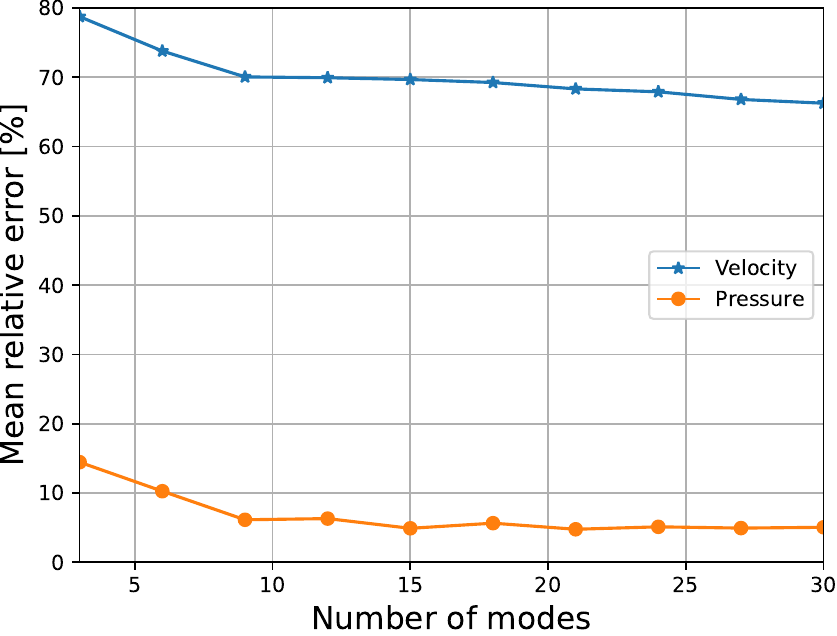}}
	\caption{(a) Cumulative energy of the eigenvalues \eqref{eq:energy} and (b) mean relative error \eqref{err} for pressure and velocity. 
 }
	\label{eig-and-err}
\end{figure}

The cumulative energy of the eigenvalues~\eqref{eq:energy} is shown in Fig.~\ref{fig:cum_eig} for velocity and pressure. While the pressure reaches $99.92\%$ of the energy with a single mode, the velocity needs 50 modes to achieve only $83.73\%$. 
The slow growth of velocity eigenvalues is reflected also in the mean relative error \eqref{err} for velocity and pressure shown in Fig.~\ref{fig:err_Nmodes}. 
We see that for both variables the use of more than 9 modes does not significantly improve the reconstruction. 

\begin{figure}[htb!]
    \centering
    \includegraphics[width=0.5\textwidth]{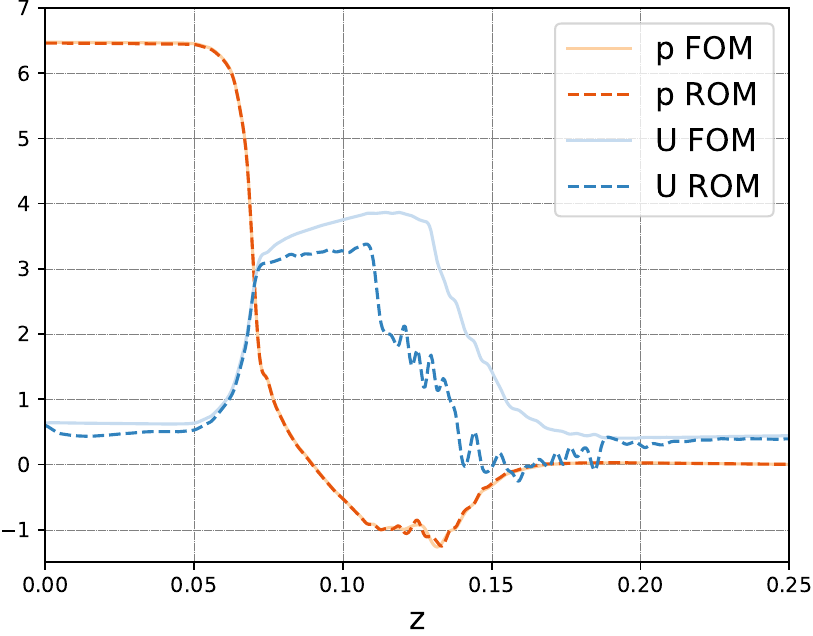}
    \caption{Comparison of FOM and ROM time-averaged axial velocity and pressure using 9 modes for both variables and no
    stabilization technique.}
    \label{fig:no-stabilization}
\end{figure}

\begin{figure}[htb!]
    \centering
    \begin{minipage}{.7\textwidth}
        \centering
        \begin{overpic}[width=1\textwidth]{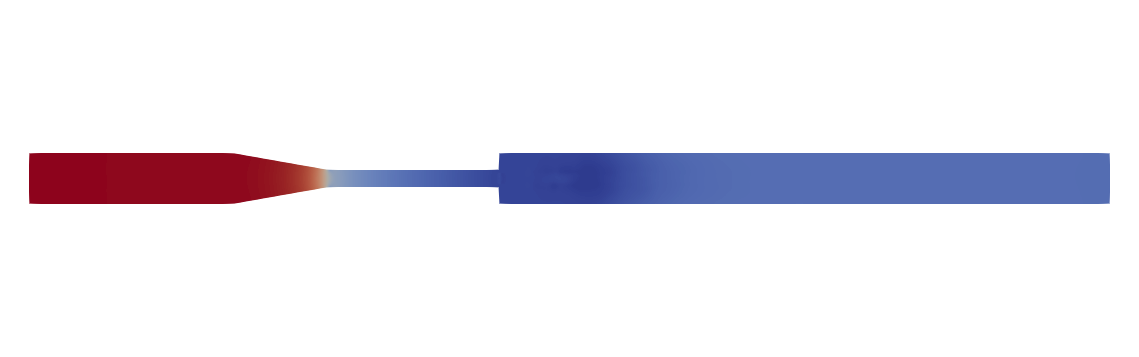} 
        \put(45,20){p FOM}
        \end{overpic}\\\vspace{-12ex}
        \begin{overpic}[width=1\textwidth]{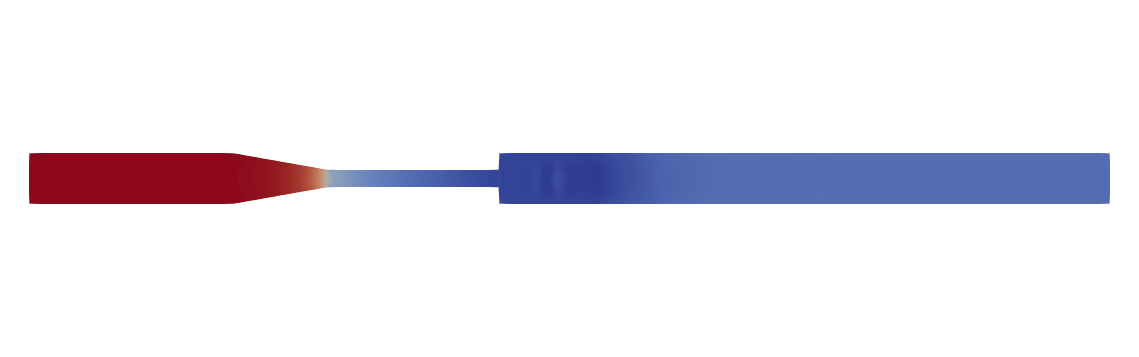} 
        \put(45,20){p ROM}
        \end{overpic}
    \end{minipage}%
    \hfill
    \begin{minipage}{0.29\textwidth}
        \centering
        \includegraphics[width=0.46\linewidth]{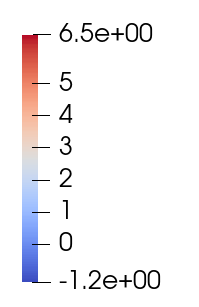}
    \end{minipage}
    \vspace{-6ex}
\caption{Qualitative comparison of the ROM time-averaged pressure without stabilization with the corresponding FOM solution.}
\label{nostabpre}
\end{figure}

\begin{figure}[!htb]
    \centering
    \begin{minipage}{.7\textwidth}
        \centering
        \begin{overpic}[width=1\textwidth]{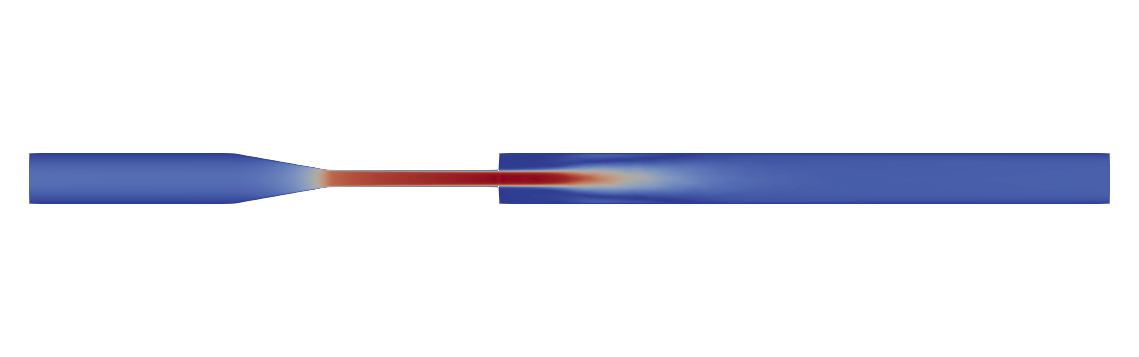} 
        \put(45,20){U FOM}
        \end{overpic}\\\vspace{-12ex}
        \begin{overpic}[width=1\textwidth]{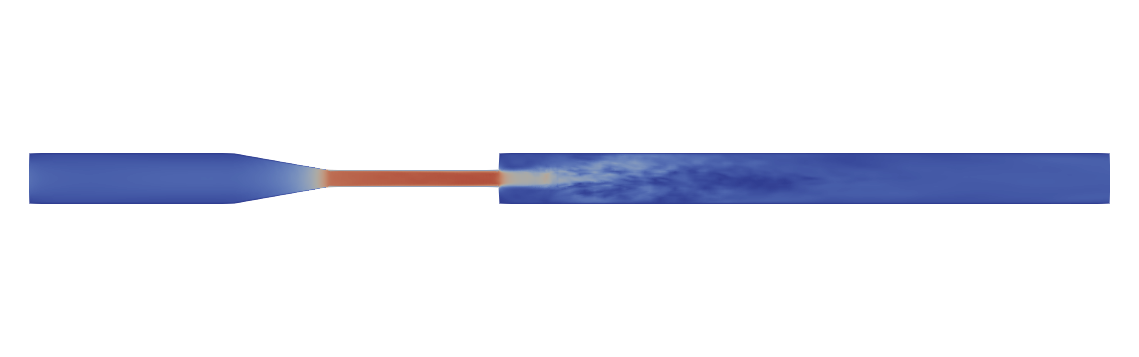} 
        \put(45,20){U ROM}
        \end{overpic}
    \end{minipage}%
    \hfill
    \begin{minipage}{0.29\textwidth}
        \centering
        \includegraphics[width=0.5\linewidth]{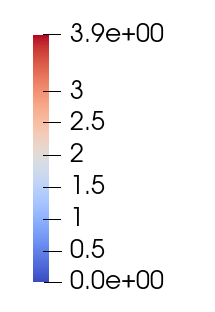}
    \end{minipage}
    \vspace{-6ex}
\caption{Qualitative comparison of the ROM time-averaged axial velocity without stabilization with the corresponding FOM solution.}
\label{nostabvel}
\end{figure}

Fig.~\ref{fig:no-stabilization} shows 
the {time-averaged} axial velocity and pressure along the $z$-axis by using 9 modes for both variables and no stabilization. 
The large distance between the FOM and ROM velocity curves reflects the large error in Fig.~\ref{fig:err_Nmodes}, while the FOM and ROM pressure curves are practically superimposed.
The qualitative comparison of the solutions is reported in Fig.~\ref{nostabpre} and \ref{nostabvel}, which confirm that the ROM reconstruction of the pressure is accurate, whereas the ROM velocity shows an evident mismatch with the FOM velocity. 



Next, we will show how the ROM solution improves with the introduction of artificial viscosity only at the reduced order level. First, we will present the results for strategy \eqref{constant_kernel} and then proceed with strategy \eqref{linear_kernel}.

\begin{figure}[htb!]
	\centering
    \subfloat[][$\nu_a=0.8$.\label{fig:kernel_const_08}]{\includegraphics[width=.48\textwidth]{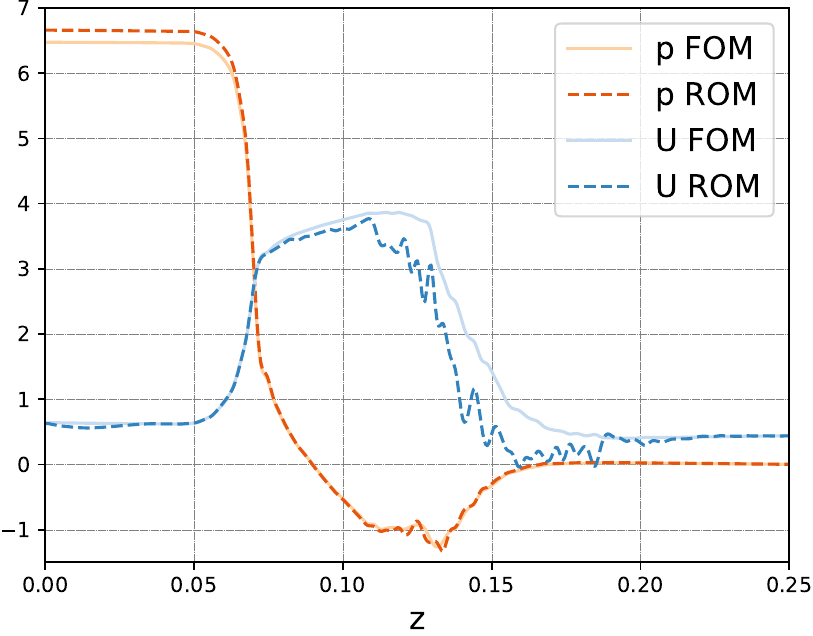}}
 	\subfloat[][$\nu_a=1.5$.\label{fig:kernel_const_15}]{\includegraphics[width=.48\textwidth]{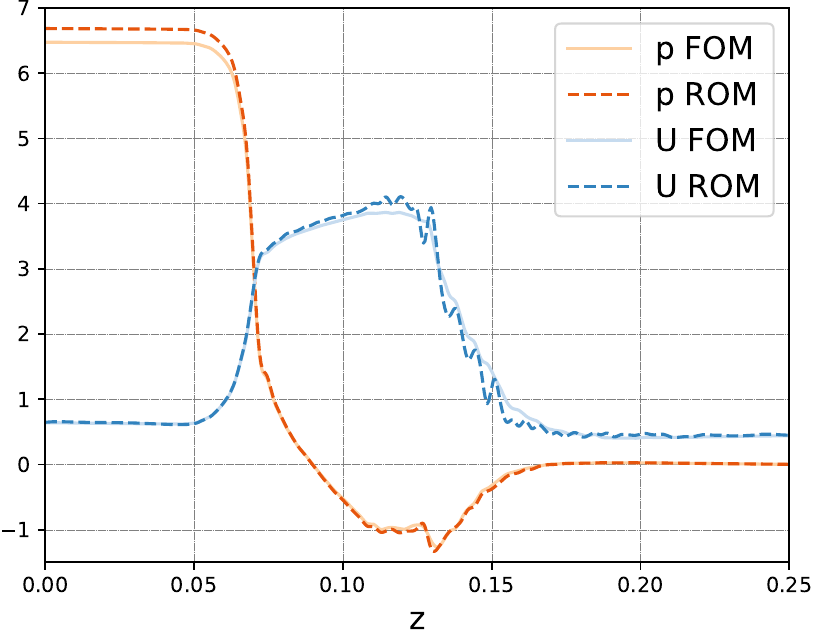}}\\
    \subfloat[][$\nu_a=20$.\label{fig:kernel_const_20}]{\includegraphics[width=.48\textwidth]{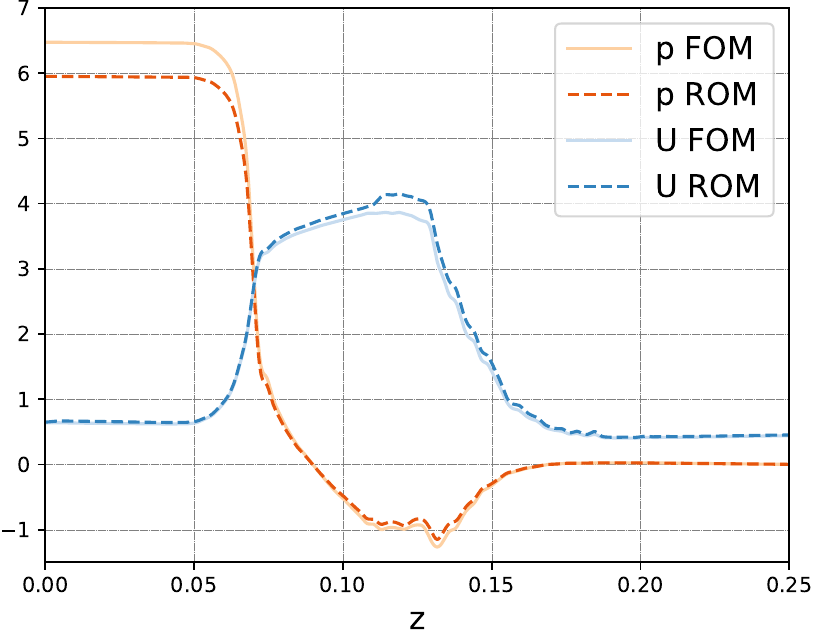}}
	\subfloat[][$\nu_a=500$.\label{fig:kernel_const_500}]{\includegraphics[width=.48\textwidth]{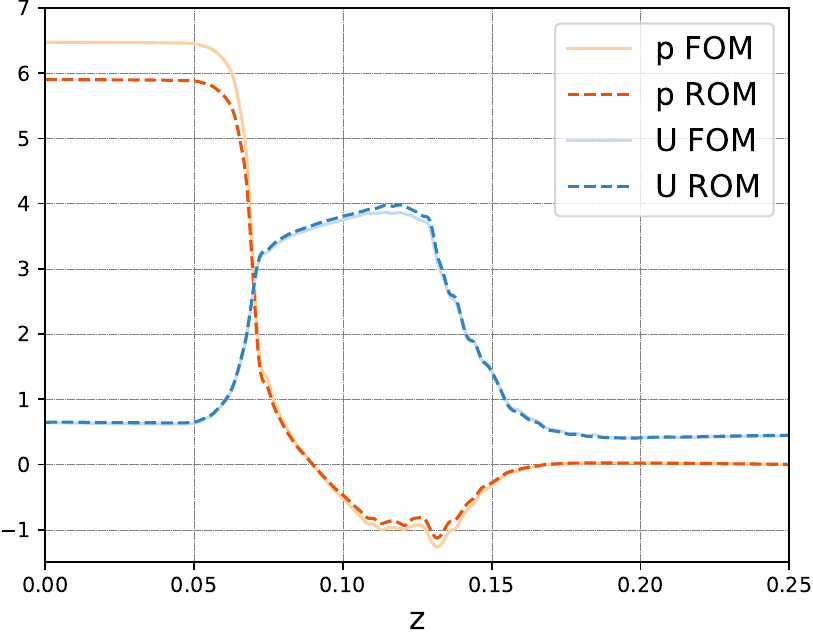}}
	\caption{Comparison of FOM and ROM time-averaged axial velocity and pressure with constant artificial viscosity for (a) $\nu_a=0.8$, (b) $\nu_a=1.5$, (c) $\nu_a=20$, and (d) $\nu_a=500$.}
	\label{kernel_const_cases}
\end{figure}

\vskip .3cm
\noindent {\bf Constant artificial viscosity.} 
Fig.~\ref{kernel_const_cases} shows the time-averaged axial velocity and pressure along the $z$-axis as the parameter $\nu_a$ increases. We see that  
the ROM reconstruction of the velocity exhibits oscillations for small values of $\nu_a$ (see Fig.~\ref{fig:kernel_const_08} and \ref{fig:kernel_const_15}). In addition, for $\nu_a=0.8$ the velocity is quite underestimated {in the expansion channel.} 
This is overcome with $\nu_a=1.5$: the ROM velocity curve gets closer to the FOM curve, however oscillations are still noticeable. Increasing $\nu_a$ to 20 removes all oscillations. However, for this value of $\nu_a$ we observe an unexpected increase in the time-averaged axial velocity in the expansion channel 
and the time-averaged ROM pressure moves away from the time-averaged FOM pressure in the entrance channel. See Fig.~\ref{fig:kernel_const_20}. The best ROM solutions is offered by $\nu_a=500$, with the exception of the ROM reconstruction of the time-averaged pressure in the entrance chamber.  See Fig.~\ref{fig:kernel_const_500}. We remark that this trial and error process to select the value of $\nu_a$ is performed during the online phase: the ROM simulation for a given value of $\nu_a$
takes only about 3 seconds. 

On important observation is in order:
even though $\nu_a=500$ gives a good approximation of the time-averaged FOM solution, the ROM solution 
becomes steady. 
Table \ref{percentage_time_dependecy} shows the percentage of time dependency (number of time dependent snapshots over the total number of snapshots) as $\nu_a$ increases. We see that 
for $\nu_a > 20$, the ROM solution 
can be considered steady.
This observation does not influence the validity of our results because we are interested in the reconstruction of a time-averaged value for pressure and velocity. However, this approach cannot be used to recover the time-dependent solution.

\begin{table}
\centering
\caption{Percentage of time dependency (number of time dependent snapshots over the total number of snapshots) of the ROM solutions, when constant artificial viscosity strategy is adopted.}
\begin{tabular}{ccccc}
\hline
\rowcolor{gray!20} $\nu_a=1$ & $\nu_a=1.5$  & $\nu_a=5$ & $\nu_a=10$ & $\nu_a=20$  \\
\hline
$\simeq74\%$  &   $\simeq54\%$    & $\simeq11\%$  & $\simeq8\%$ & $\simeq2\%$
\\
\hline
\end{tabular}
\label{percentage_time_dependecy}
\end{table}

\begin{figure}[htb!]
	\centering
    \subfloat[][Relative error for the velocity\label{fig:err_rel_kernel_const_u}]{\includegraphics[width=.48\textwidth]{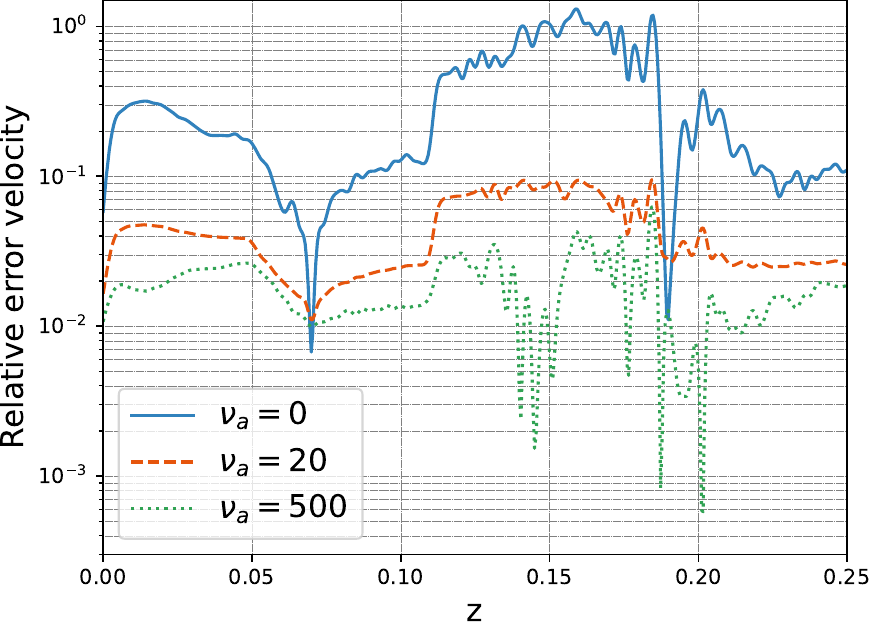}}
 	\subfloat[][Relative error for the pressure \label{fig:err_rel_kernel_const_p}]{\includegraphics[width=.48\textwidth]{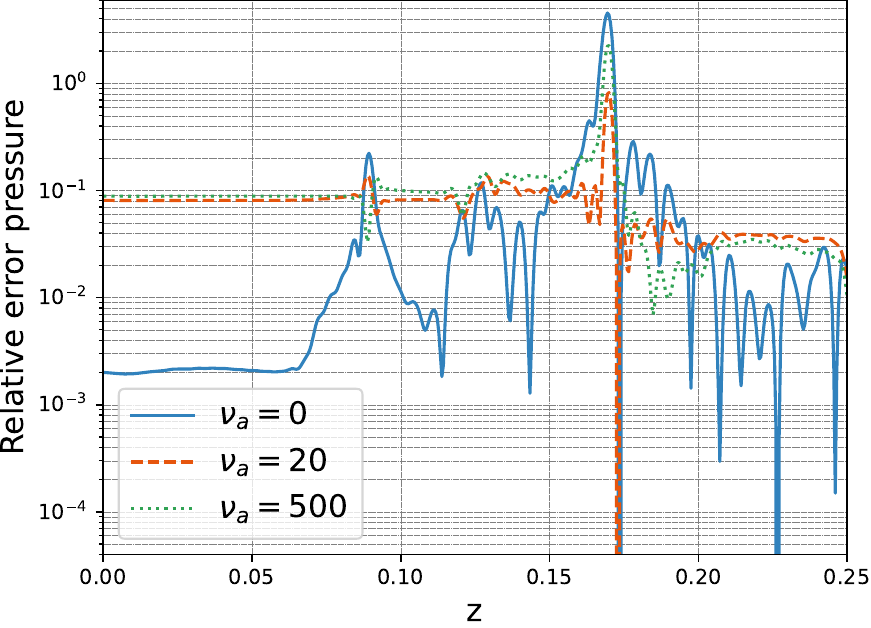}}\\
    \subfloat[][Absolute error for the velocity\label{fig:err_abs_kernel_const_u}]{\includegraphics[width=.48\textwidth]{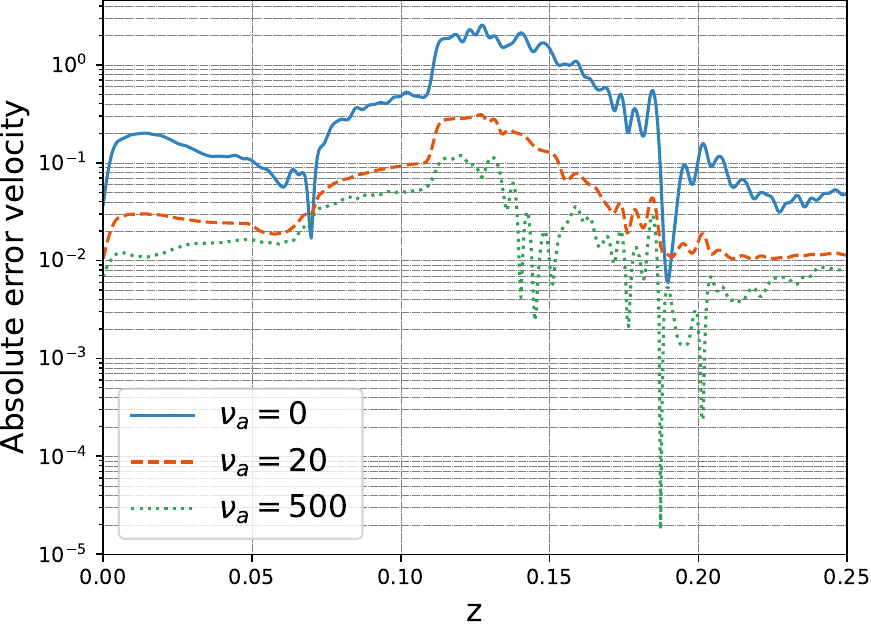}}
 	\subfloat[][Absolute error for the pressure\label{fig:err_abs_kernel_const_p}]{\includegraphics[width=.48\textwidth]{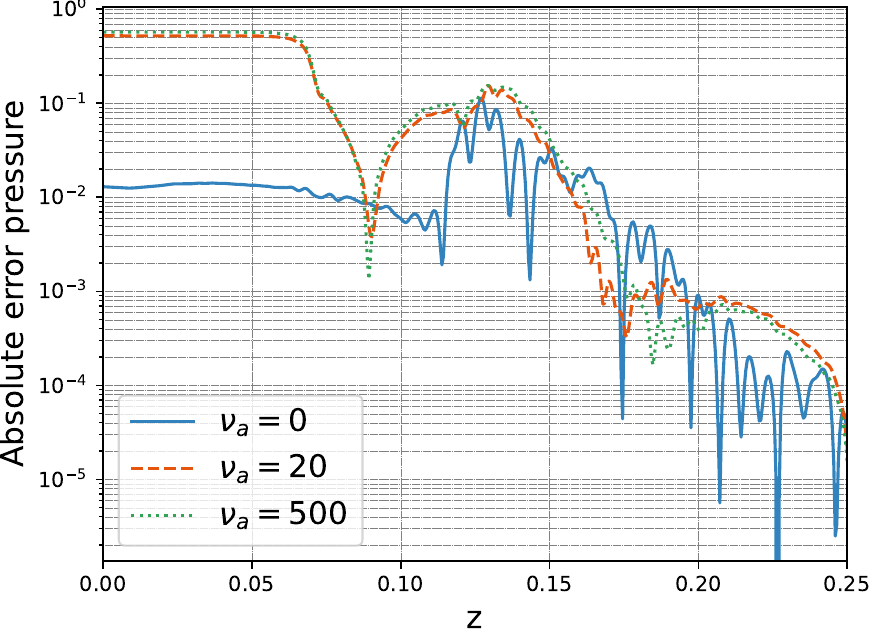}}
	\caption{Relative (top row) and absolute (bottom row) errors of the ROM axial velocity (left) and pressure (right) with respect to the FOM solution for constant artificial viscosity $\nu_a = 0, 20, 500$.}
	\label{err_abs_rel_kernel_const}
\end{figure}

In order to investigate quantitatively the effect of $\nu_a$, we present in Fig.~\ref{err_abs_rel_kernel_const}
the absolute and relative errors of the time-averaged axial velocity and pressure along the $z$-axis. The 
errors of the velocity in Fig.~\ref{fig:err_rel_kernel_const_u}
and \ref{fig:err_abs_kernel_const_u}
confirm that $\nu_a=500$ provides a very 
accurate ROM solution. 
The relative errors of the pressure in Fig.~\ref{fig:err_rel_kernel_const_p} show a significant peak around $z=0.17$, also for $\nu_a=500$. Nevertheless, the absolute errors in Fig.~\ref{fig:err_abs_kernel_const_p} show that the difference between FOM and ROM pressure is only about $10^{-2}$, therefore the peak of the relative error is purely due to the low pressure value near $z=0.17$. 
The curves in Fig.~\ref{fig:err_rel_kernel_const_p} and \ref{fig:err_abs_kernel_const_p} show a degeneration in the ROM pressure in the entrance channel as $\nu_a$ is increased. However, we note that the relative error for $\nu_a=20, 500$ is only about $10^{-1}$, which is a rather good compromise to obtain satisfactory approximations for both variables. 

\begin{figure}[!htb]
    \centering
    \begin{minipage}{.7\textwidth}
        \centering
        \begin{overpic}[width=1\textwidth]{img/p_FOM_no_stab.png} 
        \put(45,20){p FOM}
        \end{overpic}\\\vspace{-12ex}
        \begin{overpic}[width=1\textwidth]{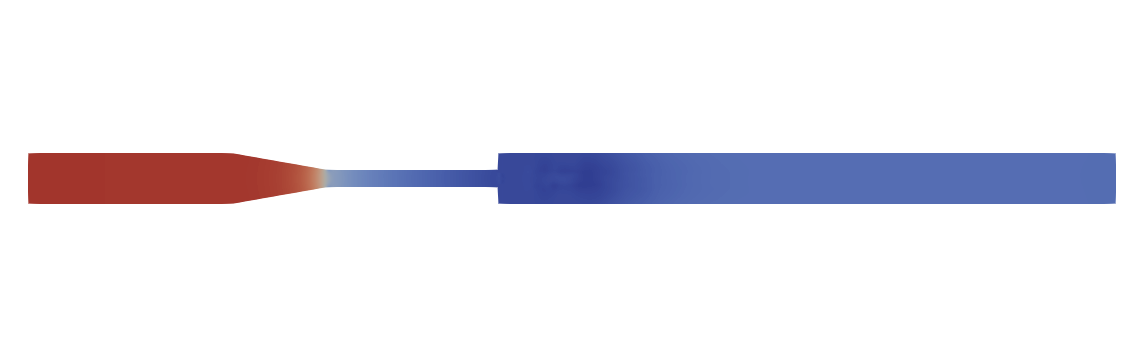} 
        \put(45,20){p ROM}
        \end{overpic}
    \end{minipage}%
    \hfill
    \begin{minipage}{0.29\textwidth}
        \centering
        \includegraphics[width=0.46\linewidth]{img/legend_p_nostab.png}
    \end{minipage}
    \vspace{-6ex}
\caption{Qualitative comparison of the ROM time-averaged pressure obtained with constant artificial viscosity $\nu_{a}=500$ with the corresponding FOM solution.}
\label{stabpre500}
\end{figure}

\begin{figure}[!htb]
    \centering
    \begin{minipage}{.7\textwidth}
        \centering
        \begin{overpic}[width=1\textwidth]{img/u_FOM_no_stab.png} 
        \put(45,20){U FOM}
        \end{overpic}\\\vspace{-12ex}
        \begin{overpic}[width=1\textwidth]{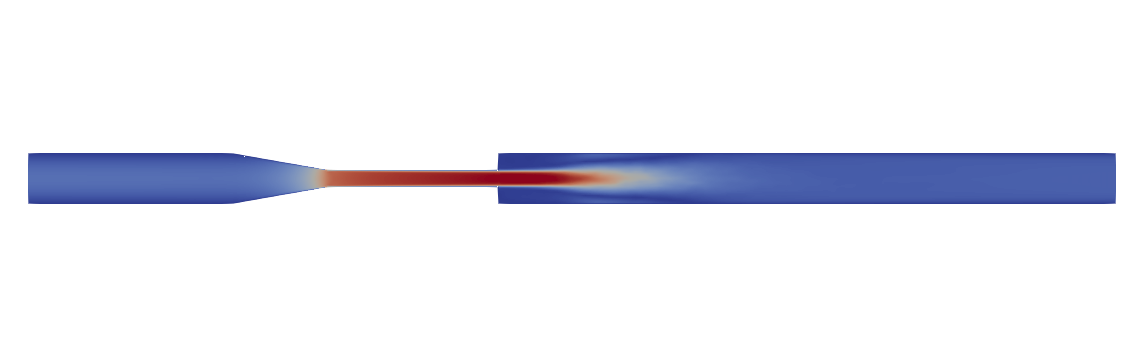} 
        \put(45,20){U ROM}
        \end{overpic}
    \end{minipage}%
    \hfill
    \begin{minipage}{0.29\textwidth}
        \centering
        \includegraphics[width=0.5\linewidth]{img/legend_u_nostab.png}
    \end{minipage}
    \vspace{-6ex}
\caption{Qualitative comparison of the ROM time-averaged axial velocity obtained with constant artificial viscosity $\nu_{a}=500$ with the corresponding FOM solution.}
\label{stabvel5001}
\end{figure}

For a visual comparison, the time-averaged pressure and velocity fields for $\nu_a=500$ are shown in Fig. \ref{stabpre500} and \ref{stabvel5001}, respectively. 

Since the pressure can be seen as a Langrange multiplier for the enforcement of the incompressibility constraint for the velocity and given the discrepancies between FOM and ROM pressure in the entrance channel, we checked if the ROM solution conserves mass using the following error \cite{passerini2013validation}: 
\begin{equation}\label{eq:e_mass}
    E^\text{mass} = \frac{Q_{\text{FOM}} (z)-Q_{\text{ROM}} (z)}{Q_{\text{FOM}}(z)},
\end{equation}
where $Q$ is the flow rate.
Table \ref{conservazione_massa} reports error \eqref{eq:e_mass} for $\nu_a=500$ and in the case of no stabilization ($\nu_a=0$) for given values of the axial coordinate $z$. 
We see that the flow rate 
computed by the ROM with $\nu_a=500$ 
is within about 1\% of the flow rate 
computed by the FOM, indicating that mass is conserved reasonably well. Instead, when $\nu_a=0$ (no stabilization) error \eqref{eq:e_mass} goes up to about 10\% in the expansion channel.


\begin{table}
\centering
\caption{Flow rate computed from the FOM solution and error \eqref{eq:e_mass} for constant artificial viscosity $\nu_a = 500, 0$ and given values of the axial coordinate $z$.}
\begin{tabular}{c|c|c|c|c}
\hline
\rowcolor{gray!20}  & $z = 0.046685 
$  & $z = 0.062685
$ & $z = 0.090685
$ & $z = 0.126685
$  \\
\hline
$Q_{\text{FOM}}$  &   3.76683e-05    & 3.64979e-05  & 3.63171e-05 & 3.62559e-05
\\
\cline{1-5}
Error \eqref{eq:e_mass} for $\nu_a = 500$ & $-0.903412\%$  & $-0.780867\%$ & $-0.844506\%$ & $-1.09748\%$  \\
\cline{1-5}
Error \eqref{eq:e_mass} for $\nu_a = 0$ & $0.596788\%$   &  $-2.06012\%$ & $-0.642948\%$ & $-9.77331\%$  \\
\hline
\end{tabular}
\label{conservazione_massa}
\end{table}

\begin{figure}
    \centering   
    \includegraphics[width=.48\textwidth]{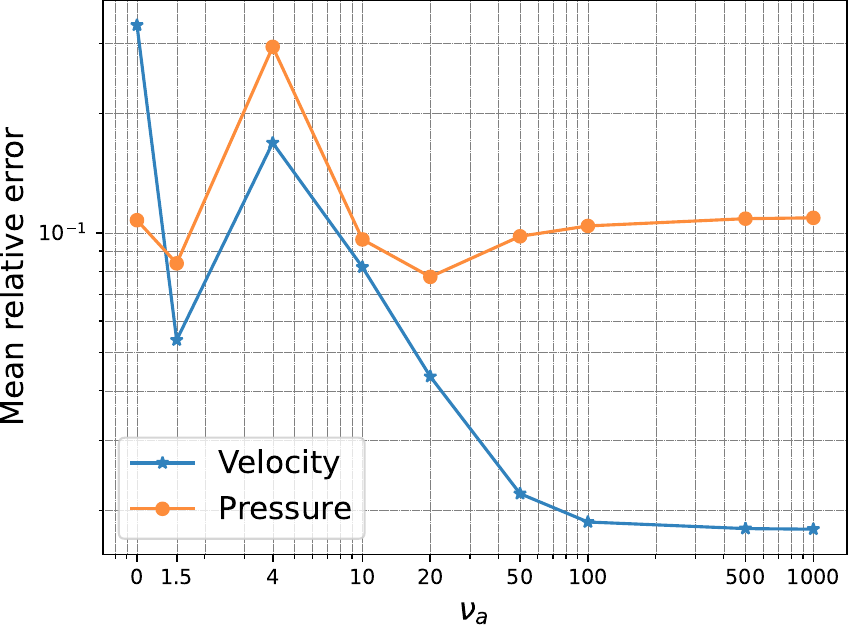}
    \caption{Time- and space-averaged relative errors of the ROM axial velocity and pressure with respect to the FOM solution as constant artificial viscosity $\nu_a$ varies.}
    \label{fig:err_nu_const}
\end{figure}

Finally, to understand if there is an ``optimal'' value of $\nu_a$, we plot in  Fig.~\ref{fig:err_nu_const} the time- and space-averages relative errors for pressure and axial velocity as $\nu_a$ increases. 
We observe that both errors peak for $\nu_a=4$ and then reach a plateau for $\nu_a>100$. For $\nu_a>4$
the error of the axial velocity decreases monotonically, whereas the error of the pressure rises slightly after attaining a minimum for $\nu_a=20$, This is consistent with the previous observation of increasing distance between the FOM and ROM pressure curves in the entrance channel in Fig. \ref{fig:kernel_const_20} and \ref{fig:kernel_const_500}.


\vskip .3cm
\noindent {\bf Modal coefficient-dependent artificial viscosity.}
Let us present the results obtained with 
the varying artificial viscosity \eqref{linear_kernel}. Our aim is to determine whether it represents an improvement over the constant 
artificial viscosity approach.

\begin{figure}[htb!]
	\centering
    \subfloat[][$\nu_a=1.5$.\label{fig:kernel_linear_15}]{\includegraphics[width=.48\textwidth]{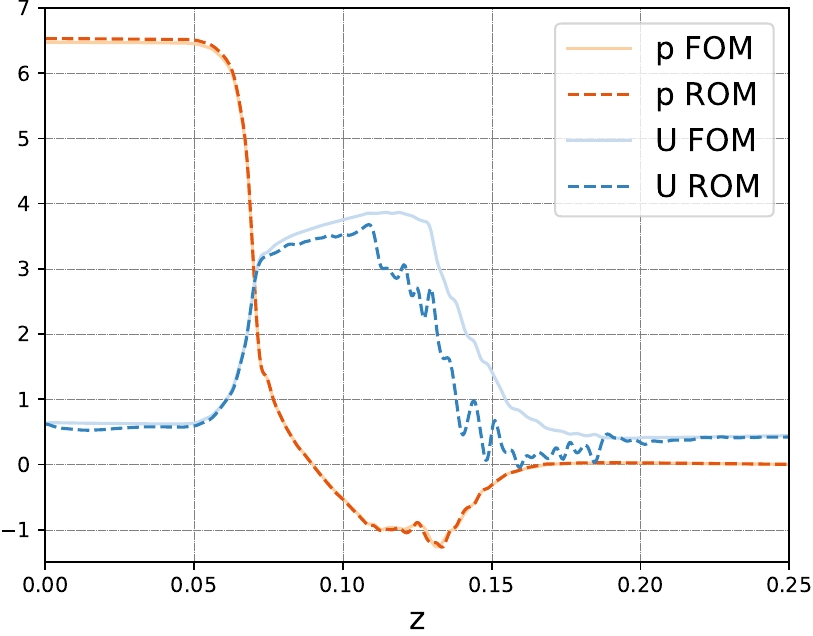}}
 	\subfloat[][$\nu_a=4$.\label{fig:kernel_linear_4}]{\includegraphics[width=.48\textwidth]{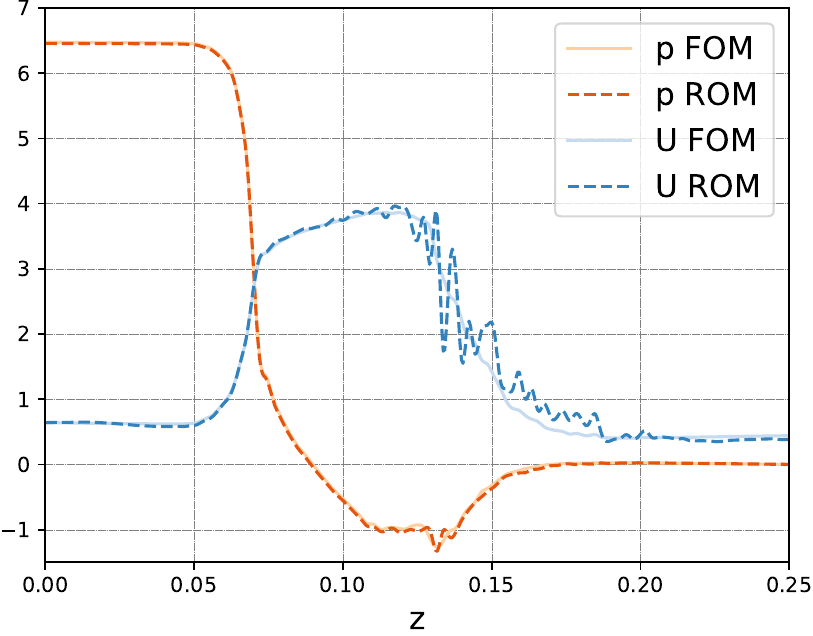}}\\
    \subfloat[][$\nu_a=20$.\label{fig:kernel_linear_20}]{\includegraphics[width=.48\textwidth]{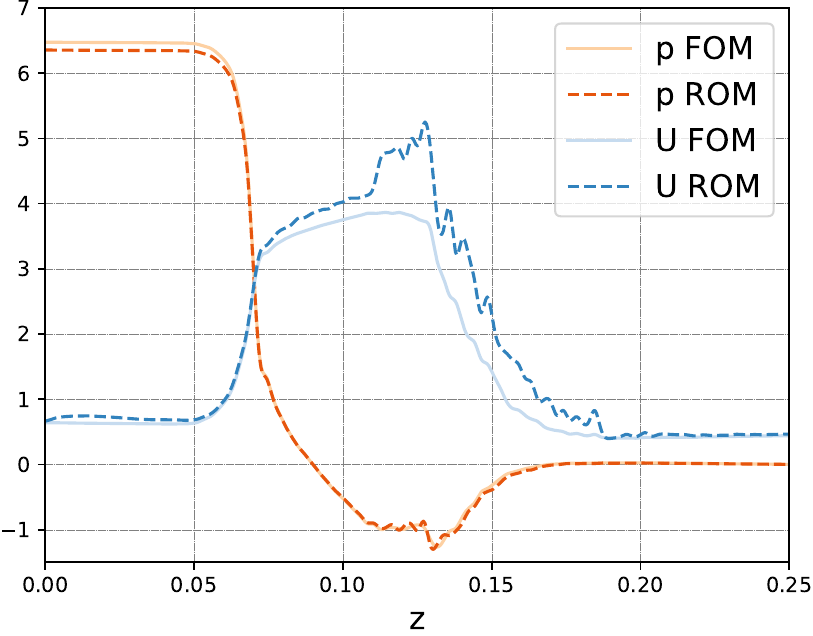}}
	\subfloat[][$\nu_a=100$.\label{fig:kernel_linear_100}]{\includegraphics[width=.48\textwidth]{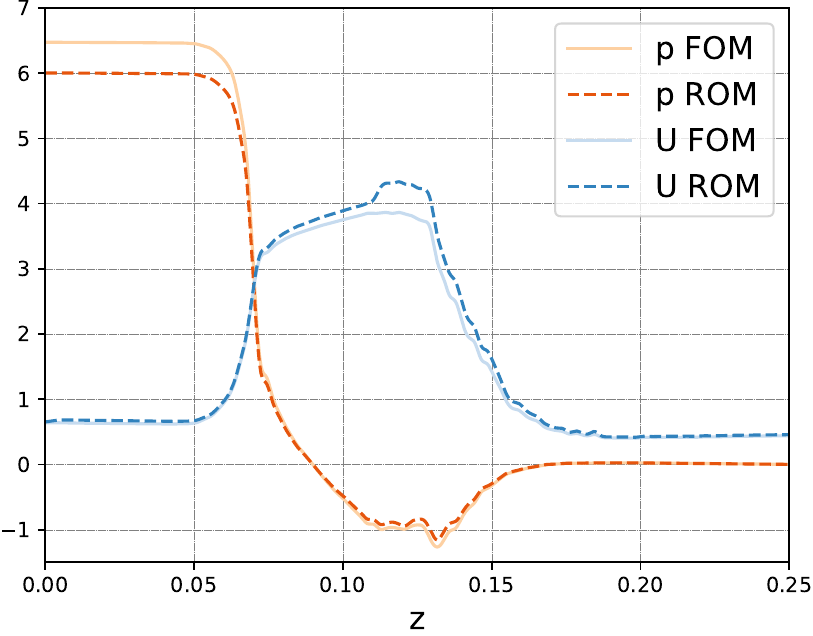}}\\
 	\subfloat[][$\nu_a=500$.\label{fig:kernel_linear_500}]{\includegraphics[width=.48\textwidth]{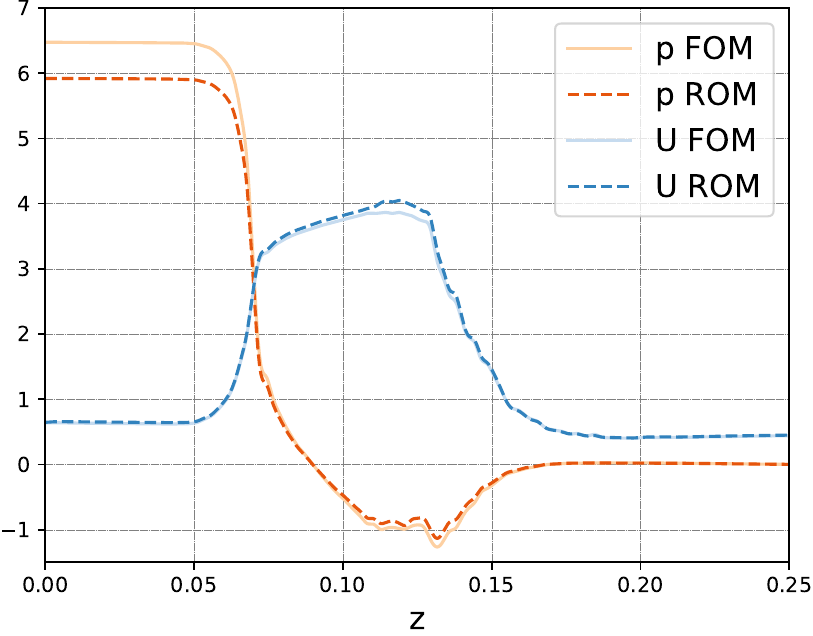}}
	\subfloat[][$\nu_a=1000$.\label{fig:kernel_linear_1000}]{\includegraphics[width=.48\textwidth]{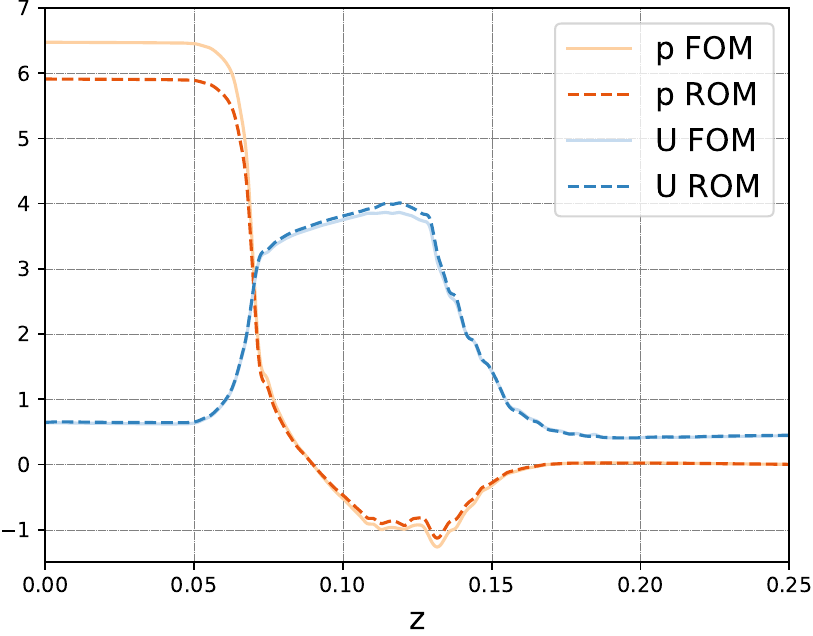}}
	\caption{Comparison of FOM and ROM time-averaged axial velocity and pressure with coefficient-dependent viscosity for (a) $\nu_a = 1.5$, (b) $\nu_a = 4$, (c) $\nu_a = 20$, (d) $\nu_a = 100$, (e) $\nu_a = 500$, and (f) $\nu_a = 1000$.}
	\label{kernel_linear_cases}
\end{figure}

Fig.~\ref{kernel_linear_cases} shows the time-averaged axial velocity and pressure along the $z$-axis for different values of $\nu_a$.
Based on the results obtained with the constant artificial viscosity, we have analyzed the ROM solutions for $\nu_a=1.5, 4, 20, 100, 500, 1000$. While $\nu_a=1.5$ provides an accurate
ROM solution in the case of constant artificial viscosity (see Fig.~\ref{fig:kernel_const_15}), in the case of coefficient-dependent artificial viscosity it leads to an underestimation of the axial velocity in the expansion channel (see Fig.~\ref{fig:kernel_linear_15}). 
By increasing the artificial viscosity to $\nu_a=4$, the ROM axial velocity profile gets closer to the FOM counterpart (see Fig.~\ref{fig:kernel_linear_4}), however it displays rather large oscillations. 
In particular, we notice larger oscillations with the coefficient-dependent artificial viscosity for $\nu_a=4$ (see Fig.~\ref{fig:kernel_linear_4}) than with a constant artificial viscosity for $\nu_a=1.5$ (see Fig.~\ref{fig:kernel_const_15}).
Furthermore, when $\nu_a$ is increased to 20, the coefficient-dependent viscosity approach gives a worse prediction than the constant viscosity approach: 
more oscillations are present in Fig.~\ref{fig:kernel_linear_20} than Fig.~\ref{fig:kernel_const_20} and the distance between FOM and ROM solutions is significantly higher. In order to obtain similar results to the constant viscosity case for $\nu_a=20$, one needs to increase
$\nu_a$ to 100 with the coefficient-dependent artificial viscosity approach. We see that in this case too the ROM axial velocity is too high in the expansion channel and the ROM pressure overestimates the FOM pressure in the entrance channel (see Fig.~\ref{fig:kernel_linear_100}). 
In order to improve the accuracy of the ROM solution, we need to further increase the artificial viscosity.
A good reconstruction of the velocity along the whole domain can be obtained with $\nu_a=500, 1000$ (see Fig.~\ref{fig:kernel_linear_500} and  \ref{fig:kernel_linear_1000}). We observe no apparent distinctions in the ROM solution 
for these two values. 

\begin{figure}[htb!]
    \centering   
    \includegraphics[width=.48\textwidth]{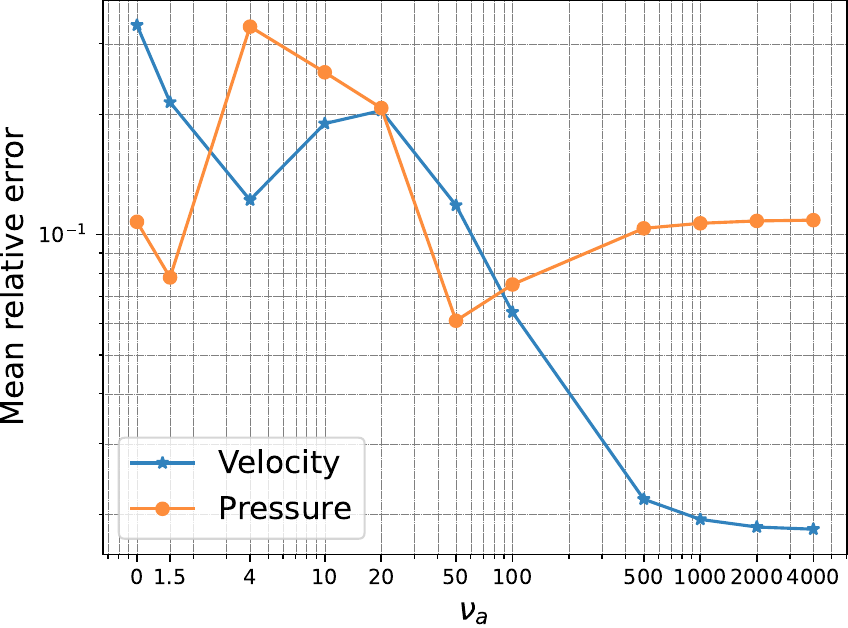}
    \caption{Time- and space-averaged relative errors of the ROM axial velocity and pressure with respect to the FOM solution as $\nu_a$ varies in the coefficient-dependent artificial viscosity approach.}
    \label{fig:err_nu_linear}
\end{figure}

We deduce that once a certain threshold is exceeded, $\nu_a$ ceases to affect the ROM solution. 
This result is confirmed in Fig.~\ref{fig:err_nu_linear}, which shows the time- and space-averaged relative errors for axial velocity and pressure as $\nu_a$ is increased. Similarly to the constant artificial viscosity approach, we see that both time- and space average- relative errors reach a plateau for $\nu_a>500$. Another similarity is the increase in the error of the pressure after reaching a minimum for $\nu_a>50$. A difference with respect to Fig.~\ref{fig:err_nu_const} is that in 
Fig.~\ref{fig:err_nu_linear} the peaks in the errors for axial velocity and pressure do not occur
for the same value of $\nu_a$.

\begin{figure}[htb!]
	\centering
    \subfloat[][Relative error for the velocity\label{fig:err_rel_kernel_linear_u}]{\includegraphics[width=.48\textwidth]{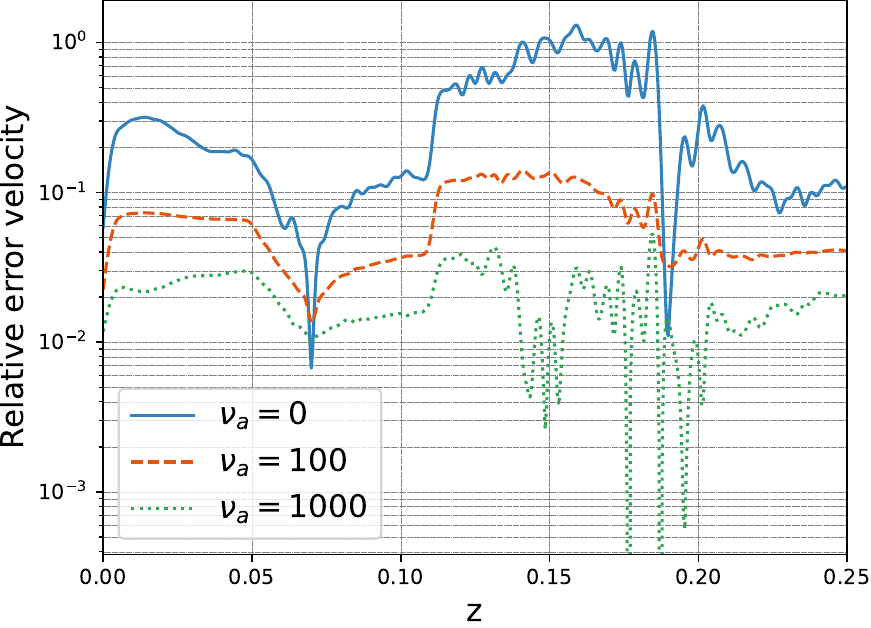}}
 	\subfloat[][Relative error for the pressure\label{fig:err_rel_kernel_linear_p}]{\includegraphics[width=.48\textwidth]{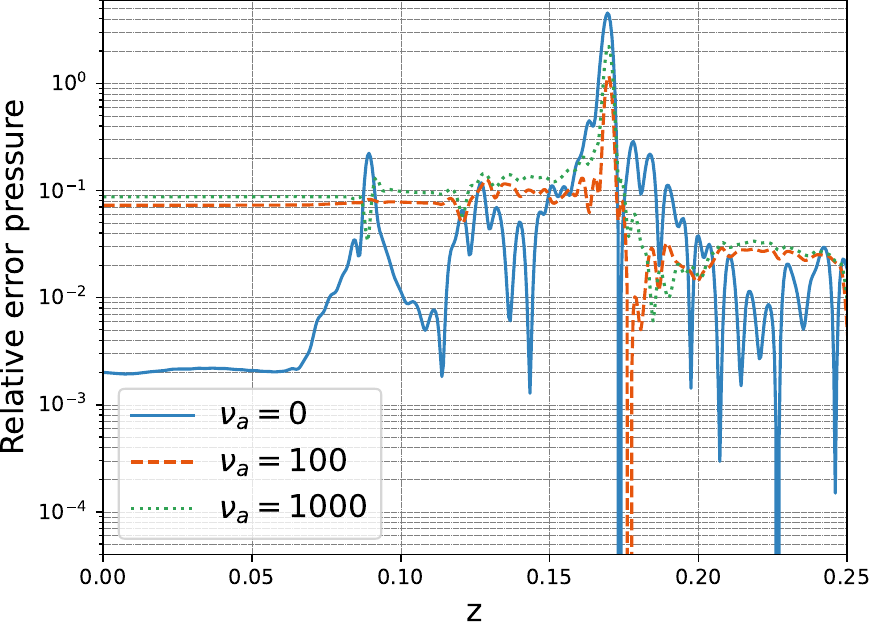}}\\
    \subfloat[][Absolute error for the velocity\label{fig:err_abs_kernel_linear_u}]{\includegraphics[width=.48\textwidth]{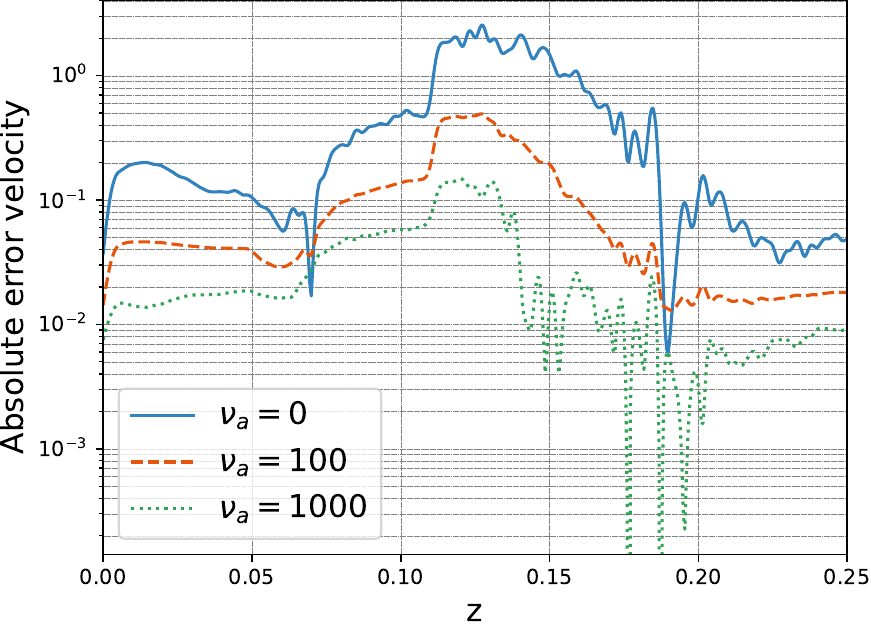}}
 	\subfloat[][Absolute error for the pressure\label{fig:err_abs_kernel_linear_p}]{\includegraphics[width=.48\textwidth]{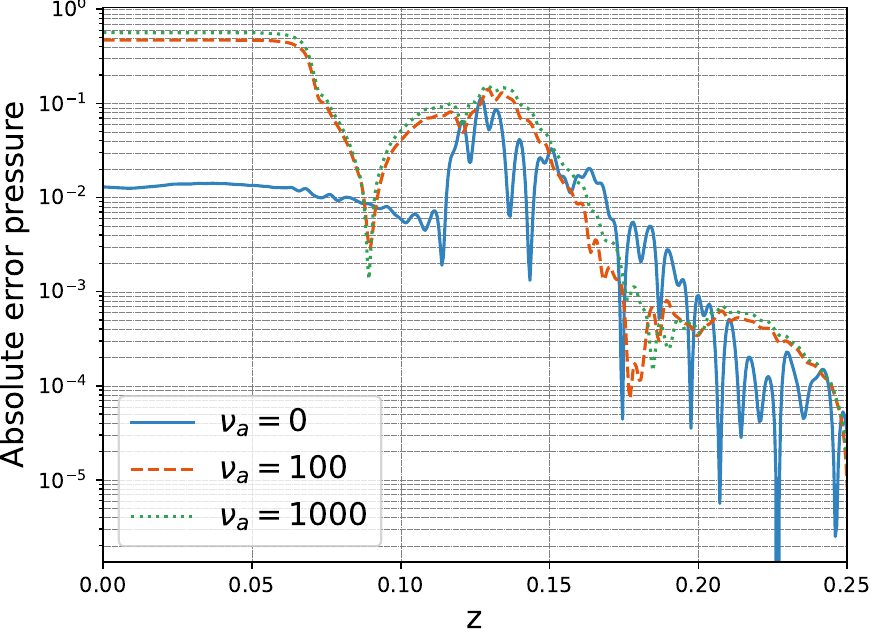}}
	\caption{Relative (top row) and absolute (bottom row) errors of the ROM velocity (left) and pressure (right) with respect to the FOM solution for $\nu_a = 0, 100, 1000$ in the coefficient-dependent viscosity approach.}
	\label{err_abs_rel_kernel_linear}
\end{figure}

Fig.~\ref{err_abs_rel_kernel_linear} reports the relative and absolute errors for axial velocity and pressure as $\nu_a$ increases.
In line with the previous considerations, the errors are similar to the constant artificial viscosity case when $\nu_a$ is  larger. Compare the curves in Fig.~\ref{err_abs_rel_kernel_const} with the curves in Fig.~\ref{err_abs_rel_kernel_linear}.
The curves for the pressure in Fig. \ref{fig:err_rel_kernel_linear_p} and \ref{fig:err_abs_kernel_linear_p} show similar features to Fig.~\ref{fig:err_rel_kernel_const_p} and \ref{fig:err_abs_kernel_const_p}: the error rises in the entrance chamber as $\nu_a$ increases, reaching an error of $10^{-1}$ for $\nu_a = 1000$, which is a good compromise to have a satisfactory ROM prediction for both variables. The peak in the relative error for the pressure at $z=0.17$ is still present, but the corresponding absolute error is small ($\approx 10^{-4}$), so the large value of the relative error is due to the pressure being close to zero. 

Given that we get a mismatch between the ROM pressure and the FOM pressure in the entrance channel in the case of coefficient-dependent viscosity too, we checked mass conservation. In Table \ref{conservazione_massa_linear}, we report error $E^\text{mass}$ \eqref{eq:e_mass} at given values of the axial coordinate $z$
for the coefficient-dependent approach at $\nu_a = 500$. Notice that the values in Table \ref{conservazione_massa_linear} are very similar to the values on 
the second row of Table \ref{conservazione_massa}. 

\begin{table}[htb]
\centering
\caption{Error \eqref{eq:e_mass} at given values of the axial coordinate $z$ for $\nu_a = 500$ with coefficient-dependent artificial viscosity.}
\begin{tabular}{c|c|c|c|c}
\hline
\rowcolor{gray!20}  & $z = 0.046685 
$  & $z = 0.062685
$ & $z = 0.090685
$ & $z = 0.126685
$  
\\
\cline{1-5}
Error \eqref{eq:e_mass} for $\nu_a = 500$ & $-0.976418\%$  & $-0.739768\%$ & $-0.843404\%$ & $-1.05969\%$  \\
\hline
\end{tabular}
\label{conservazione_massa_linear}
\end{table}

The coefficient-dependent viscosity approach does not fix the problem of losing time dependency in the ROM solution as the value of $\nu_a$ increases. Table \ref{percentage_time_dependecy_linear} reports the percentage of time dependency (number of time dependent snapshots over the total number of snapshots) as $\nu_a$ varies. We see that time evolution persists for higher value of $\nu_a$ when compared with the constant artificial viscosity case. However, for $\nu_a > 50$ the ROM reconstruction is close to being steady. 
Therefore, the coefficient-dependent viscosity approach cannot be used to predict the ROM solution in time as well. 

\begin{table}[htb!]
\centering
\caption{Percentage of time dependency (number of time dependent snapshots / total number of snapshots) of the ROM solutions when then coefficient-dependent viscosity strategy is adopted.}
\begin{tabular}{ccccc}
\hline
\rowcolor{gray!20}  $\nu_a=5$ & $\nu_a=10$ & $\nu_a=20$ & $\nu_a=30$ & $\nu_a=50$ \\
\hline
$\simeq 100\%$  & $\simeq 12 \%$ & $\simeq6\%$ & $\simeq 4 \%$  &   $\simeq 3\%$    
\\
\hline
\end{tabular}
\label{percentage_time_dependecy_linear}
\end{table}

The analysis of the results in this subsection clarifies that it is possible to use both a constant artificial viscosity \eqref{constant_kernel} and a coefficient-dependent artificial viscosity  \eqref{linear_kernel} to obtain the  time average velocity and pressure in a turbulent flow. These two approaches are equivalent, in the sense that neither features a particular advantage over the other. 
Both procedure, in combination with POD-Galerkin and PPE, are not able to recover a time dependent solution for the turbulent flow and both provide an accurate reconstruction of the time-averaged solution with a sufficient large value of $\nu_a$. 
For completeness, we tried to use both methods also in combination with the supremizer enrichment \cite{ballarin2015supremizer,gerner2012certified,rozza2007stability}. However,
the distance between FOM and ROM mean pressure increases significantly in the entrance channel when using supremizer enrichment, with a consequent higher error \eqref{eq:e_mass}.
For this reason, the results obtained with the supremizer enrichment are not shown. 

{Finally, we comment on the computational cost using an Intel(R) Core(TM) i7-7700 CPU @ 3.60GHz 2:30 16GB RAM.
Each FOM simulation takes
$5$ hours and $17$ seconds are required 
for the POD algorithm to complete, while only 1 second is needed for the evaluation phase. Therefore, the speed-up is significant.
}

\subsection{Cases $Re=2000, 6500$}\label{sec:2000_6500}

The purpose of this section is to show that 
the conclusions on the artificial viscosity and
coefficient-dependent artificial viscosity methods drawn for the flow at $Re = 3500$ can be extended to the transitional case ($Re=2000$) and the highest Reynolds number case considered by the FDA ($Re=6500$). Following the previous section, we will present both qualitative and quantitative results.

In the simulations, we set $V_{\text{mean}} = 0.3679616$ m/s to obtain $Re=2000$ and 
$V_{\text{mean}} = 1.1969$ m/s to reach $Re=6500$. Fig.~\ref{eig-and-err}
shows the 
cumulative energy of the eigenvalues~\eqref{eq:energy} 
and the average relative error \eqref{err} for velocity and pressure
for $Re=2000, 6500$.
We truncate the number of mode to 9, which is
clearly not sufficient to provide an accurate reconstruction of the flow velocity and
pressure in either case if one
uses a standard POD-Galerkin ROM.

\begin{figure}[htb!]
	\centering
 	\subfloat[][Case $Re=2000$.\label{fig:cum_eig_2000}]{\includegraphics[width=.48\textwidth]{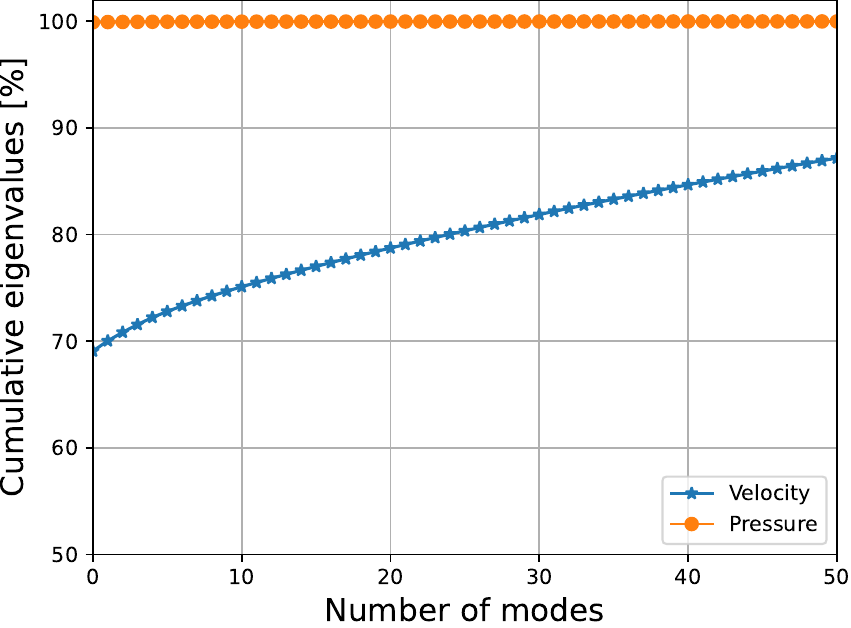}}
    \subfloat[][Case $Re=2000$.\label{fig:err_nmodes_2000}]{\includegraphics[width=.48\textwidth]{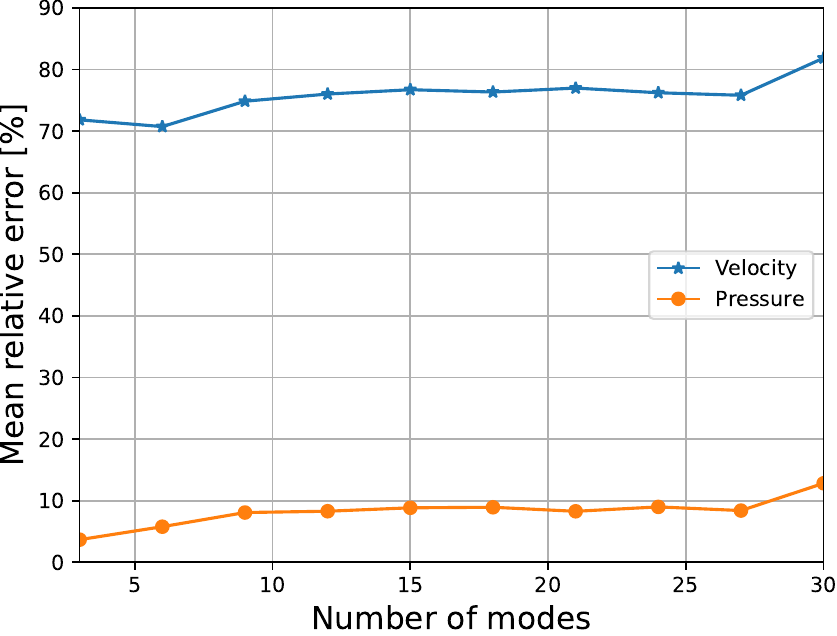}}\\
  
	\subfloat[][Case $Re=6500$.\label{fig:cum_eig_6500}]{\includegraphics[width=.48\textwidth]{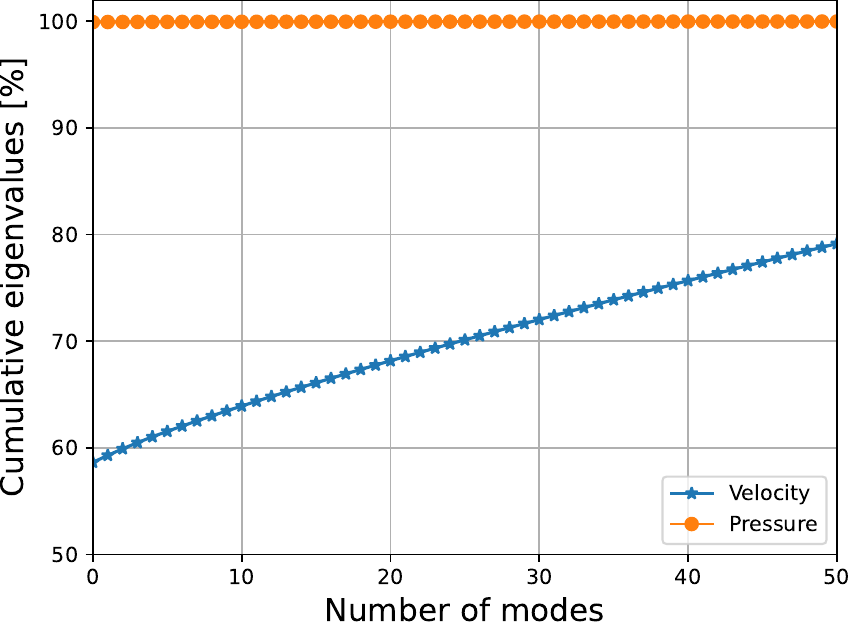}}
    \subfloat[][Case $Re=6500$.\label{fig:err_nmodes_6500}]{\includegraphics[width=.48\textwidth]{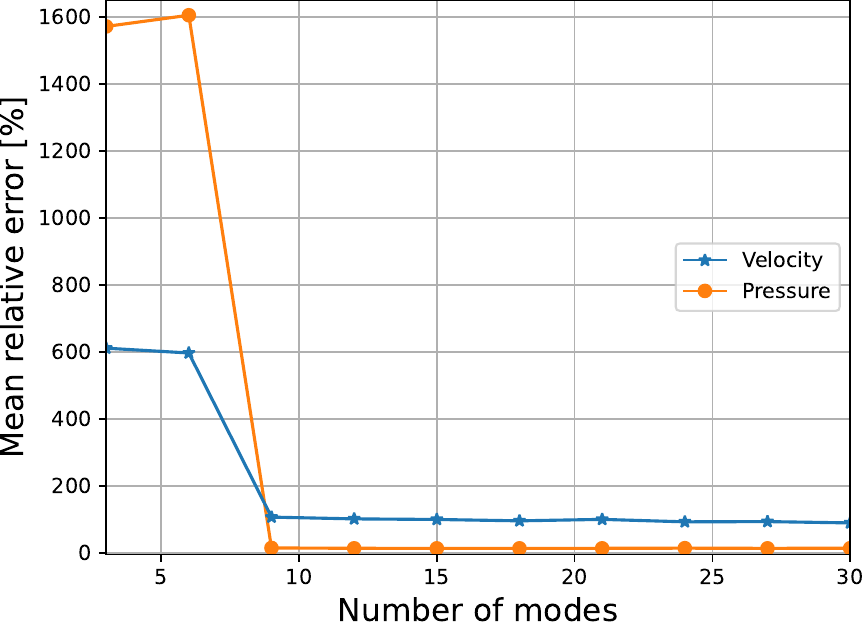}}
	
	\caption{Cumulative energy of the eigenvalues \eqref{eq:energy} at (a) $Re=2000$, (c) $Re=6500$ and mean relative error at  (b) $Re=2000$, (d) $Re=6500$ for pressure and velocity. 
 }
	\label{eig-and-err}
\end{figure}

Like in the previous section, first we will present the results for strategy \eqref{constant_kernel} and then proceed with strategy \eqref{linear_kernel}.

\begin{figure}[htb!]
	\centering
    \subfloat[][Case $Re=2000$\label{fig:err_rel_kernel_const_u_2000}]{\includegraphics[width=.48\textwidth]{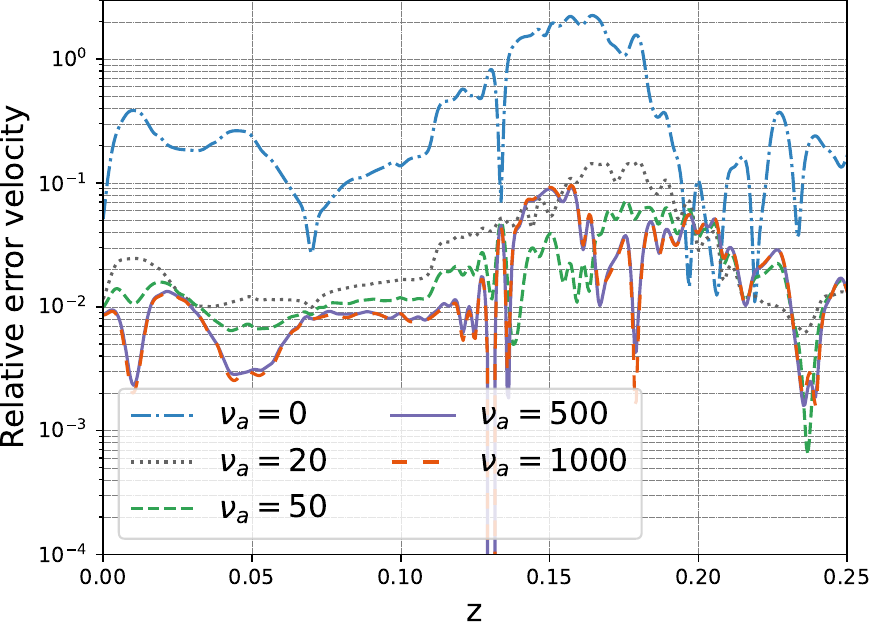}}
 	\subfloat[][Case $Re=6500$ \label{fig:err_rel_kernel_const_u_6500}]{\includegraphics[width=.48\textwidth]{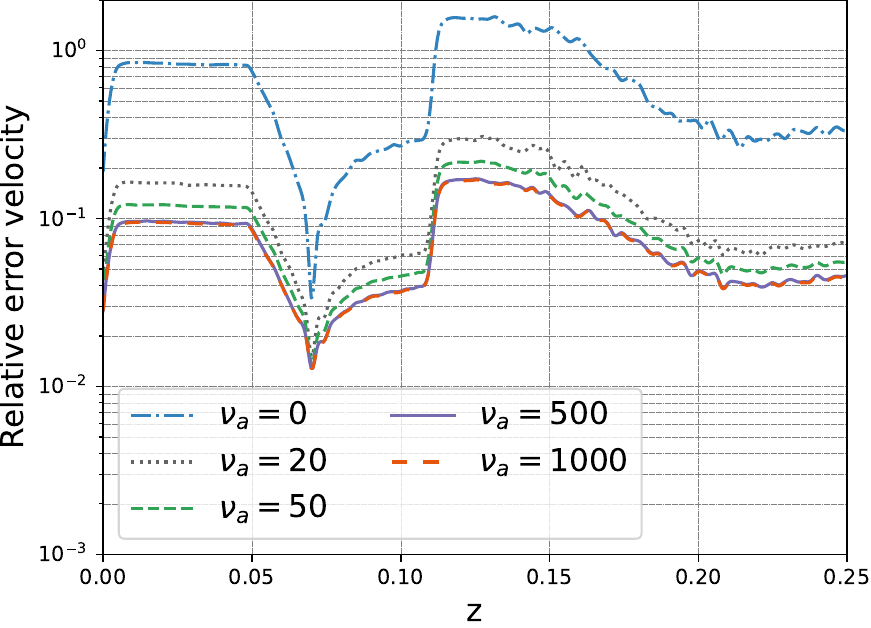}}\\

	\caption{Relative errors of the ROM axial velocity 
 with respect to the FOM solution for constant artificial viscosity
 $\nu_a = 0, 20, 50, 500, 1000$ in the case of (a) $Re=2000$ and (b)  $Re=6500$.
}
	\label{err_rel_kernel_const_2000_6500_u}
\end{figure}

\vskip .3cm
\noindent {\bf Constant artificial viscosity.}
Fig.~\ref{err_rel_kernel_const_2000_6500_u} shows the relative errors for the axial velocity for $\nu_a=0, 20, 50, 500, 1000$ in both the $Re=2000$ and $Re=6500$ case. 
Similarly to the $Re=3500$ case, the error decreases as $\nu_a$ increases. 
We observe that, upon reaching a certain value ($\nu_a=500$), higher values
of artificial viscosity do not impact the ROM solution in both cases, with the errors becoming larger in the sudden expansion.
We observe that the mean (over $z$) error in Fig.~\ref{fig:err_rel_kernel_const_u_2000} is around
$10^{-2}$ for the transitional case, whereas 
the mean  error in Fig.~\ref{fig:err_rel_kernel_const_u_6500}
is one order of magnitude larger, roughly $10^{-1}$. 
It is somewhat expected that the effectiveness of a simple
stabilization approach like \eqref{constant_kernel}
decreases as the Reynolds number is increased past a certain value. 

\begin{figure}[htb!]
	\centering
    \subfloat[][Case $Re=2000$\label{fig:err_rel_kernel_const_p_2000}]{\includegraphics[width=.48\textwidth]{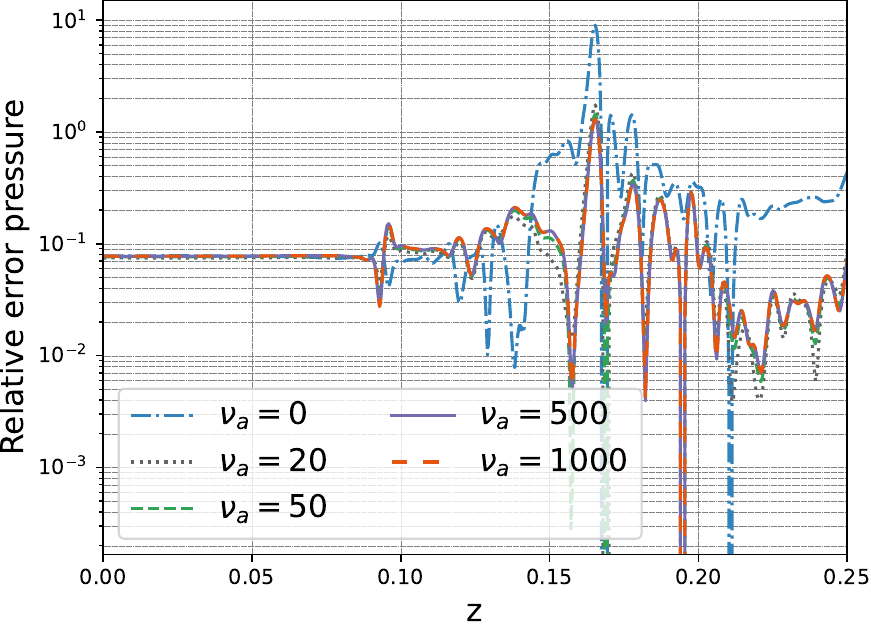}}
 	\subfloat[][Case $Re=6500$ \label{fig:err_rel_kernel_const_p_6500}]{\includegraphics[width=.48\textwidth]{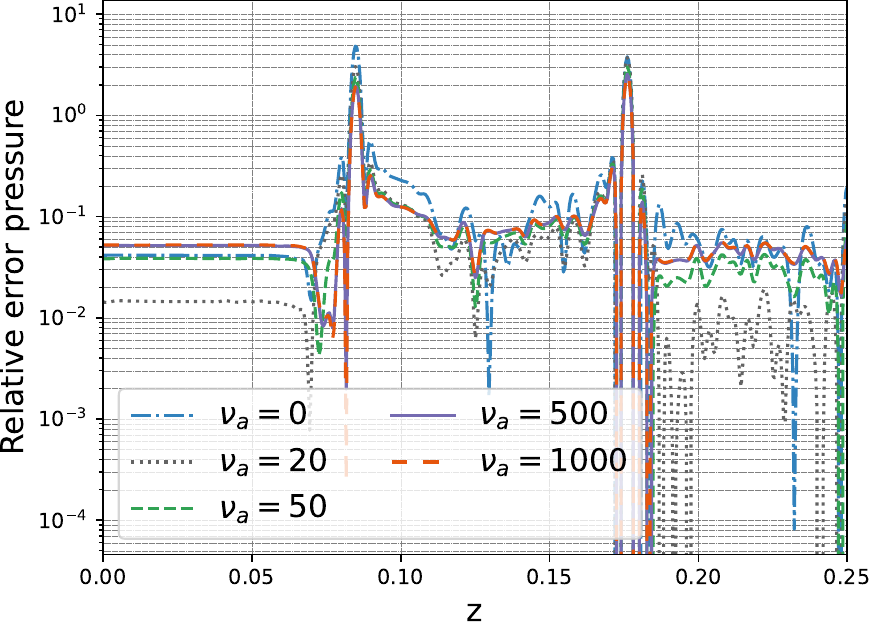}}\\

    \subfloat[][Case $Re=2000$\label{fig:err_abs_kernel_const_p_2000}]{\includegraphics[width=.48\textwidth]{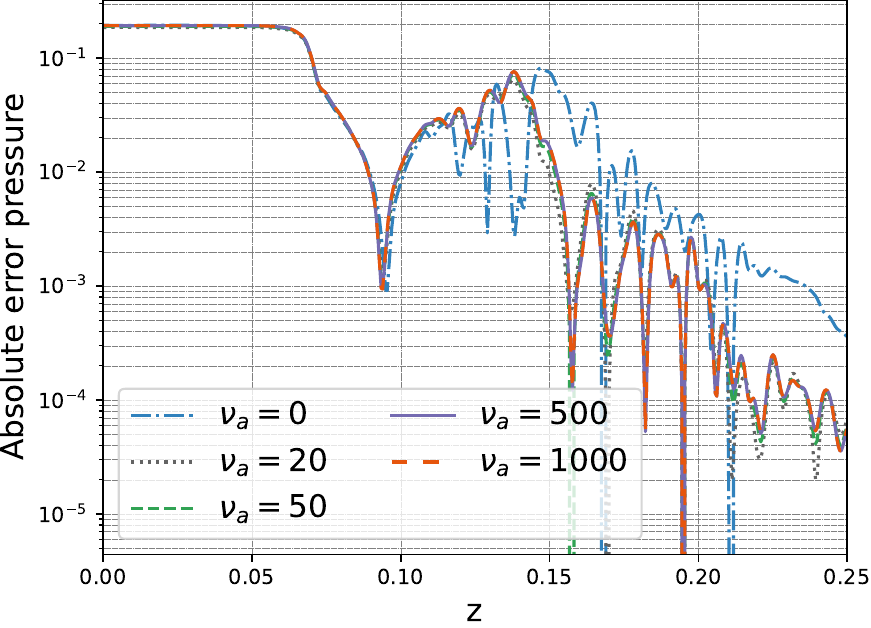}}
 	\subfloat[][Case $Re=6500$ \label{fig:err_abs_kernel_const_p_6500}]{\includegraphics[width=.48\textwidth]{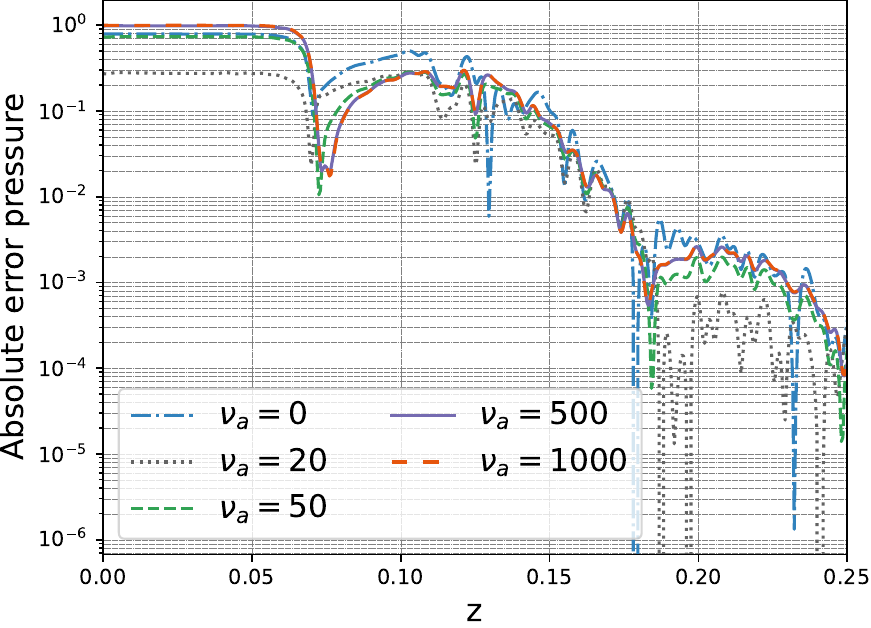}}\\
	\caption{Relative (top) and absolute (bottom) 
 errors of the ROM pressure with respect to the FOM solution for constant artificial viscosity $\nu_a = 0, 20, 50, 500, 1000$.}
	\label{err_rel_kernel_const_2000_6500_p}
\end{figure}

\begin{figure}[htb!]
	\centering
    \subfloat[][Case $Re=2000$\label{fig:kernel_const_2000_1}]{\includegraphics[height=.36\textwidth]{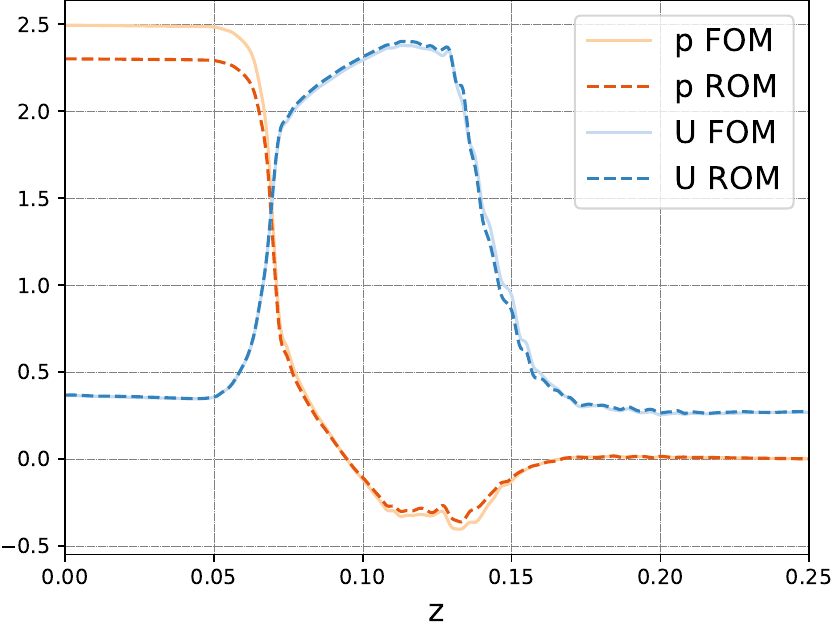}}
 	\subfloat[][Case $Re=6500$\label{fig:kernel_const_6500_1}]{\includegraphics[height=.36\textwidth]{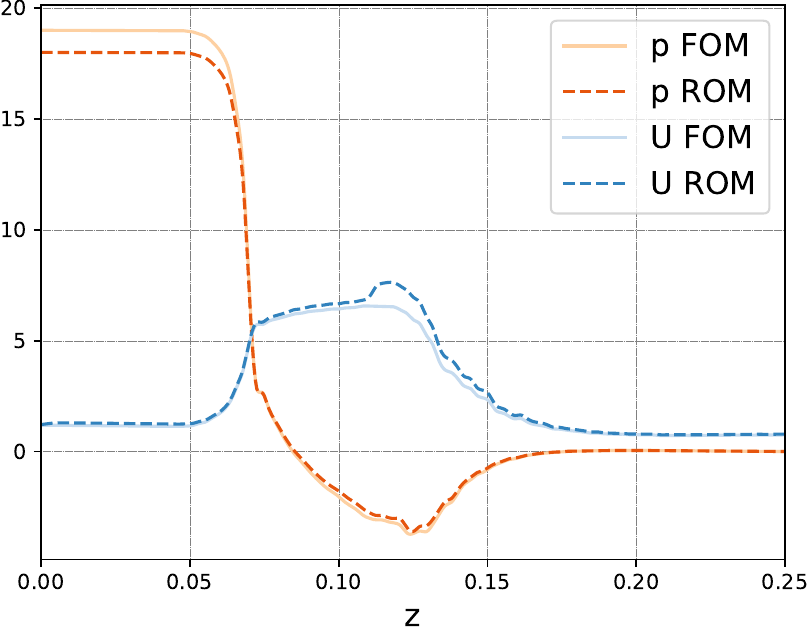}}
	\caption{Comparison of FOM and ROM time-averaged axial velocity and pressure with constant viscosity $\nu_a = 1000$ for (a) $Re=2000$ and (b) $Re=6500$.}
	\label{kernel_const_2000_6500}
\end{figure}

Fig.~\ref{err_rel_kernel_const_2000_6500_p} shows relative 
and absolute errors of the pressure  for the same values of  $\nu_a$ as in Fig.~\ref{err_rel_kernel_const_2000_6500_u}. As expected, the error of the pressure does not varies significantly as $\nu_a$ is increased. We note that the mean (over $z$) relative errors for the pressure are
around $10^{-1}$ for both $Re=2000$ and $Re=6500$, with 
isolated peaks that are due to pressure values near zero. 
In fact, the absolute errors in correspondence to those peaks
are small. See Fig. \ref{fig:err_abs_kernel_const_p_2000} and \ref{fig:err_abs_kernel_const_p_6500}. Therefore, we can conclude that overall the ROM solution has a satisfactory
level of accuracy. This is confirmed by Fig.~\ref{kernel_const_2000_6500}, which reports 
the time-averaged axial velocity and pressure along 
the $z$-axis for $\nu_a=1000$. 

\begin{figure}[htb!]
    \centering
    \begin{minipage}{.38\textwidth}
        \centering
        \begin{overpic}[width=1\textwidth]{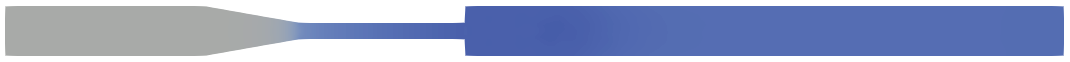} 
        \put(25,7){p FOM at $Re=2000$}
        \end{overpic}\\\vspace{4ex}
        \begin{overpic}[width=1\textwidth]{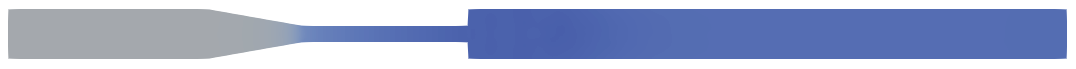} 
        \put(25,7){p ROM at $Re=2000$}
        \end{overpic}
    \end{minipage}~~
        \begin{minipage}{.38\textwidth}
        \centering
        \begin{overpic}[width=1\textwidth]{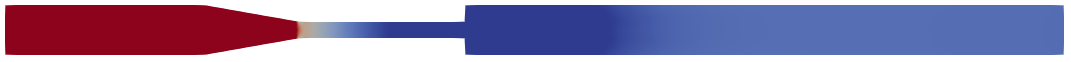} 
        \put(25,7){p FOM at $Re=6500$}
        \end{overpic}\\\vspace{4ex}
        \begin{overpic}[width=1\textwidth]{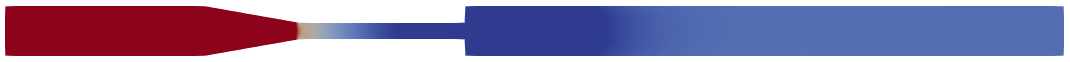} 
        \put(25,7){p ROM at $Re=6500$}
        \end{overpic}
    \end{minipage}%
    \begin{minipage}{0.2\textwidth}
        \centering
        \includegraphics[width=0.5\linewidth]{img/legend_p_nostab.png}
    \end{minipage}
\caption{
Qualitative comparison of the ROM time-averaged axial pressure obtained with constant artificial viscosity $\nu_{a}=1000$ with the corresponding FOM solution for $Re=2000$ (left)
and $Re=6500$ (right).}
\label{stabvel2000P}
\end{figure}

\begin{figure}[htb!]
    \centering
    \begin{minipage}{.38\textwidth}
        \centering
        \begin{overpic}[width=1\textwidth]{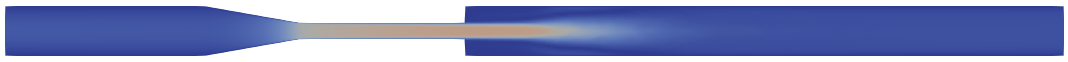} 
        \put(25,7){U FOM at $Re=2000$}
        \end{overpic}\\\vspace{4ex}
        \begin{overpic}[width=1\textwidth]{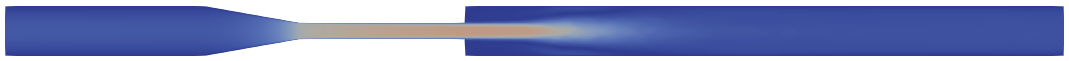} 
        \put(25,7){U ROM at $Re=2000$}
        \end{overpic}
    \end{minipage}~~
        \begin{minipage}{.38\textwidth}
        \centering
        \begin{overpic}[width=1\textwidth]{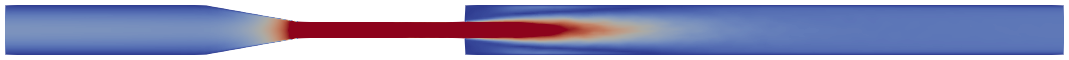} 
        \put(25,7){U FOM at $Re=6500$}
        \end{overpic}\\\vspace{4ex}
        \begin{overpic}[width=1\textwidth]{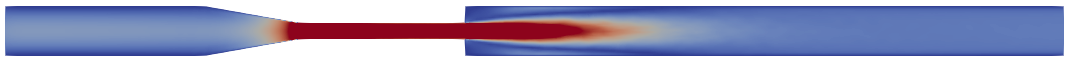} 
        \put(25,7){U ROM at $Re=6500$}
        \end{overpic}
    \end{minipage}%
    \begin{minipage}{0.2\textwidth}
        \centering
        \includegraphics[width=0.5\linewidth]{img/legend_u_nostab.png}
    \end{minipage}
\caption{
Qualitative comparison of the ROM time-averaged axial 
velocity obtained with constant artificial viscosity $\nu_{a}=1000$ with the corresponding FOM solution
for $Re=2000$ (left) and $Re=6500$ (right).}
\label{stabvel2000U}
\end{figure}

We conclude with a qualitative comparison of the 
FOM and ROM time average pressure and velocity 
in Fig. \ref{stabvel2000P} and \ref{stabvel2000U}, 
respectively. We have used the same color bar for 
all Reynolds numbers ($Re = 2000, 3500, 6500$) 
to facilitate the comparisons. We see a good match
between ROM and FOM solutions for both $Re = 2000$
and $Re = 6500$ when $\nu_a=1000$. 

\begin{figure}[htb!]
	\centering
    \subfloat[][Case $Re=2000$\label{fig:err_rel_kernel_linear_u_2000}]{\includegraphics[width=.48\textwidth]{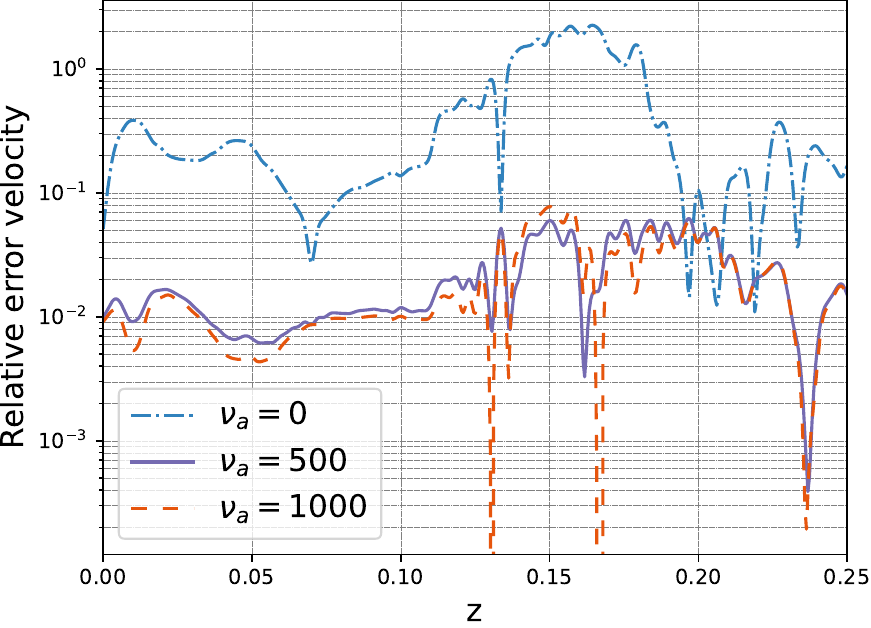}}
 	\subfloat[][Case $Re=6500$ \label{fig:err_rel_kernel_linear_u_6500}]{\includegraphics[width=.48\textwidth]{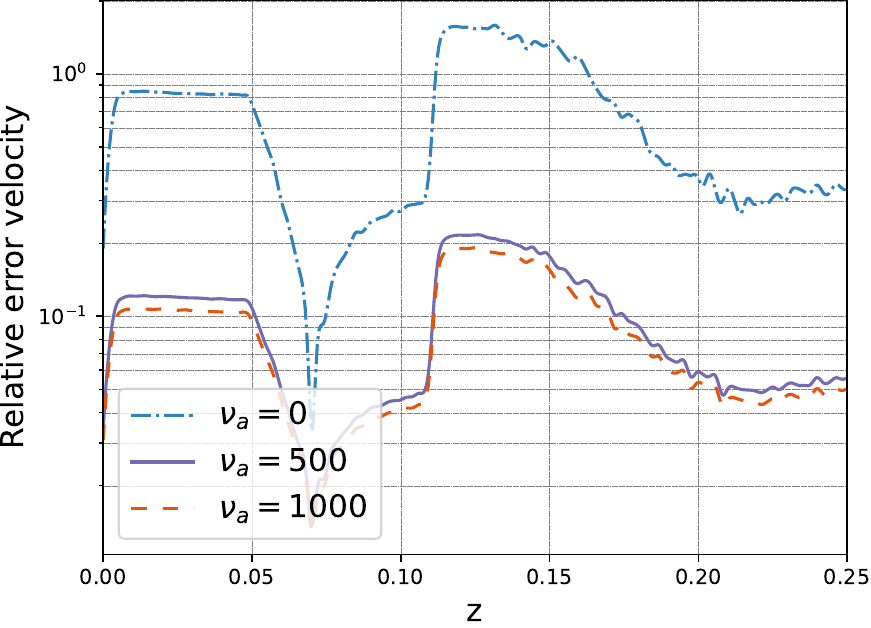}}\\

	\caption{Relative errors of the ROM axial velocity 
 with respect to the FOM solution for the coefficient dependent viscosity
 $\nu_a = 0, 500, 1000$ in the case of (a) $Re=2000$ and (b)  $Re=6500$.}
	\label{err_rel_kernel_linear_2000_6500_u}
\end{figure}

\vskip .3cm
\noindent {\bf Modal coefficient-dependent artificial viscosity.}
Fig.~\ref{err_rel_kernel_linear_2000_6500_u} shows the relative errors for the axial velocity for $\nu_a=0, 500, 1000$ when the coefficient-dependent stabilization approach \eqref{linear_kernel} is used for the $Re=2000$ and $Re=6500$. 
The results closely resemble those of the constant viscosity approach in the sense that: i) the reduced solution that remains unchanged once a given value of artificial viscosity is attained ($\nu_a\ge500$) and ii) the mean (over $z$) error in Fig.~\ref{fig:err_rel_kernel_linear_u_2000} is around
$10^{-2}$ for the transitional case, whereas 
the mean  error in Fig.~\ref{fig:err_rel_kernel_linear_u_6500}
is one order of magnitude larger, roughly $10^{-1}$.  

\begin{figure}[htb!]
	\centering
    \subfloat[][Case $Re=2000$\label{fig:err_rel_kernel_linear_p_2000}]{\includegraphics[width=.48\textwidth]{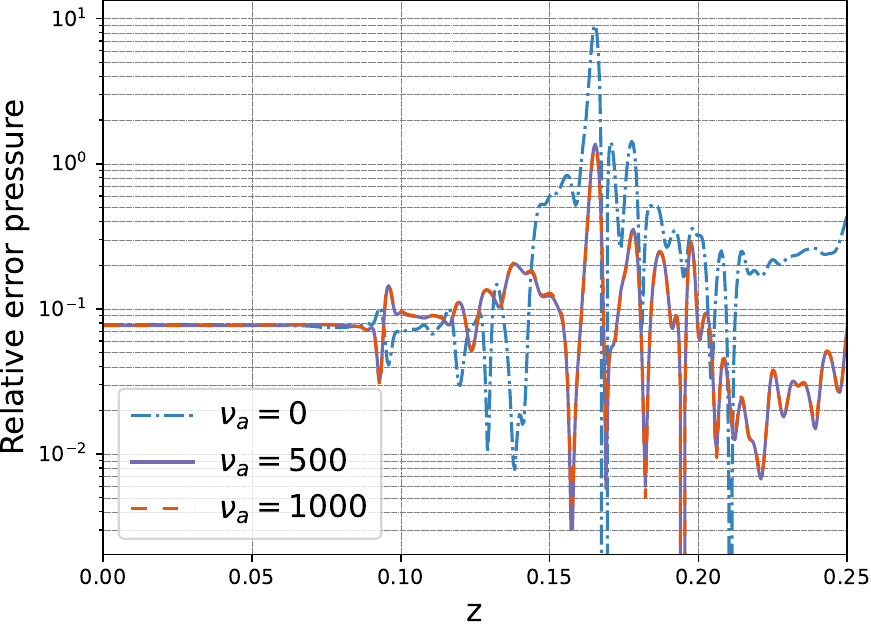}}
 	\subfloat[][Case $Re=6500$ \label{fig:err_rel_kernel_linear_p_6500}]{\includegraphics[width=.48\textwidth]{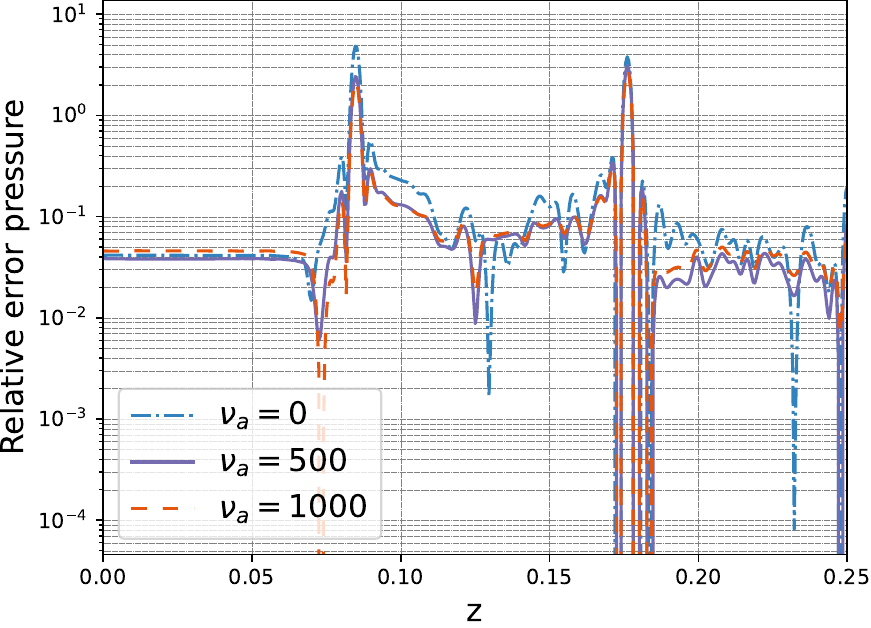}}\\

    \subfloat[][Case $Re=2000$\label{fig:err_abs_kernel_linear_p_2000}]{\includegraphics[width=.48\textwidth]{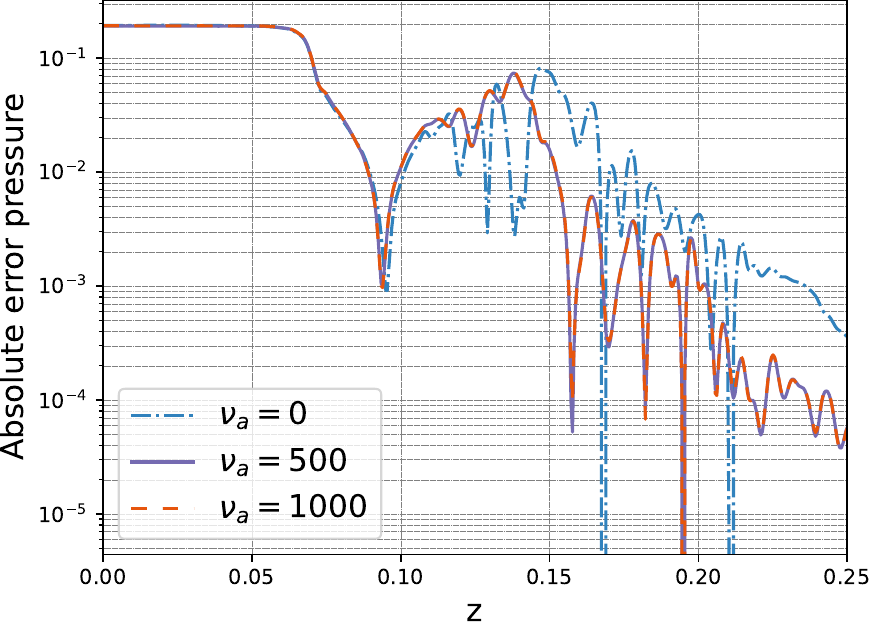}}
 	\subfloat[][Case $Re=6500$ \label{fig:err_abs_kernel_linear_p_6500}]{\includegraphics[width=.48\textwidth]{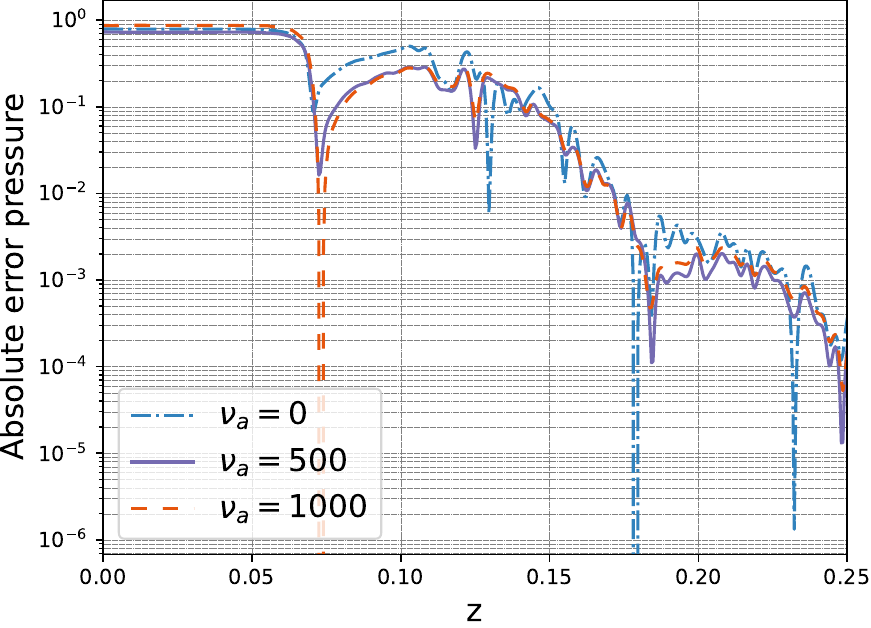}}\\
	\caption{Relative (top) and absolute (bottom) 
 errors of the ROM pressure with respect to the FOM solution for coefficient-dependent viscosity with $\nu_a = 0, 500, 1000$.}
	\label{err_rel_kernel_linear_2000_6500_p}
\end{figure}

The absolute and relative errors of the pressure for $Re=2000$ and $Re=6500$ are shown in Fig.~\ref{err_rel_kernel_linear_2000_6500_p}. A peak appears for $Re=2000$ at $z=0.17$ (see Fig.~\ref{fig:err_rel_kernel_linear_p_2000}) and two peaks appear for $Re=6500$ at $z=0.08, 0.17$ (see Fig.~\ref{fig:err_rel_kernel_linear_p_6500}).
For $Re=6500$, we observed the same peaks when using the constant viscosity strategy: compare Fig.~\ref{fig:err_rel_kernel_linear_p_6500} with Fig.~\ref{err_rel_kernel_const_2000_6500_p}. 
As explained previously, these peaks are 
due to low values of 
the pressure. This is shown also by the plots
of the absolute errors in Fig.~\ref{fig:err_abs_kernel_linear_p_2000} and \ref{fig:err_abs_kernel_linear_p_6500}.

\begin{figure}[htb!]
	\centering
    \subfloat[][Case $Re=2000$\label{fig:kernel_linear_2000_1}]{\includegraphics[height=.36\textwidth]{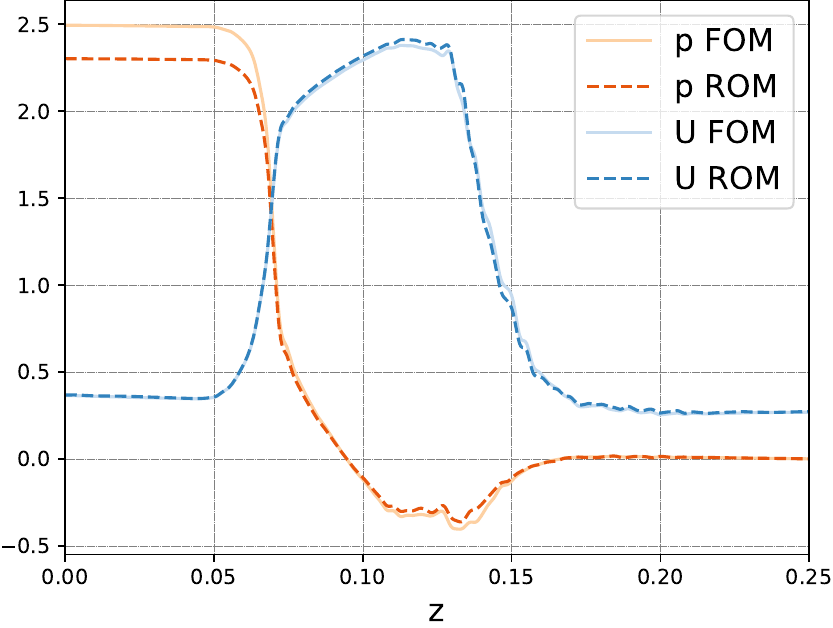}}
 	\subfloat[][Case $Re=6500$\label{fig:kernel_linear_6500_1}]{\includegraphics[height=.36\textwidth]{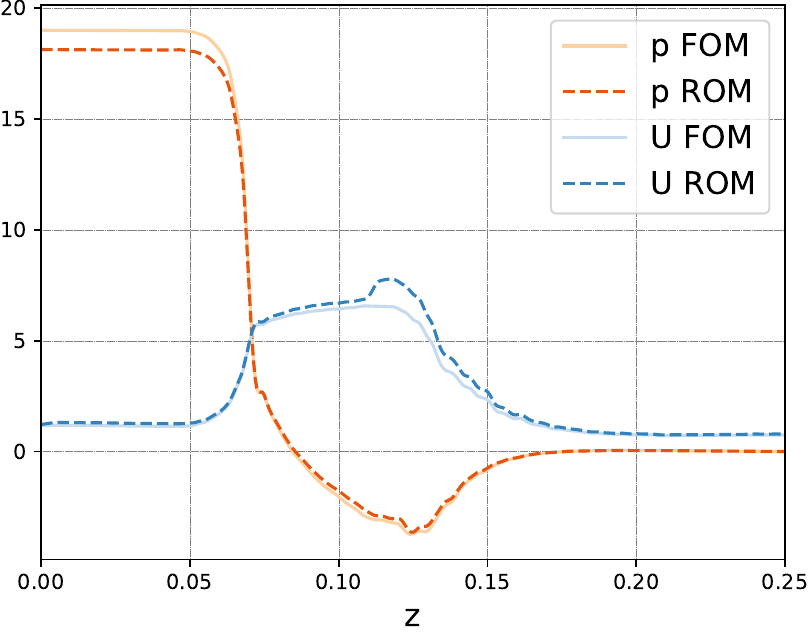}}
	\caption{Comparison of FOM and ROM time-averaged axial velocity and pressure for (a) $Re=2000$ and (b) $Re=6500$ when using the coefficient-dependent viscosity with $\nu_a = 1000$.}
	\label{kernel_linear_2000_6500}
\end{figure}

To conclude, Fig.~\ref{kernel_linear_2000_6500} displays the FOM and ROM time-averaged axial velocity and pressure distributions along the $z$-axis when we set $\nu_a=1000$. Once again, we see a great match for the pressure. For $Re=2000$ there
is also an excellent match for the axial velocity, while for $Re=6500$ the ROM axial velocity deviates from the FOM axial velocity in the expansion channel. This phenomenon gets worse with the coefficient dependent viscosity (compare Fig.~\ref{fig:kernel_linear_6500_1} with Fig.~\ref{fig:kernel_const_6500_1}) 
and as Reynolds increases (compare Fig.~\ref{fig:kernel_linear_6500_1} with Fig.~\ref{fig:kernel_linear_1000}). A further
increase of $\nu_a$ does not improve the match between FOM and ROM axial velocity for $Re=6500$.

\subsection{Physical parameterization}\label{sec:physical_param}
In this section, we will consider the Reynolds number in the throat as a physical 
parameter varying in $[3500,6500]$. Thus, our parameter domain in the
$Re$-$t$ plane is $[3500,6500] \times (0, 1.2)$ s. 
To sample this parameter domain, 
we have considered 9 equi-spaced values of $Re$ ($\Delta Re$ = 375)
and 60 equispaced snapshots in time (every 0.02 s) for each $Re$. 
The set $Re=\{3500, 3875, 4250, 4625, 5375, 5750, 6125, 6500\}$, with the 
all the corresponding snapshots in time, is used to generate the reduced basis, 
while $Re=5000$ is used for testing. 
So, the total number of snapshots employed in the POD is $8\times60=480$.
We remark that this number is a compromise choice between the accuracy 
of the ROM solution and the computational resources. 
In fact, it was not possible to perform a POD with 120 snapshots in time (used in Secs. \ref{sec:3500} and \ref{sec:2000_6500}) for each $Re$ on the machine 
available to us (see last paragraph in Sec.~\ref{sec:3500}).

\begin{figure}[htb!]
	\centering
 	\subfloat[][Cumulative eigenvalues. \label{fig:cum_eig_physical}]{\includegraphics[width=.48\textwidth]{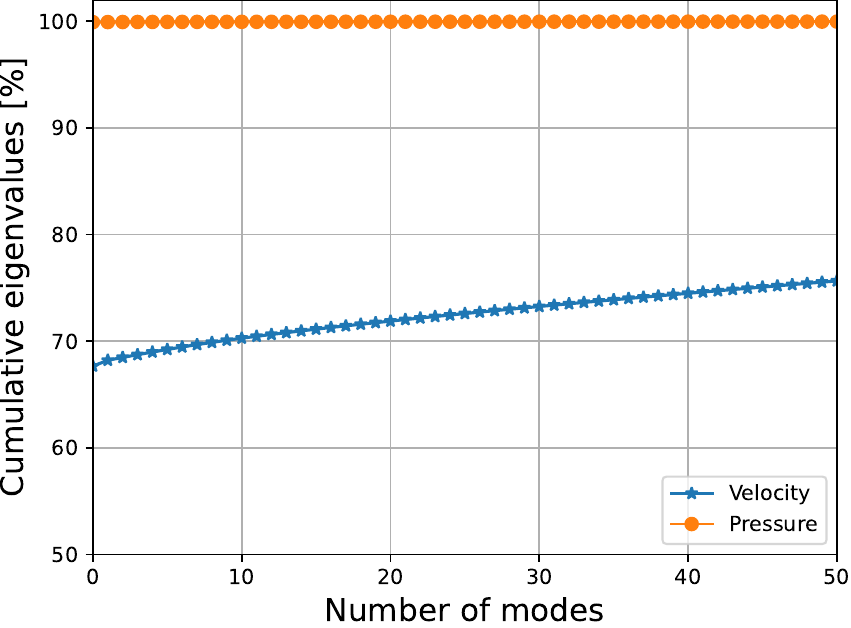}}
    \subfloat[][Mean relative error. \label{fig:err_nmodes_physical}]{\includegraphics[width=.48\textwidth]{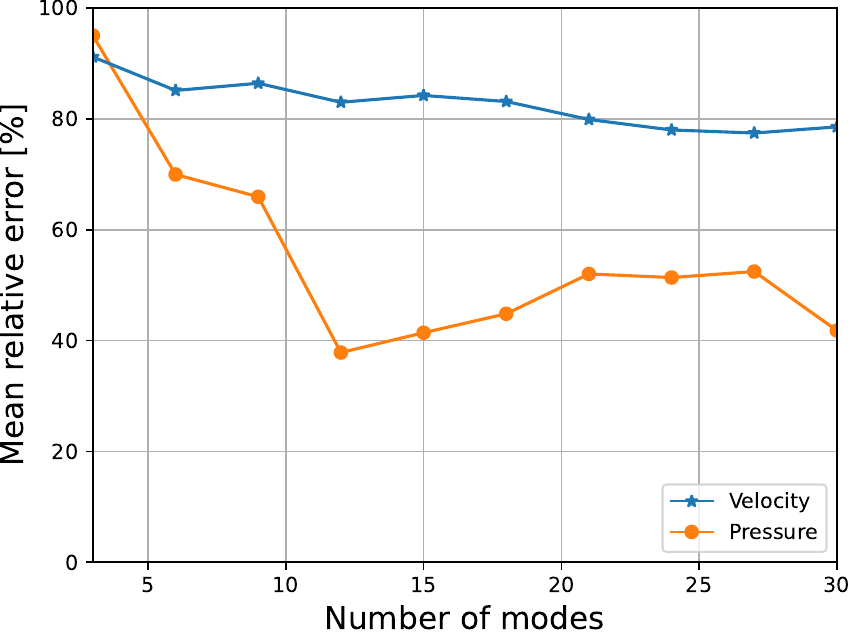}}\\
	
	\caption{(a) Cumulative energy of the eigenvalues \eqref{eq:energy} and (b) mean relative error \eqref{err} for pressure and velocity when $Re$ is a varying parameter. 
 }
	\label{eig-and-err-physical}
\end{figure}

The cumulative energy of the eigenvalues~\eqref{eq:energy} is shown in Fig.~\ref{fig:cum_eig_physical} for velocity and pressure. While the pressure reaches $99.91\%$ of the energy with a single mode, the velocity needs 50 modes to achieve only $75.64\%$. The mean relative error \eqref{err} for pressure and velocity is shown in Fig. \ref{fig:err_nmodes_physical} as the number of modes increase. 
As expected, the error decreases very slowly with the number of modes: for example, when using 30 modes
for each variable, 
we get around $80\%$ error for the velocity error and $40\%$ for the pressure. 

\begin{figure}[htb!]
    \centering
    \includegraphics[width=0.5\textwidth]{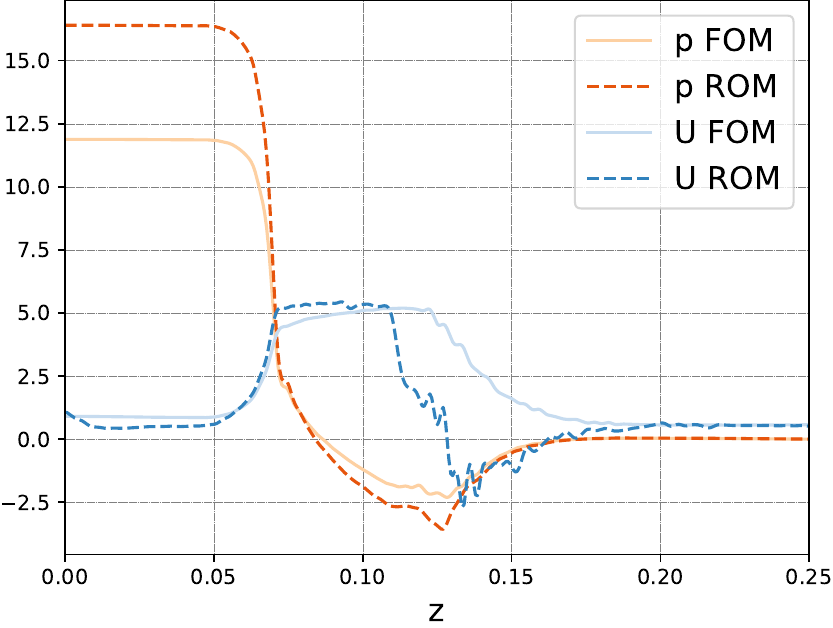}
    \caption{Comparison of FOM and ROM time-averaged axial velocity and pressure 
    at $Re = 5000$ (test value) using 12 modes for each variable and no
    stabilization technique.}
    \label{fig:no-stabilization1}
\end{figure}

To contain the computational cost, 
we cut the number of modes to 12 for each variable. Fig. \ref{fig:no-stabilization1} displays 
the reconstructed time-averaged axial velocity and pressure profiles along the $z$-axis
at $Re = 5000$ if we use 12 modes and no stabilization. 
The large difference between the FOM and ROM curves is expected
given the significant errors depicted in Fig.~\ref{fig:err_nmodes_physical}.

Like in the previous sections, we will present first the results obtained with method \eqref{constant_kernel} and then the results of approach \eqref{linear_kernel}.

\begin{figure}[htb!]
	\centering
    \subfloat[][$\nu_a=1.5$.\label{fig:kernel_const_3_physical}]{\includegraphics[width=.48\textwidth]{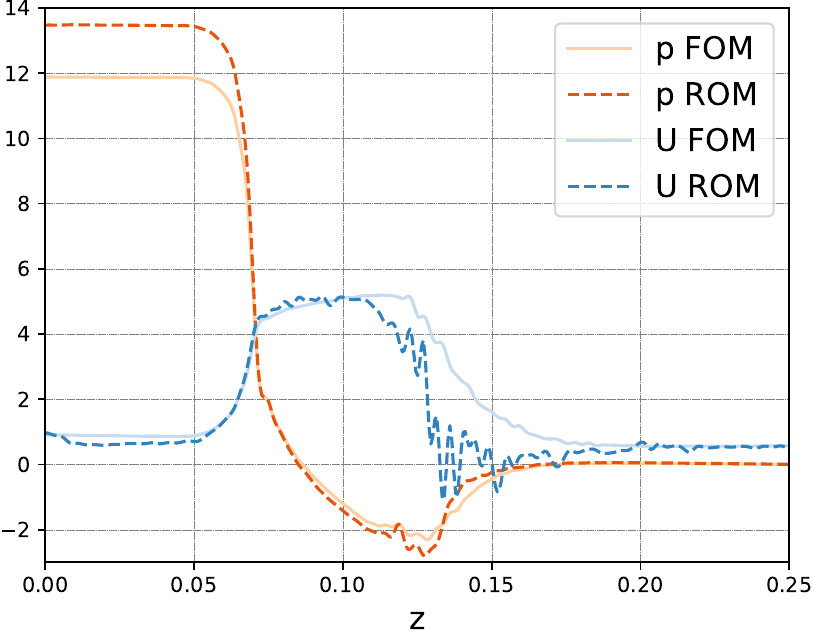}}
    \subfloat[][$\nu_a=3$.\label{fig:kernel_const_3_physical}]{\includegraphics[width=.48\textwidth]{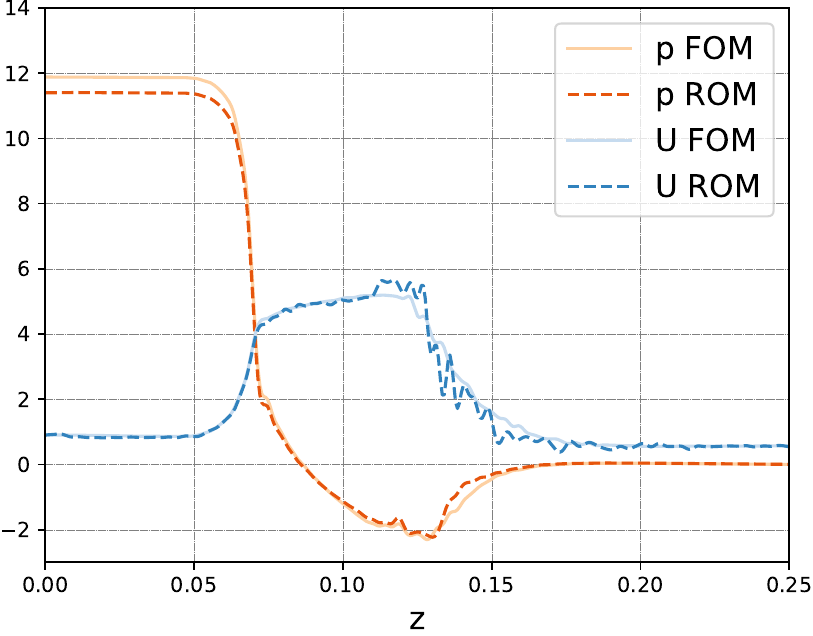}}\\
 	\subfloat[][$\nu_a=20$.\label{fig:kernel_const_20_physical}]{\includegraphics[width=.48\textwidth]{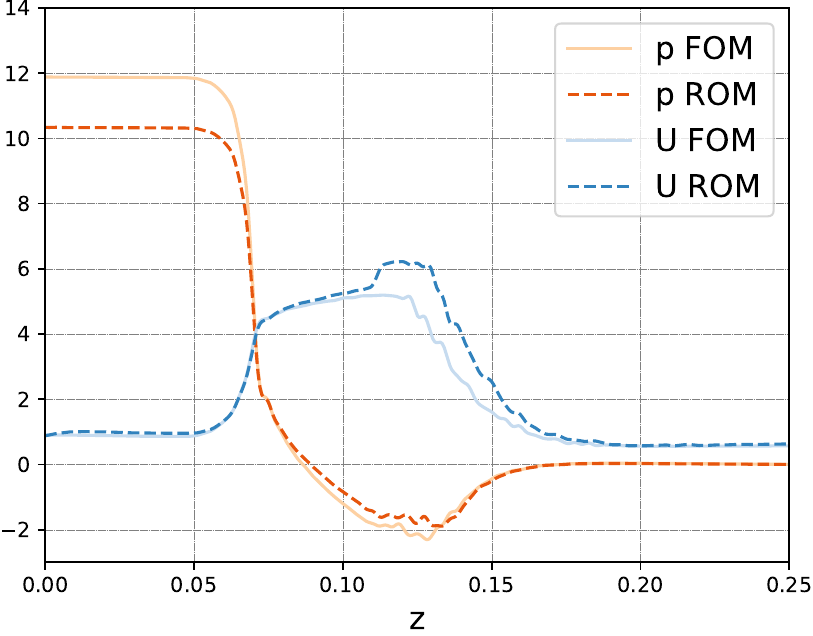}}
    \subfloat[][$\nu_a=1000$.\label{fig:kernel_const_1000_physical}]{\includegraphics[width=.48\textwidth]{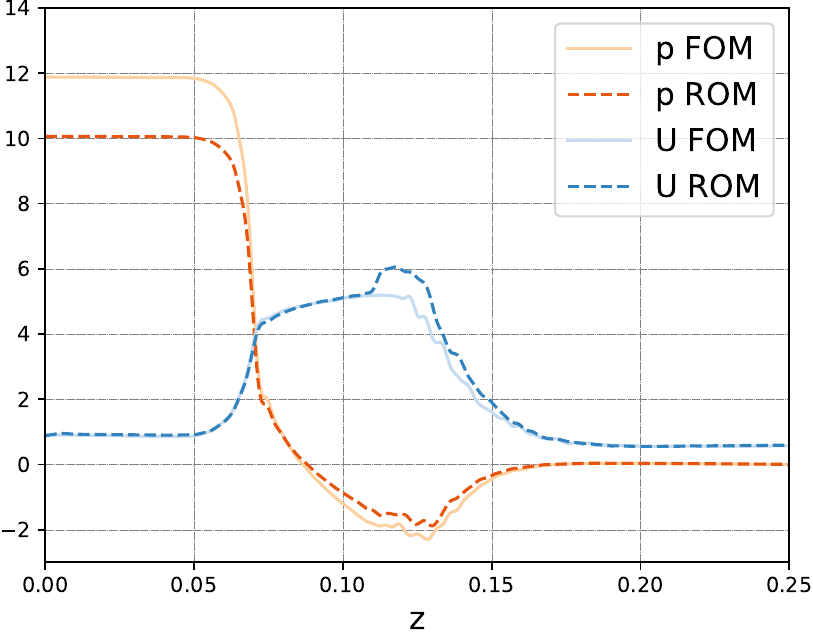}}
	
	\caption{Comparison of FOM and ROM time-averaged axial velocity and pressure 
 for $Re = 5000$ (test value) with constant artificial viscosity for (a) $\nu_a=1.5$, (b) $\nu_a=3$, (c) $\nu_a=20$, and (d) $\nu_a=1000$.}
	\label{kernel_const_cases_physical}
\end{figure}

\vskip .3cm
\noindent {\bf Constant artificial viscosity.}
Fig. \ref{kernel_const_cases_physical} displays the time-averaged axial velocity and pressure along the $z$-axis at $Re = 5000$ for
different values of $\nu_a$. 
We recall that the difference with the previous subsections is that the FOM solutions for 
$Re = 5000$ were not used to generate the reduced basis. Nonetheless, 
the observations are similar to the ones made
for the other Reynolds numbers. 
For low $\nu_a$ (i.e., $\nu_a=1.5$), the velocity is underestimated 
in the sudden expansion and it exhibits oscillations. 
The ROM reconstruction improves considerably for both pressure and velocity when $\nu_a=3$, but the fluctuations of the velocity curve in the sudden expansion are still present. 
For higher values of $\nu_a$, oscillations disappear but
this method is not able to recover properly the velocity at the sudden expansion 
and the pressure in the entrance channel.
Increasing the artificial viscosity above $\nu_a=1000$ 
does not influence the results.

\begin{figure}[htb!]
	\centering
    \subfloat[][Relative error for the velocity\label{fig:err_rel_kernel_const_u_physical}]{\includegraphics[width=.48\textwidth]{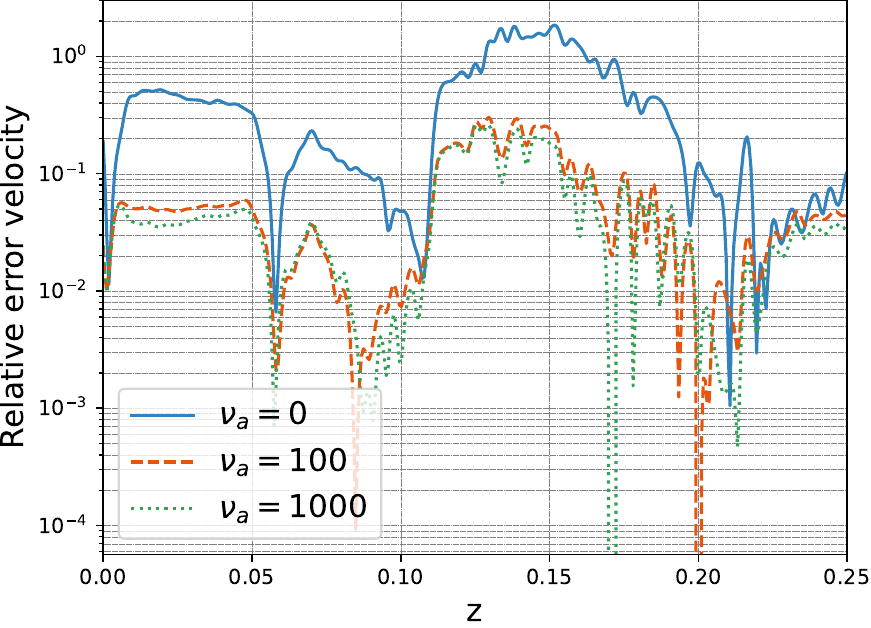}}
 	\subfloat[][Relative error for the pressure\label{fig:err_rel_kernel_const_p_physical}]{\includegraphics[width=.48\textwidth]{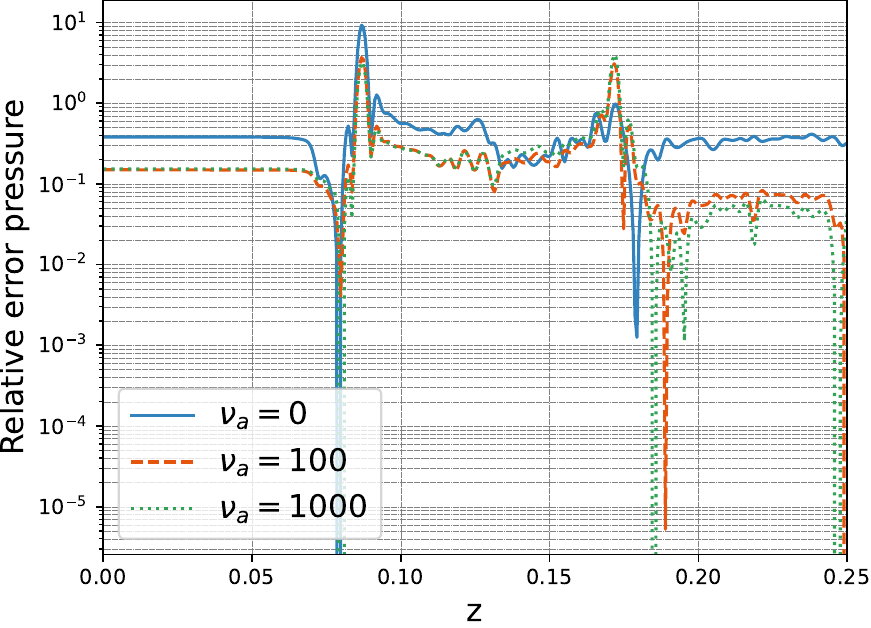}}\\

    \subfloat[][Absolute error for the velocity\label{fig:err_abs_kernel_const_u_physical}]{\includegraphics[width=.48\textwidth]{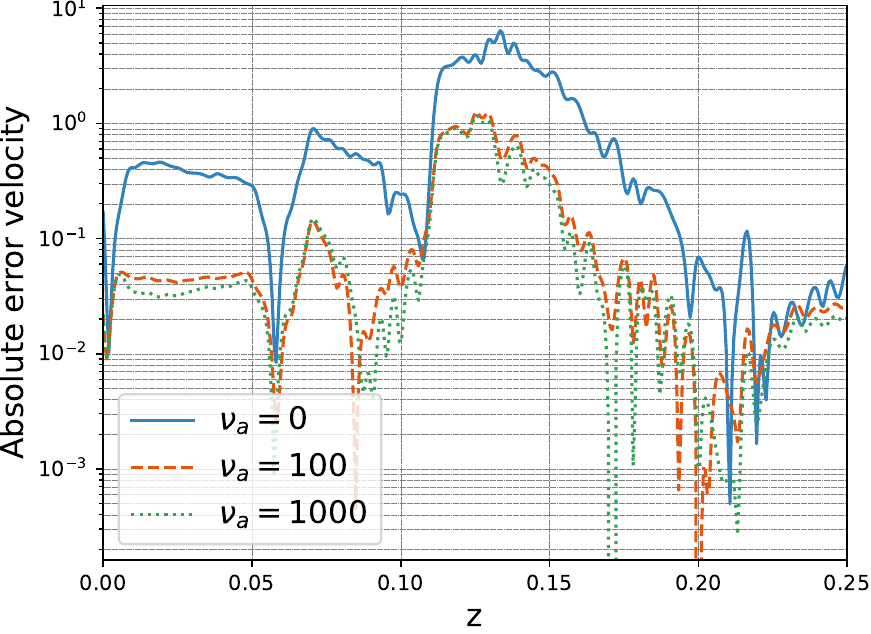}}
 	\subfloat[][Absolute error for the pressure\label{fig:err_abs_kernel_const_p_physical}]{\includegraphics[width=.48\textwidth]{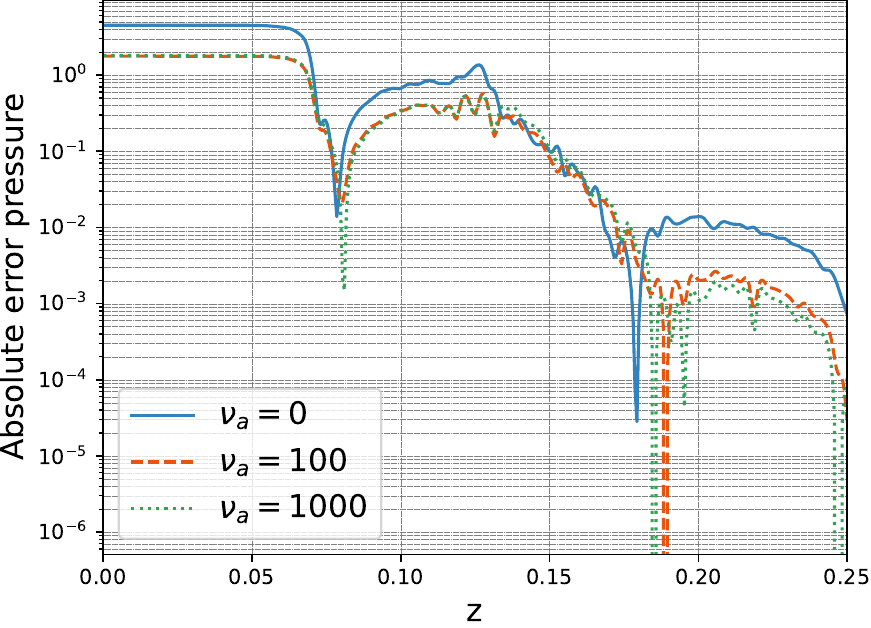}}
	\caption{Relative (top row) and absolute (bottom row) errors of the ROM velocity (left) and pressure (right) with respect to the FOM solution for $Re = 5000$ (test value) 
    when using the constant viscosity approach and $\nu_a = 0, 100, 1000$.}
	\label{err_abs_rel_kernel_const_physical}
\end{figure}

Relative and absolute errors along the $z$-axis for time-averaged pressure 
and velocity are shown in Fig. \ref{err_abs_rel_kernel_const_physical}. Overall, the 
axial velocity error decreases as $\nu_a$ increases, 
with the relative error remaining below $10^{-1}$ along most of the $z$-axis. 
The relative error for the pressure oscillates around $10^{-1}$. 

\begin{figure}[htb!]
    \centering
    \begin{minipage}{.35\textwidth}
        \centering
        \begin{overpic}[width=1\textwidth]{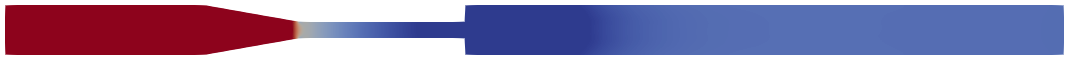} 
        \put(25,7){p FOM}
        \end{overpic}\\\vspace{4ex}
        \begin{overpic}[width=1\textwidth]{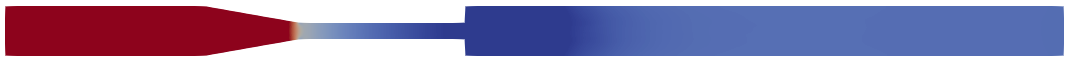} 
        \put(25,7){p ROM}
        \end{overpic}
    \end{minipage}
   \begin{minipage}{0.14\textwidth}
        \centering
        \includegraphics[width=0.85\linewidth]{img/legend_u_nostab.png}
    \end{minipage}
    \begin{minipage}{.35\textwidth}
        \centering
        \begin{overpic}[width=1\textwidth]{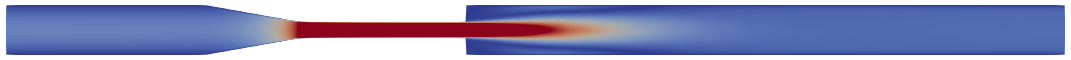} 
        \put(25,7){U FOM }
        \end{overpic}\\\vspace{4ex}
        \begin{overpic}[width=1\textwidth]{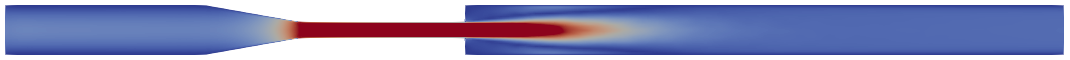} 
        \put(25,7){U ROM}
        \end{overpic}
    \end{minipage}%
    \begin{minipage}{0.14\textwidth}
        \centering
        \includegraphics[width=0.8\linewidth]{img/legend_p_nostab.png}
    \end{minipage}
\caption{
Qualitative comparison of the ROM time-averaged axial pressure (left) and velocity (right) obtained with constant artificial viscosity $\nu_{a}=1000$ with the corresponding FOM solution for $Re = 5000$ (test value).}
\label{stab_physical}
\end{figure}

A qualitative comparison of the FOM and ROM time average pressure and 
velocity for the test point $Re=5000$ is shown in Fig. \ref{stab_physical}.
Overall, we see a good reconstruction of axial velocity and pressure in the entire domain. 

\begin{figure}[htb!]
	\centering
    \subfloat[][$\nu_a=3$.\label{fig:kernel_linear_3_physical}]{\includegraphics[width=.48\textwidth]{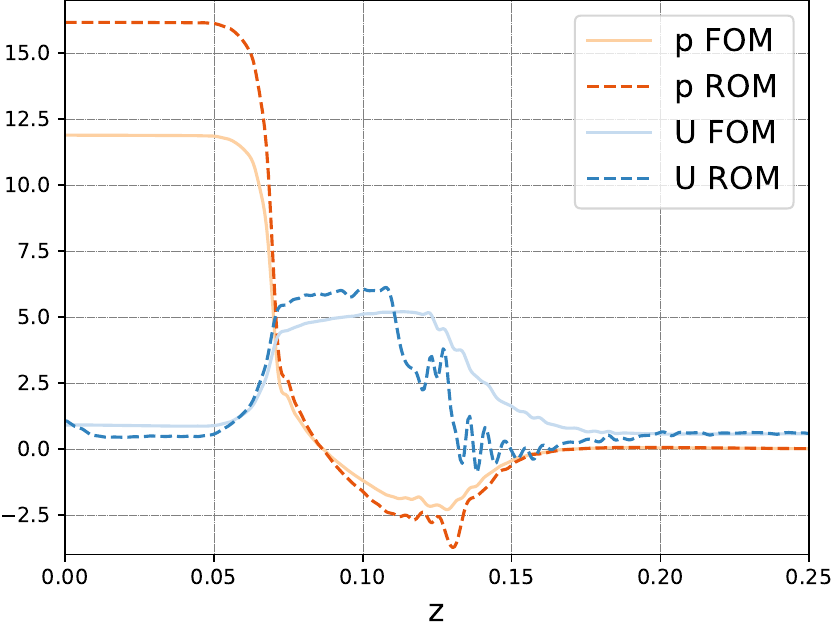}}
    \subfloat[][$\nu_a=10$.\label{fig:kernel_linear_10_physical}]{\includegraphics[width=.48\textwidth]{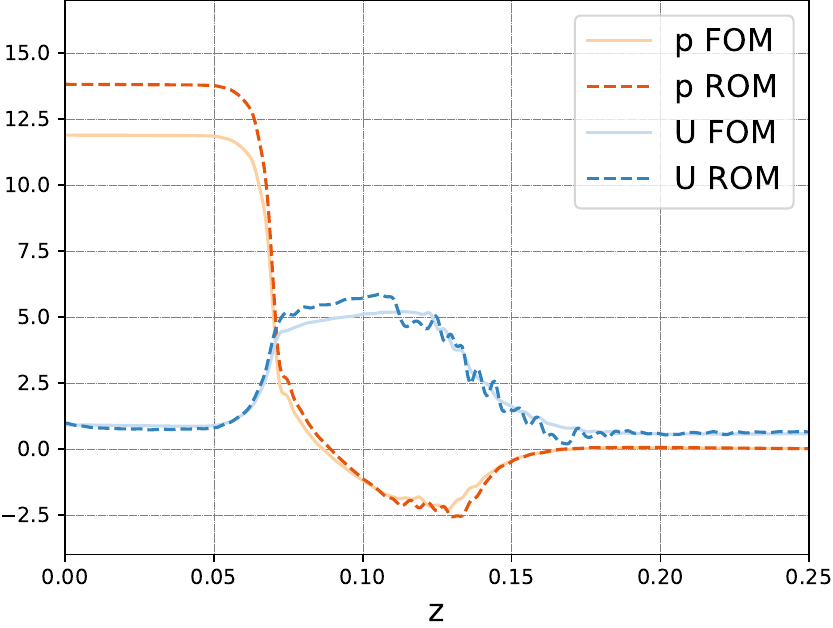}}\\
 	\subfloat[][$\nu_a=50$.\label{fig:kernel_linear_50_physical}]{\includegraphics[width=.48\textwidth]{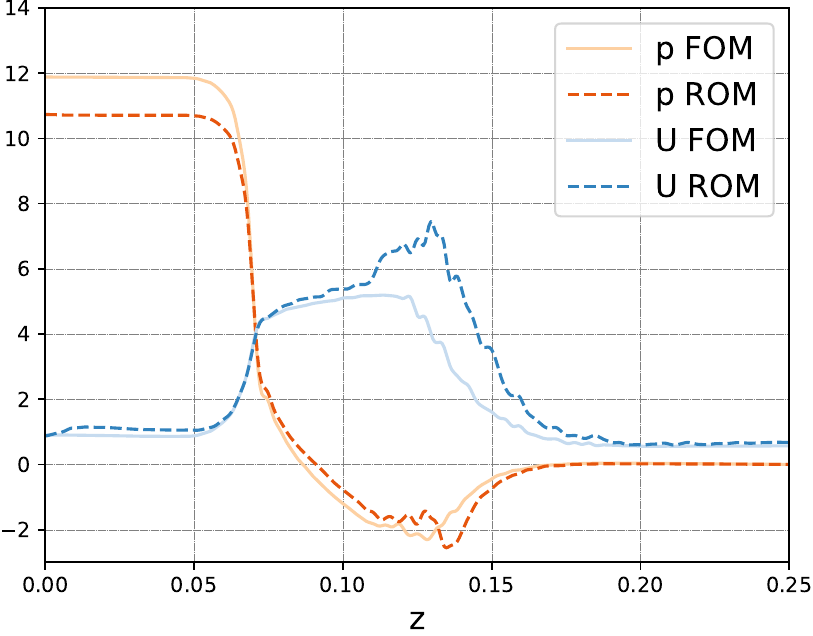}}
    \subfloat[][$\nu_a=1000$.\label{fig:kernel_linear_1000_physical}]{\includegraphics[width=.48\textwidth]{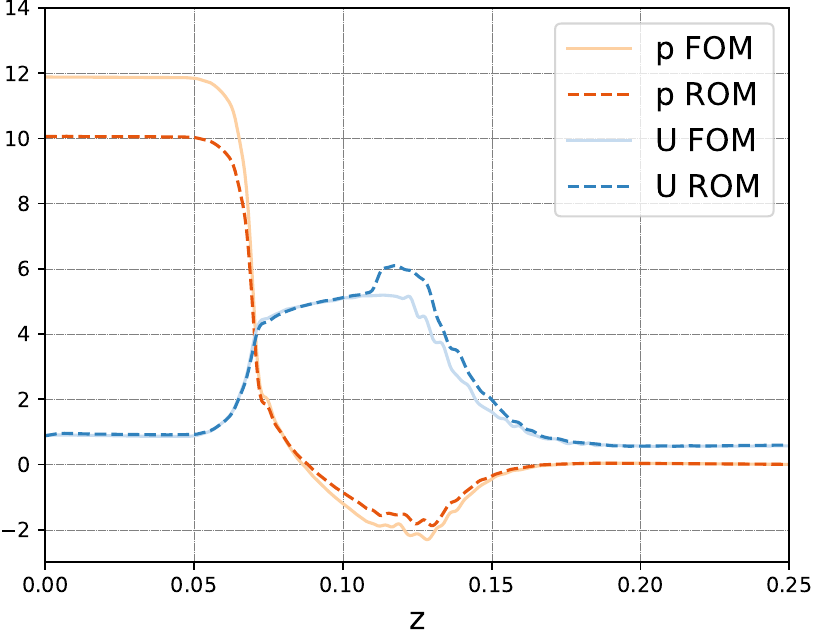}}
	
	\caption{Comparison of FOM and ROM time-averaged axial velocity and pressure at $Re = 5000$ (test value) with coefficient-dependent artificial viscosity for (a) $\nu_a=3$, (b) $\nu_a=10$, (c) $\nu_a=50$, and (d) $\nu_a=1000$.}
	\label{kernel_linear_cases_physical}
\end{figure}

\vskip .3cm
\noindent {\bf Modal coefficient-dependent artificial viscosity.}
The time-averaged axial velocity and pressure along the $z$-axis 
obtained with stabilization technique \eqref{linear_kernel} are depicted 
in Fig.~\ref{kernel_linear_cases_physical}, for $\nu_a=3, 10, 50, 1000$. 
For low values of $\nu_a$ ($\nu_a=3,10$) we see 
oscillations in the sudden expansion, like with 
the constant viscosity approach, but unlike it
we have an overestimation of the velocity in the throat. With $\nu_a=50$, we note a significantly higher ROM velocity compared to the FOM solution in the sudden expansion. Finally, with $\nu_a=1000$ we obtain 
the same reconstruction as with the constant viscosity strategy. Increasing $\nu_a$ does not lead to any improvement in the results. 

\begin{figure}[htb!]
	\centering
    \subfloat[][Relative error for the velocity\label{fig:err_rel_kernel_linear_u_physical}]{\includegraphics[width=.48\textwidth]{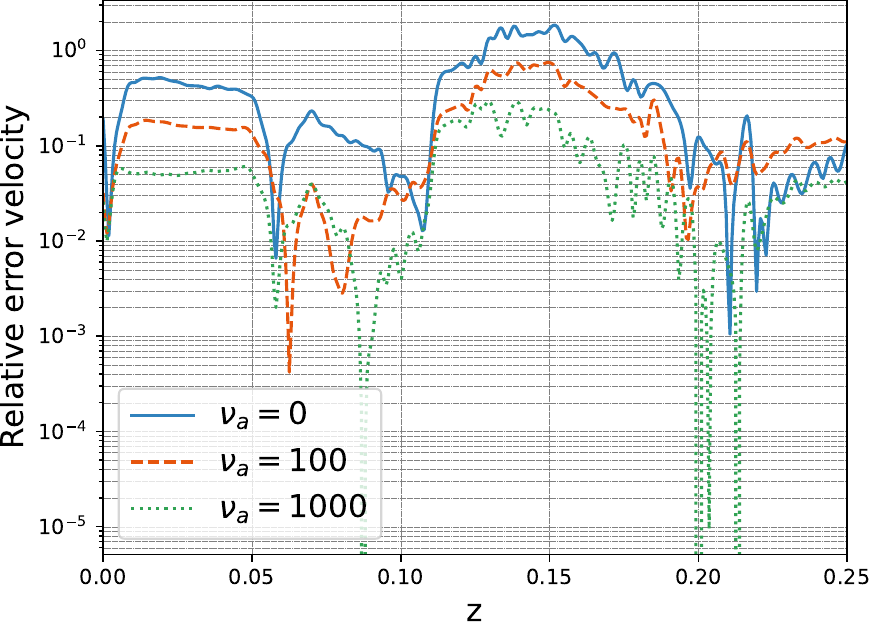}}
 	\subfloat[][Relative error for the pressure\label{fig:err_rel_kernel_linear_p_physical}]{\includegraphics[width=.48\textwidth]{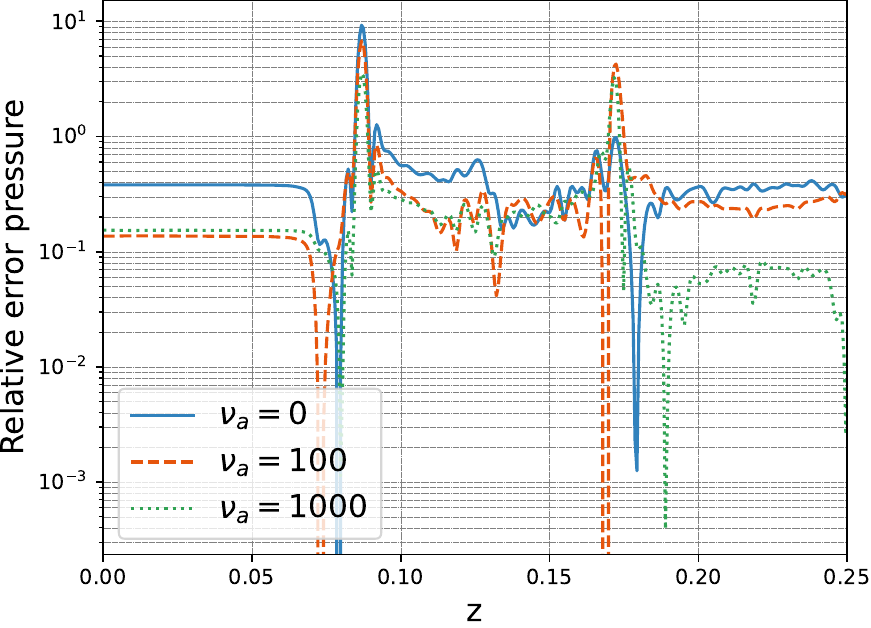}}\\

    \subfloat[][Absolute error for the velocity\label{fig:err_abs_kernel_linear_u_physical}]{\includegraphics[width=.48\textwidth]{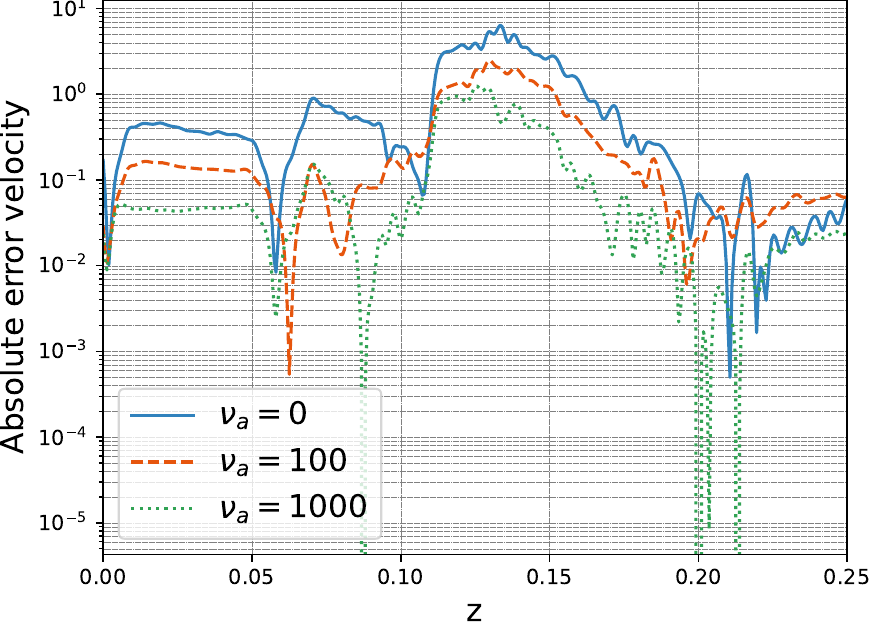}}
 	\subfloat[][Absolute error for the pressure\label{fig:err_abs_kernel_linear_p_physical}]{\includegraphics[width=.48\textwidth]{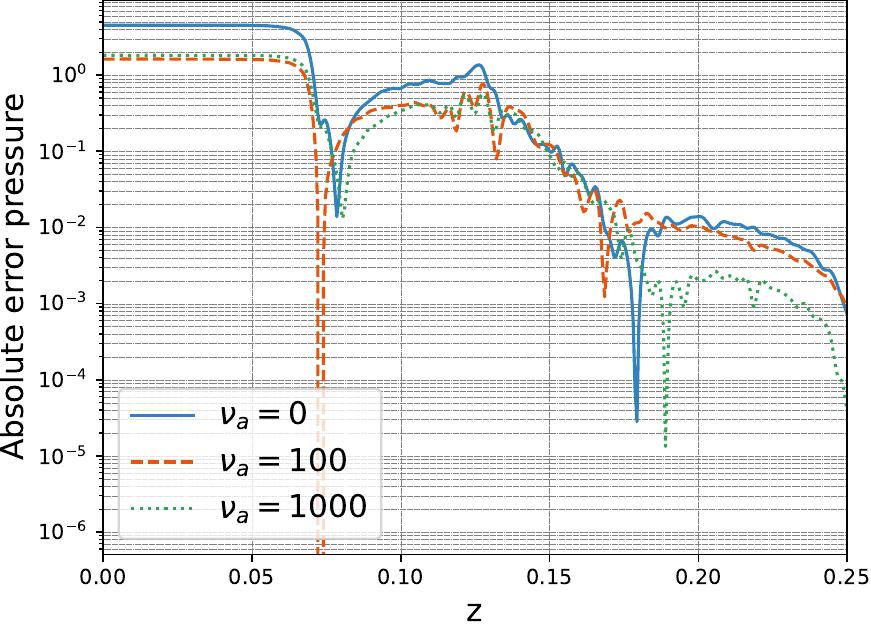}}
	\caption{Relative (top row) and absolute (bottom row) errors of the ROM velocity (left) and pressure (right) with respect to the FOM solution at $Re = 5000$ (test value) for $\nu_a = 0, 100, 1000$ in the coefficient-dependent viscosity approach.}
	\label{err_abs_rel_kernel_linear_physical}
\end{figure}

Fig.~\ref{err_abs_rel_kernel_linear_physical} shows the relative and absolute error along the $z$-axis for time-averaged pressure and velocity for increased values of $\nu_a$. The trend and magnitude of the errors are similar to the case with constant artificial viscosity, although we can observe some differences. In fact, 
while with the constant artificial viscosity method
changing $\nu_a$ from 100 to 1000 made little difference
in the errors, with the coefficient-dependent 
artificial viscosity the difference in the errors is more
significant. 

Finally, we would like to comment on the
overestimation of the ROM axial velocity in the sudden expansion. While for $Re = 2000, 3500$ it can be minimized by increasing $\nu_a$, we saw in Sec.~\ref{sec:2000_6500} and in this section that increasing $\nu_a$ 
does not work for $Re=5000, 6500$.
We experimented with augmenting the training dataset by including $Re=4813,5188$, i.e., we refined 
the sampling near the test point, but that did not help. We suspect that a much finer sampling
of the parameter domain is needed to lower
the ROM axial velocity in the the sudden expansion.
However, that comes with considerable 
computational challenges. 



\section{Conclusions}\label{sec:concl}

We presented two POD-Galerkin reduced order models that
stabilize the ROM solution with artificial diffusion. 
The first ROM introduces a constant artificial viscosity, while the second employs a mode-dependent artificial viscosity. 
We evaluated the accuracy of both ROMs in the simulation of flow 
in a nozzle spanning the transitional and
turbulent regimes (from $Re=2000$ to $Re=6500$).

First, we considered time as the only parameter.
Our numerical results showed that both stabilized ROMs accurately reproduce time-averaged FOM solutions. We obtained a high-accuracy ROM reconstruction for $Re=2000,3500$ when the added artificial viscosity is sufficiently large. In addition,
we showed that for such large values, the constant and coefficient-dependent viscosity ROMs become equivalent. The relative error for the time-averaged axial velocity is around $10^{-2}$, while the relative error for the pressure is approximately $10^{-1}$. For $Re=6500$, the ROMs solution becomes less accurate in the sudden expansion region and the relative error for the time-averaged axial velocity increases to $10^{-1}$. 

Next, we added the Reynolds number to the parameter space.
We observed that the accuracy of the ROM solution dependents on the Reynolds number:
as the Reynolds number gets higher, the ROM solution becomes progressively less accurate 
in the sudden expansion, also seen when time is the only parameter. However, for all $Re$, the relative error for axial velocity ranges between $10^{-1}$ and $10^{-2}$, while the relative error for pressure is around $10^{-1}$. 



The main limitation of these stabilized ROMs appears when the added artificial viscosity is large enough to ensure good accuracy: the ROM solution becomes steady. 
We plan to overcome the limitation through the use of more sophisticated stabilization techniques based on elliptic filters. 

\section*{Acknowledgments}
We acknowledge the support provided by PRIN “FaReX - Full and Reduced order modelling of
coupled systems: focus on non-matching methods and automatic learning” project, PNRR NGE iNEST “Interconnected Nord-Est Innovation Ecosystem” project, INdAM-GNCS 2019–2020 projects
and PON “Research and Innovation on Green related issues” FSE REACT-EU 2021 project. This
work was also partially supported by the U.S. National Science Foundation through Grant No.
DMS-1953535 (PI A. Quaini).
\bibliography{sample.bib}






\end{document}